\documentclass[a4paper,twoside,10pt]{article}

\usepackage[cmex10]{amsmath}
\usepackage{amssymb}
\usepackage{fixltx2e}
\usepackage[T1]{fontenc}
\usepackage[colorlinks,linkcolor=blue,citecolor=blue,urlcolor=blue,pdfauthor=Andreas M. Tillmann]{hyperref}
\usepackage{url}
\usepackage{dsfont}
\usepackage{mathtools}
\usepackage{color}
\usepackage{graphicx}
\usepackage{bm}
\usepackage{amsthm}
\usepackage{tabularx}
\usepackage[twoside,left=2cm,top=2cm,bottom=2cm]{geometry}
\usepackage{tikz}
\usepackage{graphicx}
\usepackage{subfigure}
\usetikzlibrary{calc}
\usepackage{pdfpages}
\usepackage{algorithm}
\usepackage[noend]{algorithmic}

\newtheorem{theorem}{Theorem}[section]

\newtheorem{corollary}[theorem]{Corollary}     		   	
\newtheorem{remark}[theorem]{Remark}

\newcommand{\R}{\mathds{R}}
\newcommand{\Z}{\mathds{Z}}

\newcommand{\Q}{\mathds{Q}}

\renewcommand{\O}{\mathcal{O}}

\newcommand{\cF}{\mathcal{F}}
\newcommand{\cB}{\mathcal{B}}
\newcommand{\cS}{\mathcal{S}}
\newcommand{\cC}{\mathcal{C}}

\newcommand{\cN}{\mathcal{N}}
\newcommand{\cM}{\mathcal{M}}
\newcommand{\cG}{\mathcal{G}}
\newcommand{\cZ}{\mathcal{Z}}

\newcommand{\cD}{\mathcal{D}}
\newcommand{\cE}{\mathcal{E}}
\newcommand{\cV}{\mathcal{V}}

\newcommand{\st}{\,:\,}
\newcommand{\define}{\coloneqq}
\newcommand{\norm}[1]{\lVert {#1} \rVert}
\newcommand{\Norm}[1]{\left\lVert {#1} \right\rVert}
\newcommand{\abs}[1]{\lvert {#1} \rvert}

\newcommand{\ownproblem}[2]{\noindent\textbf{\textsc{{#1}}: }\emph{#2}}

\renewcommand{\P}{\textsf{P}}
\newcommand{\NP}{\textsf{NP}}

\newcommand{\ones}{\mathds{1}}

\begin{document}
\def\gr{(1+sqrt(5))/2}%

%%%%%%%%%%%%%%%%%%%%%%%%%%%%%%%%%%%%%%%%%%%%%%%%%%%%%%%%%%%%%%%%%%%%%%%%%%%%%%%%%%%%%%%%%%%%%%%%%%%%%%%%%%%%%%%%%%%%%%%%%%%%%%%%%%%%%%%%%%%%%%%
%                                            Title page & abstract                                                                            %
%%%%%%%%%%%%%%%%%%%%%%%%%%%%%%%%%%%%%%%%%%%%%%%%%%%%%%%%%%%%%%%%%%%%%%%%%%%%%%%%%%%%%%%%%%%%%%%%%%%%%%%%%%%%%%%%%%%%%%%%%%%%%%%%%%%%%%%%%%%%%%%
\title{Structured Discrete Shape Approximation:\\
  Theoretical Complexity and Practical Algorithm}
\author{Andreas M. Tillmann%
\thanks{A. M. Tillmann is with the Institute for Mathematical Optimization,
  Technische Universit\"at Braunschweig, Germany (e-mail: a.tillmann@tu-bs.de.}%
\and
Leif Kobbelt%
\thanks{L. Kobbelt is with the Visual Computing Institute, RWTH Aachen
  University, Aachen, Germany (e-mail: kobbelt@cs.rwth-aachen.de)}
}%

\maketitle

\markboth
{Structured Discrete Shape Approximation}
{Structured Discrete Shape Approximation}

\begin{abstract}\noindent
  We consider the problem of approximating a two-dimensional shape contour
  (or curve segment) using discrete assembly systems, which allow to build
  geometric structures based on limited sets of node and edge types subject
  to edge length and orientation restrictions. We show that already
  deciding feasibility of such approximation problems is \NP-hard, and
  remains intractable even for very simple setups. We then devise an
  algorithmic framework that combines shape sampling with exact
  cardinality minimization to obtain good approximations using few
  components. As a particular application and showcase example, we discuss
  approximating shape contours using the classical Zometool construction
  kit and provide promising computational results, demonstrating that our
  algorithm is capable of obtaining good shape representations within
  reasonable time, in spite of the problem's general intractability. We
  conclude the paper with an outlook on possible extensions of the
  developed methodology, in particular regarding 3D shape approximation
  tasks.
\end{abstract}

%\begin{keyword}
%  shape approximation, discrete assembly systems, computational complexity, mixed-integer programming, Zometool
%\end{keyword}

%%%%%%%%%%%%%%%%%%%%%%%%%%%%%%%%%%%%%%%%%%%%%%%%%%%%%%%%%%%%%%%%%%%%%%%%%%%%%%%%%%%%%%%%%%%%%%%%%%%%%%%%%%%%%%%%%%%%%%%%%%%%%%%%%%%%%%%%%%%%%%%
%                                            Introduction & Preliminaries                                                                     %
%%%%%%%%%%%%%%%%%%%%%%%%%%%%%%%%%%%%%%%%%%%%%%%%%%%%%%%%%%%%%%%%%%%%%%%%%%%%%%%%%%%%%%%%%%%%%%%%%%%%%%%%%%%%%%%%%%%%%%%%%%%%%%%%%%%%%%%%%%%%%%%
\section{Introduction and Preliminaries}\label{sec:prelim}

We are interested in approximately representing two-dimensional shape
contours by non-self-intersecting structures that consist of a limited
variety of edge and node types which can be connected only in certain
finitely many ways.
The two main goals of this general task are simplicity of the structure
(i.e., using a small number of components) and good approximation quality
with respect to the input contour. Aiming at a good trade-off between these
conflicting goals gives rise to many possible formulations, e.g.,
minimizing the number of components while observing a given approximation
error tolerance or, conversely, minimizing the approximation error under
component budget constraints. Such problem settings are highly relevant in
technical applications where scaffolds or custom support structures are
built from a small variety of standard beams and connectors in order to
reduce fabrication costs (``rationalization''). These structures are often
made by assembling sets of (parallel) 2D profiles. This motivates to
initially focus on the case of 2D contour approximation; we will remark on
possibilities to exploit the methodology introduced here for generalized
setups in 2D and 3D in the conclusions.

To set the stage, we define a \emph{discrete assembly system (DAS)} as
$\cG\define(\cV,\cE,\cB,\cD)$, where $\cV$ is a set of \emph{node types}
($n\define\abs{\cV}$), $\cE$ a set of \emph{edge types}
($m\define\abs{\cE}$),
$\cB\define\{b_e\in\Z_+\st e\in\cE\}$ contains (optional) availability
\emph{budgets} for each edge type (which may be infinite), and
$\cD\define\{ (d_i,V_i,E_i)\st V_i\subseteq\cV, E_i\subseteq\cE,
i=1,\dots,k\}$ is a collection of~$k$ possible \emph{(edge) orientations}
$d_i$ (given, e.g., by %angles or
direction vectors in $\R^2$) along with node types~$V_i$ and edge types
$E_i$ that are \emph{compatible} with~$d_i$, i.e., to a node $v\in V_i$, an
edge $e\in E_i$ can be attached with orientation $d_i$.
(Note that this setup makes the implicit assumption that all nodes of one
type have the same orientation in space; in principle, this could be
generalized by equipping each~$V_i$ with a local coordinate system, but we
do not treat such extensions here.) In particular, generally, not every
orientation is allowed at every node, and not every edge type is allowed
for every orientation. 

Formalizing and realizing the vague task of ``finding good approximations''
of a given shape contour $\cF\subset\R^2$ (or curve segment
$\cC(p_1,p_2)\subset\R^2$ between two points $p_1,p_2\in\R^2$) by means of
a DAS~$\cG$ poses several challenges. Indeed, to even precisely define a
concrete mathematical problem already involves some nontrivial (design)
choices. Besides deciding on how to balance the aforementioned two main
goals of structural simplicity and approximation quality, one has to choose
a measure for the approximation error, deal with the combinatorial
``explosion'' induced by the discrete setup, and decide on how to handle
global positioning and rotation (fixed vs.\ variable) as well as scaling
w.r.t.\ the input shape. Furthermore, it has to be ensured that the sought
DAS construction ``follows'' the given shape contour (i.e., that it is
``\emph{$\cF$-resembling}'') and does not self-intersect in the
plane. Moreover, note also that any approximation error induced by a
$\cG$-representation of $\cF$ or even a single edge cannot be computed
without knowing its actual spatial position (relative to $\cF$), which,
however, is generally not available a priori.

To nevertheless get a grip on the task, we propose to divide the problem
into a ``feasibility part'' and an ``optimality part''. The general idea is
to first find an $\cF$-resembling $\cG$-representation and then rearrange
its edges to reduce the approximation error as far as (locally)
possible. To ensure $\cF$-resemblance, we sample the input shape and
require the DAS construction to place nodes close to the sample points and
to connect neighboring pairs of those nodes by a $\cG$-segment. We call
such a DAS construction \emph{feasible} if it adheres to given type
budgets, and we refer to it as \emph{valid} if it additionally corresponds
to a non-self-intersecting, or ``planar'', structure (i.e., its edges do
not overlap or cross one another).

A simple observation is that long $\cG$-paths (with many edges,
say~$m_\cG$) between two points will ultimately (for $m_\cG\to\infty$)
deviate strongly from any possible given curve segment between the two
points. Moreover, if the curve segment is relatively straight, it stands to
reason that it can be approximated well by a DAS construction that does not
need many struts. Therefore, after sampling the input shape contour, we use
the number of DAS edges between two neighboring nodes as a proxy for the
shape approximation error. Relating the resulting three-phase approach
(sampling being phase one) to the two conflicting goals mentioned earlier,
we thus incorporate the goal of simplicity into the second phase by means
of minimizing the number of utilized DAS components (to acquire an
$\cF$-resembling $\cG$-representation), and explicitly address the goal of
low approximation error in the third phase.

The rest of this paper is organized as follows: In the following two
subsections, we first discuss related problems from the literature, and
then describe a concrete DAS---the so-called Zometool system---that will
serve as a showcase example throughout the remainder of the paper, and for
which we demonstrate with computational experiments (presented in
Section~\ref{sec:experiments}) that our method allows to compute good
approximations within reasonable time in practice. We formally define
several shape contour approximation and point connectivity feasibility
problems and prove their respective \NP-hardness in
Section~\ref{sec:hardness}, before providing (in
Section~\ref{sec:methodology}) a detailed description of the approach that
was briefly outlined above. We conclude the paper with a discussion of some
unresolved aspects and possible ways to utilize the methodology developed
here for the even more challenging 3D shape approximation problem, thus
providing pointers for interesting future research.

\subsection{Related Work}\label{subsec:relatedwork}
In the literature, one can find many problems that are in one way or
another related to the approximation of some given shape contour or curve
segment between two points by the discrete assembly systems considered in
the present paper. We will briefly outline what appear to be the most
closely related works in the following. Notably, to the best of our
knowledge, our key feature of \emph{finitely many edge types/lengths} is
missing in all related previous works, with the exception of those
involving Zometool systems (formally described in the next subsection) or
restricted grid-graph settings, and there appears to be no clear way to
integrate it into any of the algorithmic schemes developed therein.
\begin{description}
\item[Minimum-Link Paths.] Here, the problem is to find a path within a
  given (often polygonal) shape between two specified points that uses as
  few edges as possible (equivalently, the fewest turns). Optimization is
  usually done w.r.t.\ link (edge) number and/or total path length (often
  Euclidean, but other distance measures have been considered); see, e.g.,
  \cite{HershbergerSnoeyink1994,ArkinETAL1991,Piatko1993,MitchellPiatkoArkin1992}.
  Possible constraints include, in particular, a discrete set of admissible
  edge orientations (as in, e.g.,
  \cite{Reich1991,MitchellPolishchukSysikaski2014}), but edges may be
  arbitrarily long. Related further problems include ray-shooting variants
  (turns are restricted to ``reflection'' at the shape boundary,
  cf.~\cite{KostitsynaETAL2016}), art gallery and watchman route problems,
  robot motion planning problems (finding paths that avoid obstacles as in,
  e.g., \cite{MitchellRoteWoeginger1992,Reich1991}) or VLSI routing. A comprehensive
  overview of minimum-link path related problems and algorithms is given in
  the recent survey~\cite{Mitchell2017} (and the extensive list of
  references therein), see also \cite{ChenETAL2001} as a starting point for
  corresponding path-query problems.

\item[Polygonal Approximation.] Here, the problem is often to approximate a
  given polygon by a ``simpler'' one, though other (often convex) shapes
  are sometimes also considered, and objectives may be minimizing the
  enclosed area or the perimeter of the constructed polygonal shape, see,
  for instance,
  \cite{GuibasETAL1991,NilssonETAL1992,MitchellPolishchuk2008,AgarwalSuri1998}.
  As with minimum-link paths, edge type (length) restrictions appear to
  have not been considered in this context.

  A particular subfield of polygonal approximation that, in fact, includes
  some problem classes that are special cases of (certain variants of) the
  more general problems considered in this paper, is called
  \emph{schematization}. In schematization, the goal is to approximate a
  given polygon by another in which all edges are parallel to a given set
  of orientations/directions (but, generally, not restricted in length),
  and the main motivation is to create simplified visually appealing
  representations largely for use in map-making. Thus, schematization is
  applied for the development of public transportation route maps (see,
  e.g.,~\cite{NoellenburgWolff2011,GalvaoETAL2019}) and in cartography
  (cf.~\cite{Meulemans2014,BuchinETAL2016} and references therein), but
  also in, e.g., handling electromagnetic noise in printed circuit boards
  \cite{CiceroneCermignani2012}. Discretized schematization methods embed
  the given polygon in a grid graph, wherein the output polygon then
  corresponds to a selection of edges; see, e.g.,
  \cite{Meulemans2014,BoutsETAL2016,LoefflerMeulemans2017}. These
  grid-based problem formulations can be obtained as special cases of our
  general DAS shape approximation problem by fixing the global positioning
  and rotation, allowing only edges of a single length per direction (thus
  yielding the grid graph structure), and focusing on polygonal input
  shapes. The main approximation error measures considered in discrete
  schematization are the symmetric difference, the Hausdorff distance, and
  the (possibly discretized) Fr\'{e}chet distance. While the special-case
  relationship provides some useful insights (especially w.r.t.\
  computational complexity, see Section~\ref{sec:hardness}), discretized
  schematization algorithms are inapplicable for the more general DAS
  problems treated in this paper, since they rely heavily on the grid-graph
  representation, which is unavailable for general DASs with or without
  fixed global positioning.
  
\item[Zometool Shape Approximation.] The special DAS case of Zometool
  systems (cf. Section~\ref{subsec:zome} below) and corresponding shape
  approximation problems in 3D space have been considered in,
  e.g.~\cite{Davis2007,HartPicciotto2001,ZimmerKobbelt2014,ZimmerETAL2014}. The
  latter two works use a simulated annealing approach (based on local
  configuration improvements) to approximate a shape boundary in $\R^3$
  that could be adopted to work in two dimensions as well. However, the
  simulated annealing methodology does not provide any convergence or
  approximation/optimality guarantees, and may exhibit considerable runtime
  in practice. These drawbacks partly motivated our developing the method
  proposed in Section~\ref{sec:methodology}, which differs in that it
  involves high-level heuristic model/design choices---in particular, the
  sampling phase---but can, in principle, solve the resul\-ting subproblems
  to global optimality (not restricted to local improvement sequences);
  moreover, the global positioning of the $\cG$-construction is included in
  the optimization here, but fixed in the simulated annealing approach. We
  will give some further explanation why simulated annealing would be
  problematic for the problems considered in the present work after we
  formally define these problems and describe our novel algorithm, see
  Section~\ref{subsubsec:simulatedAnnealing}.
  
  Although a future ``ultimate'' goal is to refine and improve upon solving
  shape approximation problems in three dimensions, we focus on the 2D case
  throughout this paper in order to establish theoretical hardness results
  and a novel algorithmic framework. The former carry over directly to the
  3D case, thus settling the open question of computational complexity in
  that regime as well, while we intend to utilize the latter for handling
  the 3D setting in future work (see Section~\ref{sec:conclusion} for some
  remarks in that regard). For such an extension, local improvement steps
  as employed by the simulated annealing algorithm from
  \cite{ZimmerKobbelt2014,ZimmerETAL2014} may become useful as heuristics
  or solution polishing components. 
\end{description}

There exist very many more problems that are in one way or another loosely
related to discrete assembly system shape approximation, such as similar
path-problems with prescribed angles at the turns or minimizing total turn
angle, restrictions to grid-like graph-related structures, curve smoothing
approaches, visibility polytopes, or skeleton graph computation. As these
are less relevant to the present work, we do not go into more detail
here. Similarly, the body of literature on (3D) shape approximation in
general is too vast to discuss here, and arguably not directly relevant;
also, again, the combined specifics of our DAS seem to set it apart from
other approaches.

\subsection{The Zometool System}\label{subsec:zome}
A particular discrete assembly system that exists as a real-world
construction kit is the \emph{Zome(tool) system}
$\cZ\define(\cV^\cZ,\cE^\cZ,\cB^\cZ,\cD^\cZ)$, cf., e.g.,
\cite{HartPicciotto2001}. Here, we have only one type of nodes, so
$n^\cZ=\abs{\cV^\cZ}=1$; say, $\cV^\cZ=\{z\}$. This node type is originally
defined in 3D, where it has a total of 62 slots at which edges of different
types (referred to as \emph{struts} in this context) can be attached -- 12
pentagonal slots, into which only red struts fit, 20 triangular slots for
yellow struts and 30 rectangular slots for blue struts. The most versatile
node type in 2D, derived from the 3D node type by slicing it so that as
many slots as possible remain, still has 12 slots (corresponding to
orientations in the form of 6 directions and their respective opposites),
with 4 struts per color attachable to it in total (2 orientations per
color). For each color, three lengths are available (defined as the
distance between the midpoints of the two nodes a strut can connect). Thus,
in total, there are nine types of struts
$\cE^\cZ=\cE^\cZ_r\cup\cE^\cZ_y\cup\cE^\cZ_b$, with
$\cE^\cZ_c=\{c_1,c_2,c_3\}$ for each color $c\in\cC\define\{r,y,b\}$ ($r$
for red, $y$ for yellow, $b$ for blue), where w.l.o.g.\ the length of $c_i$
is smaller than that of $c_j$ whenever $i<j$. In particular, for each strut
type, the lengths are related via the golden ratio
$\phi\define(1+\sqrt{5})/2\approx 1.618$ as follows: Identifying, for
simplicity, each $c_i$ directly with its length (so that $c_1<c_2<c_3$ for
all $c\in\cC$), the medium length is then given as $c_2=\phi c_1$ and the
longest as $c_3=\phi c_2$. Note that since $\phi^2=1+\phi$, we have
$c_3=c_1+c_2$. Further nice properties exhibited by the Zome system are
node symmetry (i.e., for each slot to plug in a strut, there is an opposite
slot of the same type in the Zome node $z$, so for each direction, the
opposite direction is also available) and a fixed node orientation (i.e.,
all nodes have the same orientation in space, defined by any one strut used
in the assembled Zome structure). Moreover, assuming w.l.o.g.\ a scaling
such that $b_1=2$ and that a Zome node is placed at the origin $(0,0)$ (or
$(0,0,0)$ for 3-dimensional setups), the coordinates of each point that can
be reached via Zometool components have the form
$(\alpha_1+\phi\beta_1,\alpha_2+\phi\beta_2)$ (or
$(\alpha_1+\phi\beta_1,\alpha_2+\phi\beta_2,\alpha_3+\phi\beta_3)$ in the
3D case), where $\alpha_i,\beta_i\in\Z$ for all $i\in\{1,2\}$ (or
$\{1,2,3\}$, respectively). It is worth emphasizing that the listed
properties are special to the Zome system and will not (in analogous form)
hold for arbitrary DASs in general. However, if we further restrict to a
\emph{Zome subsystem} by excluding certain strut lengths and/or types, the
properties are retained and, naturally, further structural properties can
be observed. Furthermore, note that any Zome path is invariant under
permutations of its elements in the sense that the two nodes connected by
the path are also connected by all other paths that contain the same
struts in an arbitrarily permuted order. (This invariance actually
\emph{does} hold for all DASs.)

\begin{figure}[t]
  \centering
    \begin{tikzpicture}[scale=1]
      \draw[color=red,line width=0.8mm] (0,0) -- (1+2*1.618,1+1.618);
      \node[anchor=north west,inner sep=0pt] at (1+2*1.618,1+1.618) {\tiny$\left(\hspace*{-2pt}\begin{array}{c}1+2\phi\\1+\phi\end{array}\hspace*{-2pt}\right)$};
      \draw[color=red,line width=1.6mm] (0,0) -- (1+1.618,1.618);
      \node[anchor=north west,inner sep=1pt] at (1+1.618,1.618) {\tiny$\left(\hspace*{-2pt}\begin{array}{c}1+\phi\\\phi\end{array}\hspace*{-2pt}\right)$};
      \draw[color=red,line width=2.4mm] (0,0) -- (1.618,1);
      \node[anchor=north west,inner sep=1pt] at (1.618,1) {\tiny$\left(\hspace*{-2pt}\begin{array}{c}\phi\\1\end{array}\hspace*{-2pt}\right)$};
      \draw[color=red!60] (0,0) -- (-0.809,-0.5);
      \draw[color=red!60] (0,0) -- (0.809,-0.5);
      \draw[color=red!60] (0,0) -- (-0.809,0.5);
      \draw[color=yellow,line width=0.8mm] (0,0) -- (1.618,1+2*1.618);
      \node[anchor=south east,inner sep=0pt] at (1.618,1+2*1.618) {\tiny$\left(\hspace*{-2pt}\begin{array}{c}\phi\\1+2\phi\end{array}\hspace*{-2pt}\right)$};
      \draw[color=yellow,line width=1.6mm] (0,0) -- (1,1+1.618);
      \node[anchor=south east,inner sep=1pt] at (1,1+1.618) {\tiny$\left(\hspace*{-2pt}\begin{array}{c}1\\1+\phi\end{array}\hspace*{-2pt}\right)$};
      \draw[color=yellow,line width=2.4mm] (0,0) -- (-1+1.618,1.618);
      \node[anchor=south east,inner sep=1pt] at (-1+1.618,1.618) {\tiny$\left(\hspace*{-2pt}\begin{array}{c}-1+\phi\\\phi\end{array}\hspace*{-2pt}\right)$};
      \draw[color=yellow!80] (0,0) -- (0.5-0.809,-0.809);
      \draw[color=yellow!80] (0,0) -- (-0.5+0.809,-0.809);
      \draw[color=yellow!80] (0,0) -- (0.5-0.809,0.809);
      \draw[color=blue,line width=0.8mm] (0,0) -- (2+2*1.618,0);
      \node[anchor=north] at (2+2*1.618,0) {\tiny$\left(\hspace*{-2pt}\begin{array}{c}2+2\phi\\0\end{array}\hspace*{-2pt}\right)$};
      \draw[color=blue,line width=1.6mm] (0,0) -- (2*1.618,0);
      \node[anchor=north] at (2*1.618,0) {\tiny$\left(\hspace*{-2pt}\begin{array}{c}2\phi\\0\end{array}\hspace*{-2pt}\right)$};
      \draw[color=blue,line width=2.4mm] (0,0) -- (2,0);
      \node[anchor=north] at (2,0) {\tiny$\left(\hspace*{-2pt}\begin{array}{c}2\\0\end{array}\hspace*{-2pt}\right)$};
      \draw[color=blue!60] (0,0) -- (-1,0);
      \draw[color=blue!60] (0,0) -- (0,1);
      \draw[color=blue!60] (0,0) -- (0,-1);		
      \draw[black!10,fill=black!10,opacity=1] (0,0) circle (5mm);
      \node at (0,0) {$z$};
    \end{tikzpicture}
    \caption{\small %
      The different Zometool struts and their lengths and orientations
      (implicitly) in the most versatile two-dimensional Zome system. Strut
      thicknesses differ for visualization purposes only. Global (node)
      orientation is~$0^\circ$.}
    \label{fig:ZometoolCheatSheet1a}
\end{figure}
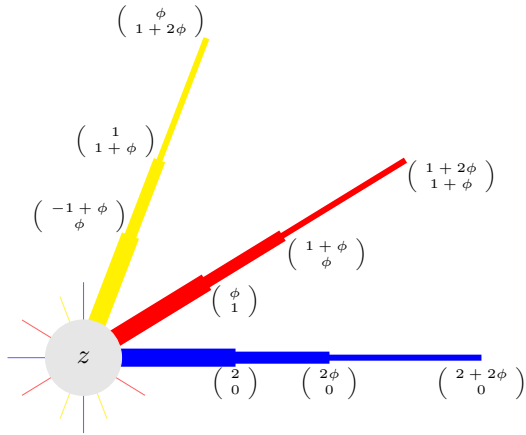

Based on this scaling ($b_1=2$), all other Zome strut lengths are
determined as well; Figure~\ref{fig:ZometoolCheatSheet1a} provides an
overview of the corresponding edge types in the 2D Zome system which offers
the largest admissible set of strut orientations. It is worth mentioning
that there also exist green Zome struts, leading from the origin to
$(2,2)$, $(2\phi,2\phi)$ and $(2+2\phi,2+2\phi)$, respectively; these are
part of an \emph{extended Zome system} which retains the properties of the
basic system but naturally allows for even more variety in
constructions. Nevertheless, for simplicity, we restrict ourselves to the
standard Zome system throughout.

Further details on the Zome system, local configurations and construction
examples can be found in \cite{Davis2007,HartPicciotto2001}; in particular,
see also the Zometool shape approximation algorithms based on simulated
annealing in \cite{ZimmerKobbelt2014,ZimmerETAL2014}.

%%%%%%%%%%%%%%%%%%%%%%%%%%%%%%%%%%%%%%%%%%%%%%%%%%%%%%%%%%%%%%%%%%%%%%%%%%%%%%%%%%%%%%%%%%%%%%%%%%%%%%%%%%%%%%%%%%%%%%%%%%%%%%%%%%%%%%%%%%%%%%%
%                                            NP-Hardness of DGSAPs                                                                            %
%%%%%%%%%%%%%%%%%%%%%%%%%%%%%%%%%%%%%%%%%%%%%%%%%%%%%%%%%%%%%%%%%%%%%%%%%%%%%%%%%%%%%%%%%%%%%%%%%%%%%%%%%%%%%%%%%%%%%%%%%%%%%%%%%%%%%%%%%%%%%%%
\section{Formal Problem Statements and Computational Complexity}\label{sec:hardness}
In this section, we provide strong intractability results for several
essential discrete (2D) shape contour and curve segment approximation
problems. In fact, it will turn out that already deciding feasibility is
\NP-hard, so our complexity results hold regardless of any objective
function. Since no theoretically efficient (i.e., polynomial-time) solution
methods exists unless \P$=$\NP, such intractability results motivate the
development of fast heuristics or, possibly, polynomial-time approximation
algorithms (with provable guarantees on the solution quality w.r.t.\
optimality), as well as dedicated exact (but exponential-time)
solvers. Thus, ultimately, they provide justification for the mixed-integer
programming based approach we propose later (see
Section~\ref{sec:methodology}) and, retrospectively, also for the simulated
annealing heuristic from~\cite{ZimmerKobbelt2014,ZimmerETAL2014} for (3D)
Zometool shape approximation.

\subsection{DAS Shape Approximation (Feasibility) Problems}\label{subsec:hardness:probdefs}
Let us begin with formal definitions of the problems considered here. For
simplicity, we state them all as decision problems; corresponding
optimization versions can be obtained by asking to optimize some objective
function over the set of feasible solutions (in particular, e.g.,
minimizing some approximate error). To that end, in the following, let
$\cF\subset\R^2$ be either a \emph{curve segment} (an arbitary curve
segment between two distinct points in the plane that does not cross
itself) or a \emph{shape contour} (an arbitrary closed curve in the plane
that encloses precisely one nonempty connected area). For a point
$p\in\R^2$, some nonnegative-definite distance measure
$\rho:\R^2\times\R^2\to\R_+$ (e.g., the Euclidean $\ell_2$-norm difference)
and some $\delta\geq 0$, let
$\cN_\delta^\rho(p)\define\{x\in\R^2 : \rho(x,p)\leq\delta\}$ denote the
$\rho$-ball of radius~$\delta$ around $p$.

For a given $\cG$ and $\cF$, the outcomes we are interested in are always
simple paths or cycles\footnote{Here, we appropriate some terminology from
  graph theory, but note that, while structurally similar to (planar)
  graphs, concrete $\cG$-constructions are essentially \emph{defined} via
  their embedding in the plane, whereas graphs are more abstract objects
  that may or may not have planar embeddings.} constructed from components
of $\cG$ that (more or less closely) ``follow'' all of $\cF$; for clarity,
we shall thus call such paths and cycles \emph{$\cF$-resembling
  $\cG$-paths/cycles}, respectively. (Also, recall that such paths/cycles
are \emph{valid} only if they are ``planar'', i.e., without
self-intersections, and obey possible edge-type budgets.)

One may think of ``following'' here as nowhere deviating from $\cF$ by more
than some constant distance, i.e., being contained in a ``corridor''
$C^\sigma_\varepsilon(\cF)\define\{x\in\R^2 :
\sigma(x,y)\leq\varepsilon~\forall y\in\cF\}=\{x :
x\in\cN^\sigma_\varepsilon(y)\text{ for some }y\in\cF\}$ for some distance
measure~$\sigma$ (possibly different from~$\rho$) and constant
$\varepsilon\geq 0$; in case of shape contours, such a corridor
should---for a suitable scaling/resolution---generally resemble a ``2D
torus'', i.e., still have a ``hole''. We note that, here, the corridor
definition serves primarily to better describe the notion of following the
contour, but otherwise has no direct impact on the main problems under
consideration and is not used algorithmically either. Nevertheless,
corridor containment could be formally required, as seems common,
e.g., in schematization; such requirements lead to variants of our problems
of interest, which will be briefly touched upon in the context of
computational complexity (Section~\ref{subsec:hardness:results}).

Rather than via corridor containment constraints, our approach to achieve
$\cF$-resemblance focusses on sampling the input contour and requiring the
$\cG$-construction to place DAS nodes in the vicinity of the sampled anchor
points and connect each such node to the next---i.e., \emph{in order
  corresponding to how the sample points occur when traversing $\cF$ in a
  fixed direction (clockwise or counter-clockwise)}---with DAS
components. (This will be made precise in the definitions below, though for
notational simplicity, we shall assume the sample points are ordered in
this fashion, and later implicitly assume this node visitation order when
talking about $\cF$-resembling $\cG$-cycles/paths.)

Our main (decision) problem of interest can be formally stated as
follows:

\medskip
\ownproblem{Discrete Contour Approximation with Sampling
  (DCA-S)}{Given a shape contour $\cF\subset\R^2$, ordered sample points
  $p_1,\dots,p_k$ on $\cF$, a DAS~$\cG=(\cV,\cE,\cB,\cD)$, a distance
  measure $\rho:\R^2\to\R_+$ and a parameter~$\delta\geq 0$, does there
  exist a valid $\cF$-resembling $\cG$-cycle with at least one node~$q_i$
  in each $\cN^\rho_\delta(p_i)$, $i\in\{1,\dots,k\}$, that consists of
  $\cG$-path segments connecting the pairs
  $(q_1,q_2),(q_2,q_3),\dots,(q_{k-1},q_k),(q_k,q_1)$?}

\medskip
Our computational method (cf. Section~\ref{sec:methodology}) will
be based on the DCA-S problem. (Indeed, the mixed-integer program we will
employ provides a way to answer the above DCA-S problem.) Moreover, the
following variant asking for a $\cG$-path approximating a curve segment
between two distinct points is both of separate interest and will resurface
as a subproblem (in a certain sense) in our algorithm.

\medskip
\ownproblem{Discrete Path Approximation with Sampling (DPA-S)}{
  Given a curve segment $\cF\subset\R^2$ between two points
  $x_1,x_2\in\R^2$, a DAS~$\cG=(\cV,\cE,\cB,\cD)$, ordered sample points
  $p_1,\dots,p_k$ on $\cF$, a distance measure $\rho:\R^2\to\R_+$ and a
  parameter $\delta\geq 0$, does there exist a valid $\cF$-resembling
  $\cG$-path connecting $x_1$ and $x_2$ (either directly, or approximately by
  placing nodes in $\cN^\rho_\delta(x_1)$ and $\cN^\rho_\delta(x_2)$) with
  at least one node $q_i$ in each $\cN^\rho_\delta(p_i)$, $i\in\{1,\dots,k\}$,
  that consists of $\cG$-path segments connecting the pairs
  $(q_1,q_2),(q_2,q_3),\dots,(q_{k-1},q_k)$?}

\medskip
In fact, the feasibility of \emph{freely} connecting two distinct
points in $\R^2$ (without regard to any curve segment resemblance
objectives, sample points or corridor containment) using components of a
DAS is still of interest as an intuitively simpler-seeming, very basic
problem:

\medskip
\ownproblem{Discrete Point Connectivity (DPC)}{Given two points
  $x_1,x_2\in\R^2$ and a DAS~$\cG$, can $x_1$ and~$x_2$ be connected by a
  valid $\cG$-path?}

\subsection{Intractability Results}\label{subsec:hardness:results}
We now turn to the computational complexity of the fundamental DAS shape
approximation problems defined above. Before providing our novel results,
we point out that the connection to discrete schematization discussed in
Section~\ref{subsec:relatedwork} allows to directly infer hardness results
for some special cases of our DAS problems. More precisely, suppose that,
in our setting, we do \emph{not} use sample points, that the global
position of the sought $\cG$-cycle is fixed, and that only horizontal and
vertical edges of the same unit length are available. The position fixing
and available edge types imply that the $\cG$-cycle is, in fact, a (simple)
cycle in a grid graph that ``overlays'' the given shape contour $\cF$. Now,
\cite[Thm.~1]{LoefflerMeulemans2017} states that, even for a simple
polygon~$\cF$, it is \NP-complete to decide whether this grid graph
contains a ($\cG$-)cycle with Fr\'{e}chet distance no larger than a given
$\epsilon>0$. The Fr\'{e}chet distance (see, e.g.,
\cite{LoefflerMeulemans2017} for a precise definition) takes the place of
an approximation error measure to quantify shape resemblance; using the
symmetric difference (the total area of those parts of the union of two
shapes that are contained in only one of them) leads to a similar
\NP-hardness results, see \cite[Thm.~2]{LoefflerMeulemans2017}. When a
corridor is defined in terms of Hausdorff distances,
\cite[Thm.~1]{BoutsETAL2016} gives \NP-hardness of an analogous
grid-graph/polygon DAS variant.

Thus, if we replace the sample-point part in DCA-S by corridor containment
requirements, and restrict further to a fixed global positioning and
grid-inducing DASs, the abovementioned results can indeed be seen as
special cases, and consequently, the \NP-hardness results carry
over. Nevertheless, these special cases are not precisely what we are
interested in here: We explicitly do \emph{not} want to fix the global
positioning a priori, and we intend to work with a DAS that does \emph{not}
induce a grid-like graph (such as the Zome system described in
Section~\ref{subsec:zome}). Furthermore, the requirement to place nodes in
sample point neighborhoods is weaker than requiring the full graph to lie
within a certain distance of the given shape contour (or curve segment).
Therefore, while no doubt closely related, the hardness results described
above do not precisely match our setting. While it may be possible to
extend the rather complicated reductions from
\cite{LoefflerMeulemans2017,BoutsETAL2016} to analogous DPA-S variants
and/or modify the corridor containment there by a requirement to have the
sought cycle traverse sample point neighborhoods, the novel proofs we
provide below are also significantly simpler.

Our intractability results will be based on the following two problems,
which are well-known to be \NP-complete (see~\cite[problems SP12 and
SP15]{GareyJohnson1979}):

\medskip
\ownproblem{Partition}{Given positive integers $a_1,\dots,a_n$, does there
  exist a subset $I^\prime$ of $I\define[n]\define\{1,\dots,n\}$ such that
  $\sum_{i\in I^\prime}a_i=\sum_{i\in I\setminus I^\prime}a_i$?}

\medskip
\ownproblem{3-Partition}{Given positive integers $a_1,\dots,a_{3m}$
  and $A$ such that $A/4<a_i<A/2$ for all $i\in I\define[3m]$ and
  $\sum_{i\in I}a_i=mA$, can $I$ be partitioned into $m$ disjoint
  ($3$-element) sets $I_1,\dots,I_m$ such that, for all
  $j\in\{1,\dots,m\}$, $\sum_{i\in I_j}a_j=A$?}

\medskip
Our first result is about DPA-S; the subsequent ones about the other
problems are derived along the same lines, with slight variations of the
same main proof idea.

\begin{theorem}\label{thm:DPAShardness}
  The DPA-S problem is \NP-hard in the strong sense, even restricted to
  piecewise linear curve segments $\cF$ with right-angle turns and
  integral-coordinate turning points and end points ($x_1,x_2$),
  integral-coordinate sample points, square ($\ell_\infty$-norm) boxes as
  sample point neighborhoods $\cN^\rho_\delta(\cdot)$, and DASs $\cG$ with
  only one node type, horizontal and vertical orientations, and
  integral-length edge types.
\end{theorem}
\begin{proof}
  We reduce from \textsc{3-Partition}: Let $(a_1,\dots,a_{3m},A)$ be a
  given \textsc{3-Par\-ti\-tion} instance; we may and do assume that $A$ is
  polynomially bounded my $m$, since \cite{GareyJohnson1979} proved that
  \textsc{3-Partition} remains \NP-complete under these restrictions and is
  thus, in fact, \NP-complete \emph{in the strong sense}\footnote{Recall
    that this means that, unless \P$=$\NP, not only does there not exist a
    polynomial algorithm to decide the problem, but also no
    pseudo-polynomial algorithm (and, for strongly \NP-hard
    \emph{optimization} problems, no fully polynomial-time approximation
    scheme (FPTAS)). We refer to~\cite{GareyJohnson1979} for details.}. We
  construct an instance of DCA-S with restrictions as specified in the
  theorem statement: Let~$k$ be the number of \emph{distinct} $a_i$-values
  in the given \textsc{3-Partition} instance and associate with each such
  value (say, $a^\prime_\ell$) an edge type~$e_\ell$ with length
  $a^\prime_\ell$; set $\cE\define\{e_\ell\st\ell=1,\dots,k\}$ and define
  the budgets 
  $\cB\define\{b_e\st e\in\cE\}$
  with 
  $b_{e_\ell}\define\abs{\{i\in[3m]\st a_i=a^\prime_\ell\}}$.  Further,
  let 
  $\cV\define\{v\}$ (the single node type), and let $\cD$ be given via the
  horizontal and vertical direction vectors, all of which are defined to be
  compatible with each node and edge type, i.e.,
  $\cD\define\{(d,\cV,\cE)\st d\in\{\pm(1,0),\pm(0,1)\}\}$. This completes
  the (obviously polynomial) construction of the DAS
  $\cG\define(\cV,\cE,\cB,\cD)$. To specify the curve segment, sample
  points and their neighborhoods, let $0\leq\delta<A/4$ (say,
  $\delta\define A/9$), $\rho(x,y)\define\norm{x-y}_\infty$, and define
  points $p_j\define(\lfloor j/2\rfloor A,(\lceil j/2\rceil\mod 2)A)$ for
  $j=0,1,\dots,m$. Set $x_1\define p_0$, $x_2\define p_m$ and define $\cF$
  to be the chain of piecewise linear segments connecting $p_j$ with
  $p_{j+1}$ for all $j=0,\dots,m-1$ (i.e.,
  $\cF\define\bigcup_{i\in[m]}\{\lambda p_{i-1}+(1-\lambda)p_i:\lambda\in
  [0,1]\}$), see Figure~\ref{fig:npproofconstruction} for a
  visualization. Clearly, by means of $p_j$, $\rho$ and $\delta$, both
  $\cF$ and each $\cN^\rho_\delta(p_i)$ can be encoded with size polynomial
  in that of the given \textsc{3-Partition} instance, and containment in
  either can be evaluated in polynomial time.
\begin{figure}[t]
  \centering
    \begin{tikzpicture}[scale=0.718]
      \fill[black!20] (-1/3,-1/3) -- (1/3,-1/3) -- (1/3,1/3) -- (-1/3,1/3);
      \fill[black!20] (-1/3,8/3) -- (-1/3,10/3) -- (1/3,10/3) -- (1/3,8/3);
      \fill[black!20] (8/3,10/3) -- (10/3,10/3) -- (10/3,8/3) -- (8/3,8/3);
      \fill[black!20] (8/3,-1/3) -- (8/3,1/3) -- (10/3,1/3) -- (10/3,-1/3);
      \fill[black!20] (17/3,-1/3) -- (19/3,-1/3) -- (19/3,1/3) -- (17/3,1/3);
      \fill[black!20] (17/3,8/3) -- (17/3,10/3) -- (19/3,10/3) -- (19/3,8/3);
      \fill[black!20] (26/3,10/3) -- (28/3,10/3) -- (28/3,8/3) -- (26/3,8/3);
      \fill[black!20] (26/3,-1/3) -- (26/3,1/3) -- (28/3,1/3);
      \fill[black!10] (26/3,-1/3) -- (28/3,1/3) -- (28/3,-1/3);
      \fill[black!10] (35/3,-1/3) -- (35/3,1/3) -- (37/3,-1/3);
      \fill[black!20] (35/3,1/3) -- (37/3,1/3) -- (37/3,-1/3);
      \fill[black!20] (35/3,8/3) -- (37/3,8/3) -- (37/3,10/3) -- (35/3,10/3);
      \fill[black!20] (44/3,10/3) -- (44/3,8/3) -- (46/3,8/3) -- (46/3,10/3);
      \draw (0,0) -- (0,3);
      \fill (0,3) circle (2pt);
      \draw (-1/2,0) -- (-2/3,0);
      \draw[dotted] (0,0) -- (-1/2,0);
      \draw[<->] (-7/12,0) --node[left]{\small $A$} (-7/12,3);
      \draw (-1/2,3) -- (-2/3,3);
      \draw[dotted] (0,3) -- (-1/2,3);
      \draw[<->] (-7/12,3) --node[left]{\small $\delta$} (-7/12,10/3);
      \draw[dotted] (-1/3,10/3) -- (-1/2,10/3);
      \draw (-1/2,10/3) -- (-2/3,10/3);
      \draw (0,3) -- (3,3);
      \fill (3,3) circle (2pt);
      \draw (3,3) --node[right]{\small $\cF$} (3,0);
      \draw (3,0) -- (6,0);
      \fill (6,0) circle (2pt);
      \fill (3,0) circle (2pt);
      \draw (6,0) -- (6,3);
      \fill (6,3) circle (2pt);
      \draw (6,3) -- (9,3);
      \fill (9,3) circle (2pt);
      \draw (9,3) -- (9,0);
      \fill (9,0) circle (2pt);
      \draw[dashed] (9,0) -- (12,0);
      \draw (12,0) -- (12,3);
      \fill (12,0) circle (2pt);
      \fill (12,3) circle (2pt);
      \draw (12,3) -- (15,3);
      \fill (0,0) circle (2pt);
      \node at (2/3,1/3) {\small $x_1$};
      \draw[->] (1/2,1/6) -- (1/9,1/24);
      \node at (15,7/3) {\small $x_2$};
      \draw[->] (15,61/24) -- (15,17/6);
      \fill (15,3) circle (2pt);
      \draw[white] (10,-2/3) circle (1pt); 
    \end{tikzpicture}
    \caption{Schematic illustration of the construction from the proof
      of Theorem~\ref{thm:DPAShardness}: The neighborhoods $\cN^\rho_\delta(p_i)$
      (shaded boxes) contain all points with $\ell_\infty$-norm distance at
      most $\delta$ from the associated sample points $p_i$ (black dots),
      respectively. The piecewise linear $x_1$-$x_2$-path $\cF$ is the
      solid curve (with turns at $p_1,\dots,p_{m-1}$); all $m$ segments of
      this path have length $A$. The dashed and lighter-shaded parts of
      $\cF$ and some $\cN^\rho_\delta(p_i)$, respectively, illustrate a
      continuation of the discernible pattern in accordance with the actual
      instance~size.}
    \label{fig:npproofconstruction}
\end{figure}
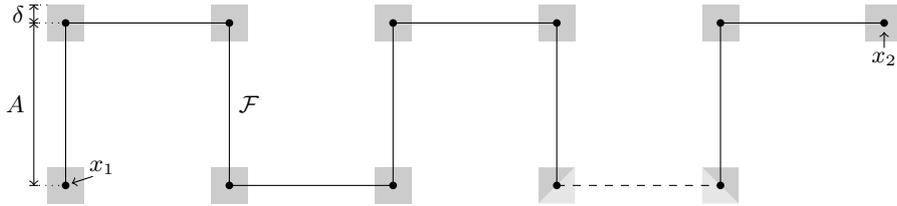
  
  Thus, the overall construction is indeed polynomial, so it remains to
  show that the given \textsc{3-Partition} instance has a positive answer
  if and only if the constructed DPA-S instance does as well:
  \begin{description}
  \item[``$\Rightarrow$'':] If $(a_1,\dots,a_{3m},A)$ is a ``yes''-instance
    of \textsc{3-Partition} with certificate $(I_1,\dots,I_m)$, a
    corresponding solution to the DPA-S instance is given by the valid
    $\cF$-resembling $\cG$-path that connects $x_1$ and $x_2$ by using
    (exactly) three edges to reach each respective next (right-angle) turn,
    thus mapping $I_1,\dots,I_m$ to the $m$ path segments of length
    $A$. Clearly, all direction constraints are obeyed, the budgets are
    sufficient (in fact, depleted with no left-overs) and the $\cG$-path
    precisely overlays $\cF$ (thus placing DAS-nodes at each turning/sample
    point, so each $\cN^\rho_\delta(p_i)$ indeed contains a node). Thus, this
    solution shows that the DPA-S instance has a ``yes'' answer.
  \item[``$\Leftarrow$'':] Conversely, let a ``yes''-certificate of the
    DPA-S instance be given, and let $e_1,\dots,e_M$ be the sequence of
    edge types used along the $\cG$-path from $x_1$ to $x_2$. Since
    $\delta<A/4<\min a^\prime_i$ and $\max a^\prime_i<A/2$, exactly three
    edges of the same orientation must be used in each
    path-segment. Moreover, the lengths of all these three-edge subsets
    must sum to $A$, because otherwise, a neighborhood
    ($\cN^\rho_\delta(p_i)$-) containment condition would be violated
    either immediately or after the next turn (recall also that the
    neighborhoods are traversed according to a given order). This shows
    that $M=3m$, and that we can construct a ``yes''-certificate
    $I_1,\dots,I_m$ for the input \textsc{3-Partition} instance by
    traversing $e_1,\dots,e_{3m}$ and mapping the encountered three-tuples
    $E_j\define (e_{3j-2},e_{3j-1},e_{3j})$, $j=1,\dots,m$, to
    $I_j\define\{j_1,j_2,j_3\}$ with
    $j_1\define \min\{i:a_i=a^\prime_{3j-2},i\notin I_k~\forall k<j\}$,
    $j_2\define\min\{i:a_i=a^\prime_{3j-1},i\notin I_k~\forall k<j,i\neq
    j_1\}$, and
    $j_3\define\min\{i:a_i=a^\prime_{3j},i\notin I_k~\forall k<j,i\neq
    j_1,i\neq j_2\}$, respectively.
    \end{description}
  This proves \NP-hardness of the DPA-S problem. Moreover, in particular,
  for the DPA-S instance constructed from the \textsc{3-Partition} input
  with the stated restrictions, all encoding lengths as well as the
  occurring numbers themselves are polynomially bounded by the instance
  \emph{size} (number of specified integers, $\O(m)$) alone. Therefore, the
  ``strong sense'' assertion of \NP-hardness carries over to DPA-S as well.

  Finally, note that the above proof goes through completely analogously if
  we require DAS nodes exactly at $x_1$ and $x_2$, so both variants from
  the DPA-S definition are covered.
\end{proof}

It is easily seen that the proof of Theorem~\ref{thm:DPAShardness} goes
through completely analogously for DPA-S variants requiring the $\cG$-path
to be contained in a corridor $C^\sigma_\varepsilon(\cF)$ and/or minimizing
a (nonnegative definite) error measure w.r.t.\ $\cF$. Thus, we immediately
obtain the following.

\begin{corollary}\label{cor:DPAShardness}
  The DPA-S problem remains \NP-hard in the strong sense (under the same
  restrictions as listed in Theorem~\ref{thm:DPAShardness}) if the
  $\cF$-resembling $\cG$-path~$P$ is required to lie within a corridor
  $C^\sigma_\varepsilon(\cF)$ around $\cF$ and/or the objective of
  minimizing an approximation error $\alpha(P,\cF)\geq 0$ is included. In
  the latter case, hardness persists even if no sample point neighborhood
  (and/or corridor) containment is required.
\end{corollary}
\begin{proof}
  We can directly extend the reduction from the proof of
  Theorem~\ref{thm:DPAShardness} by letting (for instance)
  $C_\varepsilon^\sigma(\cF)\define\{x:\norm{x-y}_\infty\leq\delta~\forall
  y\in\cF\}$ and observing that any nonnegative definite error measure
  (e.g., the area between $\cF$ and the $\cG$-path~$P$, or an arbitrary
  metric) achieves minimum value $\alpha(P,\cF)=0$ if and only if $P$
  exactly matches $\cF$.
\end{proof}

Note that as a particular consequence, testing whether the input is already
within (i.e., can be exactly overlayed by a construction from) the budgeted
DAS system is already \NP-hard. (This can be seen by considering DPA-S
without sample point or corridor requirement and the objective of
minimizing the approximation error defined, e.g., via the discrete metric
$\alpha(p,q)=0$ if $p=q$, and $1$ otherwise.)

\begin{remark}
  It is not immediately clear whether the DPA-S decision problem is
  contained in \NP. For this to hold, we would need to assert that a
  ``yes''-certificate for an (arbitrary) DPA-S instance has encoding length
  polynomially bounded by that of the instance, and that all constraints
  can be verified in polynomial time. Allowing only rational input data (in
  particular, edge lengths and point coordinates) and suitably ``simple''
  definitions of $\cF$ and $\cN^\rho_\delta(p_i)$ (and
  $\cC^\sigma_\varepsilon(\cF)$ and/or $\alpha(P,\cF)$, in the variants
  from Corollary~\ref{cor:DPAShardness}), containment in \NP\ may appear to
  be easy to demonstrate. For the variant that places DAS nodes exactly at
  $x_1,x_2\in\Q^2$, restricting to piecewise linear $\cF$ with rational
  turn points and neighborhoods based on, say, $\ell_1$- or
  $\ell_\infty$-norm differences (such as used in the reduction above),
  containment in \NP\ is indeed obtained straightforwardly, making the
  problem \NP-\emph{complete} (in the strong sense). However, allowing the
  end nodes of the $\cG$-path to deviate from $x_1$ and $x_2$,
  respectively, a certificate might be a valid $\cF$-resembling $\cG$-cycle
  that involves node coordinates of non-polynomial encoding length (e.g.,
  irrational values), so containment in \NP\ cannot be proven in
  general. Analogous arguments hold for the other problems discussed below
  (DCA-S, DPC) as well; for brevity, we mention this only once here.
\end{remark}

The above results extend straightforwardly to the DCA-S problem (and
corresponding variants):

\begin{theorem}\label{thm:DCAShardness}
  The DCA-S problem is \NP-hard in the strong sense, even restricted to
  piecewise linear curve segments $\cF$ with right-angle turns and
  integral-coordinate turning and sample points, square
  ($\ell_\infty$-norm) boxes as sample point neighborhoods
  $\cN^\rho_\delta(\cdot)$, and DASs $\cG$ with only one node type,
  horizontal and vertical orientations, and integral-length edge types.
\end{theorem}
\begin{proof}
  We modify the reduction used to show \NP-hardness of the DPA-S problem,
  so at first, let $\cV$, $\cE$, $\cD$ and $\cB$ as well as $k$, $\rho$,
  $\delta$ and $p_0,\dots,p_m$ be constructed from a given
  \textsc{3-Partition} instance as in the proof of
  Theorem~\ref{thm:DPAShardness}. Depending on whether $m$ and
  $\lceil m/2\rceil$ are even or odd, we construct further points (in order
  to specify $\cF$ and to choose as sample points) and add edge types as
  follows:
  \begin{itemize}
  \item If $m$ is odd (so $\lceil m/2\rceil$ is always even and
    $p_m=(\lfloor m/2\rfloor A,0)$,
    $p_{m-1}=(\lfloor (m-1)/2\rfloor A,A)=(\lfloor m/2\rfloor A,A)$),
    define $p_{m+1}\define((\lfloor m/2\rfloor+2)A,0)$,
    $p_{m+2}\define((\lfloor m/2\rfloor+2)A,-2A)$,
    $p_{m+3}\define(-2A,-2A)$ and $p_{m+4}\define(-2A,0)$, and add edge
    type $e_{k+1}$ with length $(\lfloor m/2\rfloor+4)A$, budget
    $b_{e_{k+1}}=1$ and compatible directions $\pm(1,0)$, as well as edge
    type $e_{k+2}$ with length $2A$, budget $b_{e_{k+2}}=4$ and compatible
    directions $\{\pm(0,1),\pm(1,0)\}$.
  \item If both $m$ and $\lceil m/2\rceil$ are even (so
    $\lceil m/2\rceil=m/2=\lfloor m/2\rfloor$, $p_m=((m/2)A,0)$ and
    $p_{m-1}=(\lfloor(m-1)/2\rfloor A,(\lceil(m-1)/2\rceil\mod
    2)A)=((m/2-1)A,0)$), define $p_{m+1}\define(\lfloor m/2\rfloor A,-2A)$,
    $p_{m+2}\define(-2A,-2A)$ and $p_{m+3}\define(-2A,0)$, and add edge
    type $e_{k+1}$ with length $(\lfloor m/2\rfloor+2) A$, budget
    $b_{e_{k+1}}=1$ and compatible directions $\pm(1,0)$, as well as edge
    type $e_{k+2}$ with length $2A$, budget $b_{e_{k+2}}=3$ and compatible
    directions $\{\pm(0,1),\pm(1,0)\}$.
  \item In the final case, $m$ is even and
    $\lceil m/2\rceil=m/2=\lfloor m/2\rfloor$ is odd (so $p_m=((m/2)A,A)$
    and $p_{m-1}=((m/2-1)A,A)$); then, we define
    $p_{m+1}\define(\lfloor m/2\rfloor A,-2A)$, $p_{m+2}\define(-2A,-2A)$
    and $p_{m+3}\define(-2A,0)$, and add edge type $e_{k+1}$ with length
    $(\lfloor m/2\rfloor+2)A$, budget $b_{e_{k+1}}=1$ and compatible
    directions $\pm(1,0)$, as well as edge types $e_{k+2}$ and $e_{k+3}$
    with lengths $3A$ and $2A$, budgets $b_{e_{k+2}}=1$ and $b_{e_{k+3}}=2$
    and compatible directions $\pm(0,1)$ and $\{\pm(0,1),\pm(1,0)\}$,
    respectively.
  \end{itemize}
  Finally, define the shape contour $\cF$ as the piecewise-linear 
  cycle $p_0\to p_1\to \cdots\to p_{m+3} \;(\to p_{m+4})\to p_0$. 
  Figure~\ref{fig:npproofconstruction2}
  illustrates the construction for the three cases (for the DCA-S variant
  with a feasible corridor of $\ell_\infty$-width~$\delta$).
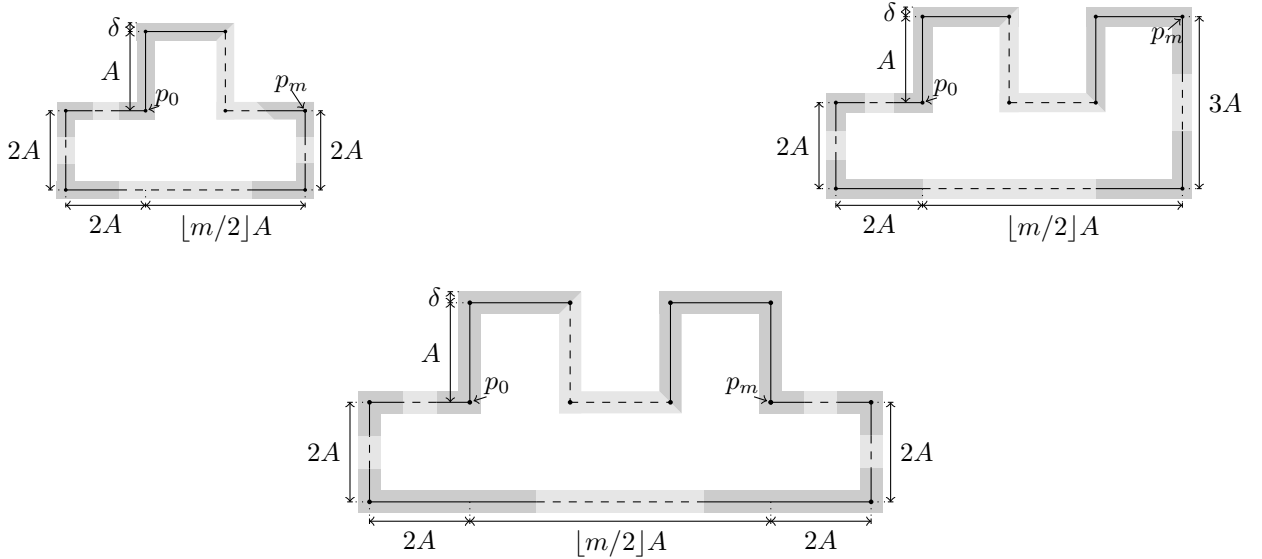
\begin{figure}[t]
  \centering
    % m even, \lceil m/2\rceil even:
    \begin{tikzpicture}[scale=0.35]
      \fill[black!20] (-1/3,-1/3) -- (1/3,-1/3) -- (1/3,10/3) -- (-1/3,10/3);
      \fill[black!20] (-1/3,8/3) -- (-1/3,10/3) -- (10/3,10/3) -- (10/3,8/3);
      \fill[black!10] (8/3,8/3) -- (10/3,10/3) -- (10/3,-1/3) -- (8/3,-1/3);
      \fill[black!10] (8/3,-1/3) -- (8/3,1/3) -- (25/6,1/3) -- (29/6,-1/3);
      \fill[black!20] (25/6,1/3) -- (29/6,-1/3) -- (19/3,-1/3) -- (19/3,1/3);
      \fill[black!20] (17/3,1/3) -- (19/3,1/3) -- (19/3,-1) -- (17/3,-1);
      \fill[black!10] (17/3,-1) -- (19/3,-1) -- (19/3,-2) -- (17/3,-2);
      \fill[black!20] (17/3,-2) -- (19/3,-2) -- (19/3,-10/3) -- (17/3,-10/3);
      \fill[black!20] (19/3,-10/3) -- (19/3,-8/3) -- (4,-8/3) -- (4,-10/3);
      \fill[black!10] (4,-10/3) -- (4,-8/3) -- (-1,-8/3) -- (-1,-10/3);
      \fill[black!20] (-1,-10/3) -- (-1,-8/3) -- (-3-1/3,-8/3) -- (-3-1/3,-10/3);
      \fill[black!20] (1/3,1/3) -- (1/3,-1/3) -- (-1,-1/3) -- (-1,1/3);
      \fill[black!10] (-1,1/3) -- (-1,-1/3) -- (-2,-1/3) -- (-2,1/3);
      \fill[black!20] (-2,1/3) -- (-2,-1/3) -- (-3-1/3,-1/3) -- (-3-1/3,1/3);
      \fill[black!20] (-1/3-3,-10/3) -- (1/3-3,-10/3) -- (1/3-3,-2) -- (-1/3-3,-2);
      \fill[black!10] (-1/3-3,-2) -- (1/3-3,-2) -- (1/3-3,-1) -- (-1/3-3,-1);
      \fill[black!20] (-1/3-3,-1) -- (1/3-3,-1) -- (1/3-3,0) -- (-1/3-3,0);
      \draw (0,0) -- (0,3);
      \draw[<->] (-7/12,0) --node[left]{$A$} (-7/12,3);
      \draw (-1/2,3) -- (-2/3,3);
      \draw[dotted] (0,3) -- (-1/2,3);
      \draw[<->] (-7/12,3) --node[left]{$\delta$} (-7/12,10/3);
      \draw[dotted] (-1/3,10/3) -- (-1/2,10/3);
      \draw (-1/2,10/3) -- (-2/3,10/3);
      \draw (0,3) -- (3,3);
      \fill (0,3) circle (2pt);      
      \fill (3,3) circle (2pt);
      \draw[dashed] (3,3) -- (3,0);
      \draw[dashed] (3,0) -- (9/2,0);
      \draw (9/2,0) -- (6,0);
      \fill (0,0) circle (2pt);
      \fill (6,0) circle (2pt);
      \node at (5/6,1/2) {$p_0$};
      \draw[->] (1/2,1/6) -- (1/9,1/24);
      \node at (11/2,3/3) {$p_m$};
      \draw[->] (11/2,6/8) -- (95/16,1/12);
      \fill (6,0) circle (2pt);
      \draw[white] (10,-14/3) circle (1pt);
      \draw (6,0) -- (6,-1);
      \draw[dashed] (6,-1) -- (6,-2);
      \draw (6,-2) -- (6,-3);
      \fill (6,-3) circle (2pt);
      \draw (-3,-3) -- (-1,-3);
      \fill (-3,-3) circle (2pt);
      \draw[dashed] (-1,-3) -- (4,-3);
      \draw (4,-3) -- (6,-3);
      \draw (0,0) -- (-1,0);
      \draw[dashed] (-1,0) -- (-2,0);
      \draw (-2,0) -- (-3,0);
      \fill (-3,0) circle (2pt);
      \fill (3,0) circle (2pt);
      \draw (-3,-3) -- (-3,-2);
      \draw[dashed] (-3,-2) -- (-3,-1);
      \draw (-3,-1) -- (-3,0);
      \draw (-3-1/2,0) -- (-3-2/3,0);
      \draw[dotted] (-3,0) -- (-3-1/2,0);
      \draw[<->] (-3-7/12,0) --node[left]{$2A$} (-3-7/12,-3);
      \draw (-3-1/2,-3) -- (-3-2/3,-3);
      \draw[dotted] (-3,-3) -- (-3-1/2,-3);
      \draw (13/2,0) -- (20/3,0);
      \draw[dotted] (6,0) -- (13/2,0);
      \draw[<->] (6+7/12,0) --node[right]{$2A$} (6+7/12,-3);
      \draw[dotted] (6,-3) -- (13/2,-3);
      \draw (13/2,-3) -- (20/3,-3);
      \draw (-3,-7/2) -- (-3,-11/3);      
      \draw[dotted] (-3,-3) -- (-3,-7/2);
      \draw[<->] (-3,-43/12) --node[below]{$2A$} (0,-43/12);
      \draw (0,-7/2) -- (0,-11/3);      
      \draw[dotted] (0,-3) -- (0,-7/2);
      \draw[<->] (0,-43/12) --node[below]{$\lfloor m/2\rfloor A$} (6,-43/12);
      \draw (6,-7/2) -- (6,-11/3);
      \draw[dotted] (6,-3) -- (6,-7/2);
    \end{tikzpicture}
    \hfill
    % m even, \lceil m/2\rceil odd:
    \begin{tikzpicture}[scale=0.38]
      \fill[black!20] (-1/3,-1/3) -- (1/3,-1/3) -- (1/3,10/3) -- (-1/3,10/3);
      \fill[black!20] (-1/3,8/3) -- (-1/3,10/3) -- (10/3,10/3) -- (10/3,8/3);
      \fill[black!10] (8/3,8/3) -- (10/3,10/3) -- (10/3,-1/3) -- (8/3,-1/3);
      \fill[black!10] (8/3,-1/3) -- (8/3,1/3) -- (17/3,1/3) -- (19/3,-1/3);
      \fill[black!20] (17/3,1/3) -- (19/3,-1/3) -- (19/3,10/3) -- (17/3,10/3);
      \fill[black!20] (17/3,8/3) -- (17/3,10/3) -- (28/3,10/3) -- (28/3,8/3);
      \fill[black!20] (28/3,10/3) -- (26/3,10/3) -- (26/3,1) -- (28/3,1);
      \fill[black!10] (28/3,1) -- (26/3,1) -- (26/3,-1) -- (28/3,-1);
      \fill[black!20] (28/3,-1) -- (26/3,-1) -- (26/3,-10/3) -- (28/3,-10/3);
      \fill[black!20] (28/3,-10/3) -- (28/3,-8/3) -- (6,-8/3) -- (6,-10/3);
      \fill[black!10] (6,-10/3) -- (6,-8/3) -- (0,-8/3) -- (0,-10/3);
      \fill[black!20] (0,-10/3) -- (0,-8/3) -- (-3-1/3,-8/3) -- (-3-1/3,-10/3);
      \fill[black!20] (1/3,1/3) -- (1/3,-1/3) -- (-1,-1/3) -- (-1,1/3);
      \fill[black!10] (-1,1/3) -- (-1,-1/3) -- (-2,-1/3) -- (-2,1/3);
      \fill[black!20] (-2,1/3) -- (-2,-1/3) -- (-3-1/3,-1/3) -- (-3-1/3,1/3);
      \fill[black!20] (-1/3-3,-10/3) -- (1/3-3,-10/3) -- (1/3-3,-2) -- (-1/3-3,-2);
      \fill[black!10] (-1/3-3,-2) -- (1/3-3,-2) -- (1/3-3,-1) -- (-1/3-3,-1);
      \fill[black!20] (-1/3-3,-1) -- (1/3-3,-1) -- (1/3-3,0) -- (-1/3-3,0);
      \draw (0,0) -- (0,3);
      \draw[<->] (-7/12,0) --node[left]{$A$} (-7/12,3);
      \draw (-1/2,3) -- (-2/3,3);
      \draw[dotted] (0,3) -- (-1/2,3);
      \draw[<->] (-7/12,3) --node[left]{$\delta$} (-7/12,10/3);
      \draw[dotted] (-1/3,10/3) -- (-1/2,10/3);
      \draw (-1/2,10/3) -- (-2/3,10/3);
      \draw (0,3) -- (3,3);
      \draw[dashed] (3,3) -- (3,0);
      \draw[dashed] (3,0) -- (6,0);
      \draw (6,0) -- (6,3);
      \draw (6,3) -- (9,3);
      \fill (0,0) circle (2pt);
      \node at (5/6,1/2) {$p_0$};
      \draw[->] (1/2,1/6) -- (1/9,1/24);
      \node at (17/2,7/3) {$p_m$};
      \draw[->] (17/2,61/24) -- (143/16,35/12);
      \fill (0,3) circle (2pt);
      \fill (3,3) circle (2pt);
      \fill (3,0) circle (2pt);
      \fill (6,0) circle (2pt);
      \fill (6,3) circle (2pt);
      \fill (9,3) circle (2pt);
      \fill (9,-3) circle (2pt);
      \fill (-3,-3) circle (2pt);
      \fill (-3,0) circle (2pt);
      \draw[white] (10,-14/3) circle (1pt); 
      \draw (9,3) -- (9,1);
      \draw[dashed] (9,1) -- (9,-1);
      \draw (9,-1) -- (9,-3);
      \draw (0,0) -- (-1,0);
      \draw[dashed] (-1,0) -- (-2,0);
      \draw (-2,0) -- (-3,0);
      \draw (-3,-3) -- (0,-3);
      \draw[dashed] (0,-3) -- (6,-3);
      \draw (6,-3) -- (9,-3);
      \draw (-3,-3) -- (-3,-2);
      \draw[dashed] (-3,-2) -- (-3,-1);
      \draw (-3,-1) -- (-3,0);
      \draw (-3-1/2,0) -- (-3-2/3,0);
      \draw[dotted] (-3,0) -- (-3-1/2,0);
      \draw[<->] (-3-7/12,0) --node[left]{$2A$} (-3-7/12,-3);
      \draw (-3-1/2,-3) -- (-3-2/3,-3);
      \draw[dotted] (-3,-3) -- (-3-1/2,-3);
      \draw (19/2,3) -- (29/3,3);
      \draw[dotted] (9,3) -- (19/2,3);
      \draw[<->] (9+7/12,3) --node[right]{$3A$} (9+7/12,-3);
      \draw[dotted] (9,-3) -- (19/2,-3);
      \draw (19/2,-3) -- (29/3,-3);
      \draw (-3,-7/2) -- (-3,-11/3);      
      \draw[dotted] (-3,-3) -- (-3,-7/2);
      \draw (0,-7/2) -- (0,-11/3);            
      \draw[<->] (-3,-43/12) --node[below]{$2A$} (0,-43/12);
      \draw[dotted] (0,-3) -- (0,-7/2);
      \draw[<->] (0,-43/12) --node[below]{$\lfloor m/2\rfloor A$} (9,-43/12);
      \draw (9,-7/2) -- (9,-11/3);
      \draw[dotted] (9,-3) -- (9,-7/2);
    \end{tikzpicture}

    \vspace*{1em}
    % m odd => \lceil m/2\rceil even:
    \begin{tikzpicture}[scale=0.44]
      \fill[black!20] (-1/3,-1/3) -- (1/3,-1/3) -- (1/3,10/3) -- (-1/3,10/3);
      \fill[black!20] (-1/3,8/3) -- (-1/3,10/3) -- (10/3,10/3) -- (10/3,8/3);
      \fill[black!10] (8/3,8/3) -- (10/3,10/3) -- (10/3,-1/3) -- (8/3,-1/3);
      \fill[black!10] (8/3,-1/3) -- (8/3,1/3) -- (17/3,1/3) -- (19/3,-1/3);
      \fill[black!20] (17/3,1/3) -- (19/3,-1/3) -- (19/3,10/3) -- (17/3,10/3);
      \fill[black!20] (17/3,8/3) -- (17/3,10/3) -- (28/3,10/3) -- (28/3,8/3);
      \fill[black!20] (-3-1/3,-10/3) -- (-3-1/3,-8/3) -- (2,-8/3) -- (2,-10/3);
      \fill[black!10] (2,-10/3) -- (2,-8/3) -- (7,-8/3) -- (7,-10/3);
      \fill[black!20] (7,-10/3) -- (7,-8/3) -- (12,-8/3) -- (12,-10/3);
      \fill[black!20] (1/3,1/3) -- (1/3,-1/3) -- (-1,-1/3) -- (-1,1/3);
      \fill[black!10] (-1,1/3) -- (-1,-1/3) -- (-2,-1/3) -- (-2,1/3);
      \fill[black!20] (-2,1/3) -- (-2,-1/3) -- (-3-1/3,-1/3) -- (-3-1/3,1/3);
      \fill[black!20] (-3-1/3,-10/3) -- (1/3-3,-10/3) -- (1/3-3,-2) -- (-1/3-3,-2);
      \fill[black!10] (-3-1/3,-2) -- (1/3-3,-2) -- (1/3-3,-1) -- (-1/3-3,-1);
      \fill[black!20] (-3-1/3,-1) -- (1/3-3,-1) -- (1/3-3,0) -- (-1/3-3,0);
      \fill[black!20] (28/3,10/3) -- (26/3,10/3) -- (26/3,-1/3) -- (28/3,-1/3);
      \fill[black!20] (26/3,-1/3) -- (26/3,1/3) -- (10,1/3) -- (10,-1/3);
      \fill[black!10] (10,-1/3) -- (10,1/3) -- (11,1/3) -- (11,-1/3);
      \fill[black!20] (11,-1/3) -- (11,1/3) -- (12,1/3) -- (12,-1/3);
      \fill[black!20] (37/3,1/3) -- (35/3,1/3) -- (35/3,-1) -- (37/3,-1);
      \fill[black!10] (37/3,-1) -- (35/3,-1) -- (35/3,-2) -- (37/3,-2);
      \fill[black!20] (37/3,-2) -- (35/3,-2) -- (35/3,-10/3) -- (37/3,-10/3);
      \draw (0,0) -- (0,3);
      \fill (0,3) circle (2pt);
      \draw[<->] (-7/12,0) --node[left]{$A$} (-7/12,3);
      \draw (-1/2,3) -- (-2/3,3);
      \draw[dotted] (0,3) -- (-1/2,3);
      \draw[<->] (-7/12,3) --node[left]{$\delta$} (-7/12,10/3);
      \draw[dotted] (-1/3,10/3) -- (-1/2,10/3);
      \draw (-1/2,10/3) -- (-2/3,10/3);
      \draw (0,3) -- (3,3);
      \fill (3,3) circle (2pt);
      \draw[dashed] (3,3) -- (3,0);
      \draw[dashed] (3,0) -- (6,0);
      \fill (6,0) circle (2pt);
      \draw (6,0) -- (6,3);
      \fill (6,3) circle (2pt);
      \draw (6,3) -- (9,3);
      \fill (9,3) circle (2pt);
      \draw (0,0) -- (-1,0);
      \draw[dashed] (-1,0) -- (-2,0);
      \draw (-2,0) -- (-3,0);
      \fill (-3,0) circle (2pt);
      \fill (0,0) circle (2pt);
      \node at (5/6,1/2) {$p_0$};
      \draw[->] (1/2,1/6) -- (1/9,1/24);
      \node at (49/6,1/2) {$p_m$};
      \draw[->] (17/2,1/6) -- (80/9,1/24);
      \fill (9,0) circle (2pt);
      \draw[white] (10,-14/3) circle (1pt); 
      \draw (9,3) -- (9,0);
      \fill (9,0) circle (2pt);
      \draw (9,0) -- (10,0);
      \draw[dashed] (10,0) -- (11,0);
      \draw (11,0) -- (12,0);
      \fill (12,0) circle (2pt);
      \draw (-3,-3) -- (2,-3);
      \fill (-3,-3) circle (2pt);
      \fill (3,0) circle (2pt);
      \draw[dashed] (2,-3) -- (7,-3);
      \draw (7,-3) -- (12,-3);
      \fill (12,-3) circle (2pt);
      \draw (-3,-3) -- (-3,-2);
      \draw[dashed] (-3,-2) -- (-3,-1);
      \draw (-3,-1) -- (-3,0);
      \draw (12,0) -- (12,-1);
      \draw[dashed] (12,-1) -- (12,-2);
      \draw (12,-2) -- (12,-3);
      \draw (-1/2-3,0) -- (-2/3-3,0);
      \draw[dotted] (-3,0) -- (-1/2-3,0);
      \draw[<->] (-3-7/12,0) --node[left]{$2A$} (-3-7/12,-3);
      \draw (-1/2-3,-3) -- (-2/3-3,-3);
      \draw[dotted] (-3,-3) -- (-1/2-3,-3);
      \draw (25/2,0) -- (38/3,0);
      \draw[dotted] (12,0) -- (25/2,0);
      \draw[<->] (12+7/12,0) --node[right]{$2A$} (12+7/12,-3);
      \draw[dotted] (12,-3) -- (25/2,-3);
      \draw (25/2,-3) -- (38/3,-3);
      \draw (-3,-7/2) -- (-3,-11/3);      
      \draw[dotted] (-3,-3) -- (-3,-7/2);
      \draw[<->] (-3,-43/12) --node[below]{$2 A$} (0,-43/12);
      \draw (0,-7/2) -- (0,-11/3);      
      \draw[dotted] (0,-3) -- (0,-7/2);
      \draw[<->] (0,-43/12) --node[below]{$\lfloor m/2\rfloor A$} (9,-43/12);
      \draw (9,-7/2) -- (9,-11/3);
      \draw[dotted] (9,-3) -- (9,-7/2);
      \draw[<->] (9,-43/12) --node[below]{$2 A$} (12,-43/12);
      \draw (12,-7/2) -- (12,-11/3);
      \draw[dotted] (12,-3) -- (12,-7/2);
    \end{tikzpicture}
    \caption{Schematic illustration of the construction from the proof of
      Theorem~\ref{thm:DCAShardness} (and Corollary~\ref{cor:DCAShardness})
      in the cases $m$ and $\lceil m/2\rceil$ even (top left), $m$ even and
      $\lceil m/2\rceil$ odd (top right), or $m$ odd (bottom): The corridor
      $C^\rho_\delta(\cF)$ (shaded region) contains all points with
      $\ell_\infty$-norm distance at most $\delta$ from the piecewise
      linear shape contour $\cF$ (solid curve). All $m$ segments of the
      path $p_0$-$p_1$-$\cdots$-$p_m$ have length $A$; the corridor is
      $2\delta$ wide. The dashed and lighter-shaded parts of $\cF$ and
      $\cC^\rho_\delta(\cF)$, respectively, illustrate a continuation of
      the discernible pattern (or of long auxiliary segments) in accordance
      with the actual instance~size.}
    \label{fig:npproofconstruction2}
\end{figure}

It is now easily seen, analogously to the proof of
Theorem~\ref{thm:DPAShardness}, that a solution to a ``yes''-instance of
\textsc{3-Partition} is in one-to-one correspondence with (the core part
of) that of a ``yes''-instance of the constructed DCA-S instance (in any
case); in particular, the additional edge types serve simply to ``close the
loop'' and cannot be used in any other parts of the $\cG$-cycle (note that,
since $\delta<A/4$, $\lfloor m/2\rfloor A\geq 3A>2A>A+2\delta$, so even the
shortest new edge does not fit anywhere else except in the accordingly
designed new segments). Thus, the previously crucial problem of connecting
$p_0$ and $p_m$ by a valid $\cG$-path is subsumed by that of finding a
valid $\cF$-resembling $\cG$-cycle here. (Unlike in the fixed end-point
variant of DPA-S, here, a feasible DCA-S solution never needs to overlay
$\cF$ exactly, but due to the choice of $\delta$, there still can be only
exactly three edges of type $e_\ell$, $\ell\in[k]$, per segment of the
shape contour subpath from $p_0$ to $p_m$.)  Furthermore, since the
previous construction was only modified by adding a constant number of new
polynomially-bounded elements but is otherwise completely analogous, it can
be carried out in polynomial time with all occurring numbers and their
encoding lengths being polynomially bounded by the instance size (i.e.,
polynomial in~$m$). Thus, the above extended reduction proves \NP-hardness
in the strong sense of the DCA-S problem.
\end{proof}

\begin{remark}
  In the above proof of Theorem~\ref{thm:DCAShardness}, the budget
  constraints for auxiliary edges added to the original construction are
  not essential and could just as well be omitted (i.e., using infinite
  budgets).
\end{remark}

Analogously to the corresponding results for DPA-S, we immediately also
obtain intractability of the DCA-S variants involving a feasible corridor
(cf. Figure~\ref{fig:npproofconstruction2}) and/or approximation error
minimization:
\begin{corollary}\label{cor:DCAShardness}
  The DCA-S problem is \NP-hard in the strong sense (under the same
  restrictions as listed in Theorem~\ref{thm:DCAShardness}) if the
  $\cF$-resembling $\cG$-cycle $C$ is required to lie within a corridor
  $C^\sigma_\varepsilon(\cF)$ around $\cF$ and/or the objective of
  minimizing an approximation error $\alpha(C,\cF)\geq 0$ is included. In
  the latter case, hardness persists even if no sample point neighborhood
  (and/or corridor) containment is required.
\end{corollary}
\begin{proof}
  We can extend the proof of Theorem~\ref{thm:DCAShardness} along the same
  lines as in that of Corollary~\ref{cor:DPAShardness}.
\end{proof}

A more basic problem simply asks whether a given DAS allows to connect two
given points in the plane, without asking for any resemblance to some curve
segment or shape contour. This is precisely the DPC problem defined
earlier; as it turns out, even this fundamental problem is already (weakly)
\NP-hard.

\begin{theorem}\label{thm:DPChardness}
  The DPC problem is \NP-hard, even if restricting the DAS to one node
  type, horizontal and vertical orientations, and integral-length edge
  types. For rational input, DPC is contained in \NP\ (and thus \NP-complete).
\end{theorem}
\begin{proof}
  We now reduce from \textsc{Partition}, which is a weakly \NP-hard
  problem~\cite{GareyJohnson1979}. Given an instance $(a_1,\dots,a_n)$ of
  \textsc{Partition}, let $A\define\tfrac{1}{2}\sum_{i=1}^n a_i$ and
  construct an instance of DPC as follows (along the same lines as in the
  previous proof for DPA-S): Let~$k$ be the number of distinct $a_i$-values
  and associate with each such value (say, $a^\prime_\ell$) an edge
  type~$e_\ell$ with length $a^\prime_\ell$ to obtain
  $\cE\define\{e_\ell\st\ell=1,\dots,k\}$ and budgets
  $\cB\define\{b_e\st e\in\cE\}$, where
  $b_{e_\ell}\define\abs{\{i\in[3m]\st a_i=a^\prime_\ell\}}$. We again use
  a single node type $\cV\define\{v\}$ and let
  $\cD\define\{(d,\cV,\cE)\st d\in\{\pm(1,0),\pm(0,1)\}\}$, i.e., the
  horizontal and vertical directions are to be compatible with each node
  and edge type. This defines the DAS
  $\cG\define(\cV,\cE,\cB,\cD)$. Finally, we set $x_1\define(0,0)$ and
  $x_2\define(A,A)$. Clearly this construction can be carried out in
  polynomial time and space.

  It remains to demonstrate that the given \textsc{Partition} instance has
  a ``yes''-answer if and only if the constructed DPC instance does. To
  that end, note that there is a one-to-one correspondence between a
  partition $I',[n]\setminus I'$ such that
  $\sum_{i\in I'}a_i=\sum_{i\notin I}a_i(=A)$ and the edges with horizontal
  or vertical orientation, respectively: To connect $x_1$ and $x_2$, we
  need to ``go'' $A$ in both directions, which is possible if and only if
  the total of $n$ edges can be partitioned into vertical and horizontal
  edges with respective length sums~$A$.

  While this shows \NP-hardness of DPC, when restricting to rational input
  (coordinates, orientations and edge lengths), containment in \NP\ is
  obvious, proving the second claim and thus completing the proof.
\end{proof}

The above result has several consequences that may be of further,
separate interest:
\begin{corollary}\label{cor:DPCl0minHardness}
  Given a DAS $\cG$ and two points $x_1,x_2\in\R^2$, it is \NP-hard to
  find a shortest valid $\cG$-path connecting $x_1$ and $x_2$, i.e., one
  with the smallest number of edges. 
\end{corollary}
\begin{proof}
  The proof of Theorem~\ref{thm:DPChardness} can be directly adapted
  to the corresponding decision problem ``Given $\cG$, $x_1$, $x_2$
  and a positive integer $k$, can $x_1$ and $x_2$ be connected by a
  valid $\cG$-path using at most $k$ edges?'', by letting $k\define n$ (the
  number of elements of the original input \textsc{Partition}
  instance).
\end{proof}

Furthermore, we could reintroduce a curve segment and sample points (and/or
a feasible corridor) into the DPC problem and the proof of
Theorem~\ref{thm:DPChardness}, thereby obtaining another proof of
\NP-hardness for the DPA-S problem. Although Theorem~\ref{thm:DPAShardness}
gives a stronger result than a reduction from \textsc{Partition}, this
alternative proof shows that the DPA-S problem and its variants (from
Corollary~\ref{cor:DPAShardness}) still remain (weakly) \NP-hard \emph{even
  if the curve segment to be approximated is piecewise linear with a single
  right-angle turn}. Moreover, similarly extending the construction along
the same lines as in the proof of Theorem~\ref{thm:DCAShardness}, we also
obtain that the DCA-S variants remain (weakly) \NP-hard \emph{even when
  restricted to simple square shapes}. We omit the straightforward details
of the constructions, which mirror that in the previous proofs
(cf.~Figures~\ref{fig:npproofconstruction}
and~\ref{fig:npproofconstruction2}, in particular).

In the remainder of the paper, we will demonstrate that in spite of the
theoretical intractability, good solutions to the DAS shape contour
approximation problems can be obtained in practice within reasonable
time. For simplicity, we will mostly restrict our discussion to the contour
approximation problem (DCA-S) and omit an explicit treatment of the curve
segment approximation task (DPA-S). Indeed, it is straightfoward to adapt
the developed approaches and methods to the DPA-S problem.

%%%%%%%%%%%%%%%%%%%%%%%%%%%%%%%%%%%%%%%%%%%%%%%%%%%%%%%%%%%%%%%%%%%%%%%%%%%%%%%%%%%%%%%%%%%%%%%%%%%%%%%%%%%%%%%%%%%%%%%%%%%%%%%%%%%%%%%%%%%%%%%
%                                            Methodology & Algorithm                                                                          %
%%%%%%%%%%%%%%%%%%%%%%%%%%%%%%%%%%%%%%%%%%%%%%%%%%%%%%%%%%%%%%%%%%%%%%%%%%%%%%%%%%%%%%%%%%%%%%%%%%%%%%%%%%%%%%%%%%%%%%%%%%%%%%%%%%%%%%%%%%%%%%%
\section{Methodology \& Algorithm}\label{sec:methodology}
Computing a ``best'' $\cF$-resembling $\cG$-cycle (w.r.t.\ some quality
measure) is clearly a non-trivial task posing many challenges, as outlined
in the introduction. Aside from the theoretical intractability of even
deciding feasibility established in the previous section, evaluating error
measures is generally not possible without knowing the actual
$\cG$-cycle and therefore hard to directly incorporate into an
optimization scheme.

In our general setting, graph-based methods (such as those working on
grid-graphs found in the schematization context, cf., e.g.,
\cite{Meulemans2014,BoutsETAL2016}) are inadequate since it is not known
(or computationally infeasible to set up) a priori where possible struts
may be located, and location is crucial for assigning weights that reflect
closeness to $\cF$. Indeed, the shape similarity of two contours is not
trivial to formalize. From a perceptual point of view, distance, relative
orientation, and even curvature distribution would have to be taken into
account, which is quite difficult in our DAS setting with discrete strut
lengths and orientations. For simplicity, we therefore resort to a purely
distance-based metric and leave generalizations to future work. However,
pure max-norms such as the Hausdorff-distance or the Fr\'{e}chet-distance
are not well-suited for our purposes since they would only provide guidance
for the optimization in those regions of the contour where the deviation is
maximal while changes in other regions of the contour do not affect the
metric at all. Thus, we have to find a well-balanced compromise between a
prescribed (desired) maximum approximation tolerance and an averaged
deviation norm that minimizes, or allows to control, the distance
everywhere regardless of the global tolerance threshold.

We find this compromise in the following way: We propose a three-phase
approach that first obtains a set of sample points along $\cF$, then solves
a mixed-integer program (MIP) to obtain a solution to a DCA-S problem with
minimal number of struts, and finally explores the possible arrangements of
the selected struts in order to find the best approximation to $\cF$ that
can be constructed using them. For the DCA-S subproblem, we prescribe a
``global'' desired approximation tolerance $\delta$ and, by solving a MIP
in the second stage, then find a set of points $q_i$ which lie within a
$\delta$-neighborhood of the sample points and which can be connected (in a
given order) by $\cG$-path segments that form a $\cG$-cycle. In the final
stage of the algorithm we then compute, for each segment, the permutation
of the struts of that segment that leads to the minimal averaged deviation,
i.e., to the minimal area between the two contours ($\cF$ and the computed
$\cG$-cycle). This third stage is, of course, not guaranteed to actually
respect the desired tolerance $\delta$ (except around the sampled points),
but it usually provides good shape similarity if the anchor points were
chosen such that the contour segments between them are sufficiently simple
(i.e., relatively straight).

Focussing on $\cG$-paths (between regions around the sample points) with
small number of struts avoids the problem of inaccessibility of suitable
placement-dependent strut/path-quality measures on the one hand, and on the
other hand, quite intuitively implements the fact that desirable solutions
will not stray too far from $\cF$ (or contain odd zigzagging parts) and
hence necessarily have relatively few struts. It also makes
edge-crossings/overlaps less likely -- although we do not explicitly
enforce this planarity of the sought $\cG$-paths or cycles; it turned out
to be very often fulfilled automatically in our experiments, and can
probably often easily be remedied by postprocessing in practice\footnote{It
  must be noted that there may be instances for which solution
  self-intersection issues \emph{cannot} be fixed at all. This is more
  likely to occur if the resolution is too low, i.e., struts are quite long
  compared to the overall global scale of the input. At higher resolutions,
  this should no longer be problematic, by virtue of minimizing the number
  of struts per segment. Unfortunately, an adequate choice of
  scale/resolution is not always clear a priori.}. Once the number of
struts is known (for each segment of the $\cG$-cycle to be computed), we
can exploit the fact that struts in a $\cG$-path can be arbitrarily
permuted without disconnecting the end points, and simply enumerate all
such permutations and evaluate the above-mentioned averaged deviation
approximation error to $\cF$ for each one in order to identify the best
arrangement using the computed strut collection. (Should full enumeration
become too expensive, which may happen if a segment contains many struts,
one may also resort to a greedy construction that determines a permutation
by sequentially picking locally optimal struts.)

In the remainder of this section, we will first introduce mixed-integer
programming (MIP) formulations for some of the DAS shape approximation
problems, and then describe in detail the overall algorithm to compute
$\cF$-resembling $\cG$-cycles of high quality. In order to avoid
repetitions when turning to computational experiments later, we directly
discuss the MIPs and algorithms in the context of \emph{Zometool} shape
approximation; thus, $\cG$ will henceforth always refer to the (standard)
Zometool system described in the introduction (see
Section~\ref{subsec:zome}). This concretization notwithstanding, we
emphasize that all ingredients straightforwardly generalize to arbitrary
other DASs.

\subsection{MIP Formulations}\label{subsec:MIPs}  
Recall from the introduction that, assuming a Zometool node is placed at
the origin $(0,0)$ and that the shortest blue strut length is scaled to
$b_1=2$, every point reachable by Zometool components has coordinates of
the form $(\alpha_1+\phi\beta_1,\alpha_2+\phi\beta_2)$ with
$\alpha_i,\beta_i\in\Z$ for $i=1,2$.
For every different strut in the Zometool system, there are four admissible
directions, cf. Figure~\ref{fig:ZometoolCheatSheet1a}, where two of them
are the respective opposites of the other two. Thus, for each strut color
$e\in\{b,r,y\}$ (blue, red, yellow) and length $i\in\{s,m,\ell\}$ (short,
medium, long), we may collect the corresponding translation vectors as
columns of the following matrices: 
\begin{align*}
  M_s^b = \left(\begin{array}{cc}2 & 0\\0 & 2\end{array}\right),                    &\quad& M_m^b = \left(\begin{array}{cc}2\phi & 0\\0 & 2\phi\end{array}\right),   &\quad& \\
   &\quad& M_\ell^b = \left(\begin{array}{cc}2+2\phi & 0\\0 & 2+2\phi\end{array}\right),  &\quad& \\
  M_s^r = \left(\begin{array}{cc}\phi & \phi\\-1 & 1\end{array}\right),             &\quad& M_m^r = \left(\begin{array}{cc}1+\phi & 1+\phi\\-\phi & \phi\end{array}\right), &\quad& \\
  &\quad& M_\ell^r = \left(\begin{array}{cc}1+2\phi & 1+2\phi\\-1-\phi & 1+\phi\end{array}\right),&\quad& \\
  M_s^y = \left(\begin{array}{cc}-1+\phi & -1+\phi\\-\phi & \phi\end{array}\right), &\quad& M_m^y = \left(\begin{array}{cc}1 & 1\\-1-\phi& 1+\phi\end{array}\right),        \\
&\quad& M_\ell^y = \left(\begin{array}{cc}\phi & \phi\\-1-2\phi & 1+2\phi\end{array}\right).&\quad& \\
\end{align*}
For example, starting at a node $(x,y)$, taking a ``step'' along a strut
$e$ of length~$i$ in positive or negative/opposite direction of, say, the
second orientation thus leads to target node coordinates
$(x+(M^e_i)_{12},y+(M^e_i)_{22})$ or $(x-(M^e_i)_{12},y-(M^e_i)_{22})$,
respectively. 

\subsubsection{Connecting Points by Zome-Paths}\label{subsubsec:MIPsZomeDPC} 
Now, let us first consider the (Zome-)DPC problem: Can we connect two given
points $x_1,x_2\in\R^2$ by a $\cG$-path? This question can be answered by
solving the following integer feasibility problem (IFP):
\begin{align}\label{mip:DPCifp1}
  \text{find}\quad& \gamma_{e,i}^{+},\gamma_{e,i}^{-}\in\Z^2_+\quad\forall
                    e\in\{b,r,y\}, i\in\{s,m,\ell\}\\
\nonumber  \text{s.t.}\quad& x_2 = x_1 + \sum_{e\in\{b,r,y\}}\sum_{i\in\{s,m,\ell\}}M_i^e\Big(\gamma_{e,i}^{+}-\gamma_{e,i}^{-}\Big)\\
\nonumber                  & \left(~(\gamma_{e,i}^{+})_j\cdot(\gamma_{e,i}^{-})_j=0\quad\forall e\in\{b,r,y\}, i\in\{s,m,\ell\}, j\in\{1,2\}~\right)
\end{align}
Here, for each $e\in\{b,r,y\}$ and $i\in\{s,m,\ell\}$, the variables
$(\gamma_{e,i}^{+})_j\in\Z_+$ and $(\gamma_{e,i}^{-})_j\in\Z_+$ specify the
number of times the $j$-th corresponding strut (i.e., the $j$-th column
vector of matrix $M_i^e$) is used in positive direction (directly as given
in $M_i^e$, superscript ``$+$'') or in negative direction (superscript
``$-$''), respectively. The equality constraint models the desired
connectivity by expressing point $x_2$ as $x_1$ plus an integer linear
combination of the available struts. Thus, the DPC question can be answered
in the affirmative if and only if a feasible variable assignment for the
above IFP can be found. The last, optional complementarity constraint may
be used to exclude paths that contain segments of same-color same-length
struts being used in opposite directions (which merely shift the Zome-path
part in between in the plane but serve no further purpose regarding
endpoint connectivity). To avoid nonlinearity, it could also be
formulated as a special ordered set type 1 (SOS-1) constraint\footnote{
  SOS-1$(\gamma_{e,i}^{+}, \gamma_{e,i}^{-})$ constraints enforce that only
  one of $\gamma_{e,i}^{+}, \gamma_{e,i}^{-}$ may be nonzero; they can be
  realized, e.g., by introducing auxiliary binary variables $z^{\pm}_{e,i}$ and
  linear constraints $\gamma_{e,i}^{\pm}\leq \cM z^{\pm}_{e,i}$ and
  $z^+_{e,i}+z^-_{e,i}\leq 1$, where $\cM$ is a sufficiently large constant
  (here, the respective strut budget sizes suffice, if available). Modern
  MIP solvers can handle SOS-1 constraints efficiently.}, or be
omitted entirely for suitable objective functions. Moreover, note that we
could assume w.l.o.g.\ that $x_1=(0,0)$ (otherwise, simply subtract $x_1$
from $x_1$ and $x_2$) and then may also check a priori whether $x_2$ even
has Zometool coordinates, i.e., whether
$x_2=(\alpha_1+\phi\beta_1,\alpha_2+\phi\beta_2)$ with some
$\alpha_i,\beta_i\in\Z$ for $i=1,2$.

The above feasibility problem may become numerically instable because of the
explicit occurrence of the golden ratio~$\phi$ (in all but one $M_i^e$). However,
we can make use of the special form of Zometool coordinates and, by lifting
the problem into $4$-dimensional space, obtain an equivalent IFP that only
contains small integral coefficients. To that end, let
$\hat{x}\define(\alpha^x_1,\beta^x_1,\alpha^x_2,\beta^x_2)\in\Z^4$ be the
integer representation of a point $x$ with Zometool coordinates
$x=(\alpha^x_1+\beta^x_1\phi,\alpha^x_2+\beta^x_2\phi)$. Analogously, let
$\hat{M}_i^e$ be the $4\times 2$ matrix obtained from $M_i^e$ by replacing
its columns with the respective integer representations of the
corresponding Zometool coordinate vectors. Now, we can reformulate the DPC
integer feasibility problem, where w.l.o.g.\ $x_1=(0,0)$, as:
\begin{align}\label{mip:DPCifp2}
  \text{find}\quad& \gamma_{e,i}^{+},\gamma_{e,i}^{-}\in\Z^2_+\quad\forall
                    e\in\{b,r,y\}, i\in\{s,m,\ell\}\\
\nonumber  \text{s.t.}\quad&
                    \sum_{e\in\{b,r,y\}}\sum_{i\in\{s,m,\ell\}}\hat{M}_i^e\Big(\gamma_{e,i}^{+}-\gamma_{e,i}^{-}\Big)  = \hat{x}_2 \\
\nonumber                  & \left(~(\gamma_{e,i}^{+})_j\cdot(\gamma_{e,i}^{-})_j=0\quad\forall e\in\{b,r,y\}, i\in\{s,m,\ell\}, j\in\{1,2\}~\right)
\end{align}
Note that $\abs{(\hat{M}_i^e)_{j,k}}\in\{0,1,2\}$ for all $e\in\{b,r,y\}$
and $i\in\{s,m,\ell\}$ ($j\in\{1,2,3,4\}$, $k\in\{1,2\}$).

Based on this integer formulation of Zome connectivity, we can, in
particular, formulate the DPC variant that asks for a shortest Zome-path,
i.e., one that requires the smallest number of struts to connect the given
input points:
\begin{align}\label{mip:DPCl0min}
  \min\quad       & \sum_{e\in\{b,r,y\}}\sum_{i\in\{s,m,\ell\}}\sum_{j=1}^{2}(\gamma_{e,i}^{+})_j+(\gamma_{e,i}^{-})_j\\
\nonumber  \text{s.t.}\quad& \sum_{e\in\{b,r,y\}}\sum_{i\in\{s,m,\ell\}}\hat{M}_i^e\Big(\gamma_{e,i}^{+}-\gamma_{e,i}^{-}\Big)  = \hat{x}_2 \\
\nonumber                  & \gamma_{e,i}^{+},\gamma_{e,i}^{-}\in\Z^2_+\quad\forall e\in\{b,r,y\}, i\in\{s,m,\ell\}
\end{align}
As alluded to earlier, here, minimizing the sum of all variables (a linear
function) will automatically induce the complementarity conditions
$(\gamma_{e,i}^{+})_j\cdot(\gamma_{e,i}^{-})_j=0$, which are therefore
omitted.

Finally, note that strut budget restrictions can be straightforwardly
integrated into all of the above models by adding constraints of the form
\begin{equation}\label{eq:DPCifp_strutbudgets}
  \ones^\top\gamma_{e,i}^{+}+\ones^\top\gamma_{e,i}^{-} = \sum_{j=1}^{2}(\gamma_{e,i}^{+})_j+(\gamma_{e,i}^{-})_j\leq B^e_i,
\end{equation}
where $B^e_i\in\Z_+$ is the maximal allowed number of struts of color~$e$
and length~$i$ (and $\ones$ is the all-ones vector).

\subsubsection{Zome-Cycles Resembling a Shape Contour}\label{subsubsec:MIPsZomeDCA-S}
Let us now turn to the problem of approximating a shape contour by a
Zometool cycle. We handle the nontrivial task of $\cF$-resemblance by
requiring our $\cG$-cycle to pass through the respective neighborhoods of
predefined sample points along $\cF$, i.e., we consider the DCA-S
problem. Concrete options for choosing the sample points will be discussed
in the next section; for now, assume they are already given.

Again, we focus on the objective of finding such a $\cG$-cycle that
consists of as few struts as possible (per segment between nodes placed in
sample point neighborhoods). As mentioned before, this property can
generally be seen as a loose proxy to the ultimate goal of minimizing some
measure of deviation from the actual shape contour, since a good
$\cG$-cycle w.r.t.\ the latter will neither contain undesirable detours
nor zigzagging segments, both of which are also intuitively avoided by
few-strut $\cG$-cycles.

Moreover, the basic idea for our MIP model for DCA-S builds on the
DPC-IP~\eqref{mip:DPCl0min}: We desire (at least) one Zometool node to lie
in the vicinity of each sample point (i.e., in the corresponding
$\cN^\rho_\delta(p_i)$), and require each such Zometool node to be
connected by a $\cG$-path to the Zometool node in the next sample point
neighborhood encountered when moving along $\cF$ in a fixed ``direction''
(clockwise or counter-clockwise). To that end, we can employ point
connectivity constraints similar to those employed in the IP models
discussed in Section~\ref{subsubsec:MIPsZomeDPC}. Additionally, we label
the variables that determine the numbers of struts to be used by segment,
and introduce new variables that pertain to the Zometool nodes to be put
into each sample point neighborhood. Finally, by using two continuous
coordinate variables ($g\in\R^2$), we can model the global shift of the
Zometool construction, so that the first Zometool node can be interpreted
w.l.o.g.\ to lie at the origin $(0,0)$. This shift is then used to constrain
the Zometool nodes at the end of each $\cG$-path ($\cG$-cycle segment) to
lie in the predefined sample point neighborhoods $\cN^\rho_\delta(p_i)$,
which we define as $\ell_\infty$-norm boxes of uniform width $\delta\geq 0$
around each respective sample point. Putting this all together yields the
following MIP (with optional budget and SOS-1 constraints), where
$p_1,\dots,p_k\in\R^2$ are the given sample points on $\cF$:

\begin{align}\label{mip:DCA-Sl0min}
  \min\quad       & \sum_{\kappa=1}^{k}\sum_{e\in\{b,r,y\}}\sum_{i\in\{s,m,\ell\}}\sum_{j=1}^{2}(\gamma_{e,i}^{\kappa,+})_j+(\gamma_{e,i}^{\kappa,-})_j\\
\nonumber  \text{s.t.}\quad& \hat{x}_1=(0,0,0,0)\\
\nonumber                  & -\delta\ones\leq \big((\hat{x}_\kappa)_1+(\hat{x}_\kappa)_2\phi,(\hat{x}_\kappa)_3+(\hat{x}_\kappa)_4\phi)+g-p_\kappa\leq\delta\ones\quad\forall \kappa\in[k]\\
\nonumber                  & \hat{x}_{(\kappa\,\text{mod}\,k)+1} = \hat{x}_{\kappa}+\sum_{e\in\{b,r,y\}}\sum_{i\in\{s,m,\ell\}}\hat{M}_i^e\Big(\gamma_{e,i}^{\kappa,+}-\gamma_{e,i}^{\kappa,-}\Big)\quad\forall \kappa\in[k]\\
\nonumber                  & g\in\R^2\\
\nonumber                  & \hat{x}_\kappa\in\Z^4\quad\forall \kappa\in[k]\\
\nonumber                  & \gamma_{e,i}^{\kappa,+},\gamma_{e,i}^{\kappa,-}\in\Z^2_+\quad\forall e\in\{b,r,y\}, i\in\{s,m,\ell\}, \kappa\in[k]\\
\nonumber                  & \left(~\sum_{\kappa=1}^{k}\Big(\ones^\top\gamma_{e,i}^{\kappa,+}+\ones^\top\gamma_{e,i}^{\kappa,-}\Big) \leq B^e_i\quad\forall e\in\{b,r,y\}, i\in\{s,m,\ell\}~\right)\\
\nonumber                  &
                             \left(~\text{SOS-1}\left(\gamma_{e,i}^{\kappa,+},\gamma_{e,i}^{\kappa,-}\right)\quad\forall e\in\{b,r,y\}, i\in\{s,m,\ell\}, \kappa\in[k]~\right),
\end{align}
where we again abbreviate $[k]\define\{1,\dots,k\}$.

We remark that~\eqref{mip:DCA-Sl0min} may take a prohibitively long time to
fully solve (i.e., compute a certifiably optimal solution or detect model
infeasibility) even with state-of-the-art commercial MIP solvers like
Gurobi or CPLEX. In fact, this observation already holds true for the
considerably simpler DPC problem~\eqref{mip:DPCl0min}. It may be worth
mentioning that this should generally be attributed to the respective
problems' \NP-hardness (cf. Section~\ref{sec:hardness}), not the
mixed-integer programming approach itself. 
Nevertheless, we empirically observed that if a good (``short'') solution
exists, \eqref{mip:DPCl0min} is solved to optimality by modern MIP solvers
very quickly (typically within a few seconds). This motivates, on the one
hand, the extension from DPC to DCA-S (i.e., to~\eqref{mip:DCA-Sl0min}),
and on the other hand, suggests a simple time-limit-heuristic to judge
model infeasibility or (sufficient for our purposes) undesirability of
possible ``long'' solutions: If the solution process takes more than (say)
a few minutes to either find any feasible solution or one with acceptable
optimality gap, terminate and declare that no good solutions exist or at
least can be found quickly (within the current model parameters/design
decisions). When finding relatively good solutions is not a problem, but
progress stalls and it takes too long to determine exact optimality, we may
of course also terminate prematurely and simply use the $\cG$-cycle from
the current best known solution. The MIP optimality gap then allows to
judge how ``far'' from optimal the solution thus retrieved still is, i.e.,
how much improvement (w.r.t.\ the number of utilized struts) could still be
possible by letting the solver continue. Often, good MIP solutions are
indeed found relatively quickly, and most of the remaining solver running
time is spent proving optimality (by improving dual bounds until the
optimality gap is closed). Later, in Section~\ref{subsec:slack}, we will
also describe an extension of the above model that is more successful on
instances where the solver struggles to find a feasible solution.

Finally, it is worth noting that we also considered penalizing arclength
deviation as a different objective, or in a weighted combination with the
cardinality objective from~\eqref{mip:DCA-Sl0min}, and furthermore
considered (lower and/or upper) bounding the segment arclengths to a factor
of the associated input contour segment's arclength. The arclength
difference can be seen as another loose proxy for the approximation error
(being small if the latter is, though not necessarily vice versa) that can
be evaluated regardless of the final global positioning of the
DAS-structure. While such model modifications might yield slight
improvements (w.r.t.\ the visual perception of the resulting shape
approximation quality) in some cases, at least the objective variations
appear to make the MIPs significantly harder to solve, and the obtained
solutions seemed more likely to lead to self-intersections
(``non-planarity''). Therefore, we do not consider these modifications any
further in the following.

\subsection{Algorithm for Zometool Shape Contour Approximation}\label{subsec:ZomeDCAalgo}
In order to obtain a practical method able to produce good results, two
aspects remained open so far: The \emph{(model) design decision} of how to
choose appropriate sample points (and the neighborhood radius~$\delta$),
and the \emph{solution construction}, i.e., how to determine the ``best''
(valid) $\cF$-resembling $\cG$-cycle among all those sharing the same
combination of struts per segment that was computed by the above
MIP. (Recall that any DAS path can be rearranged without disconnecting its
endpoints simply by permuting its struts/edges.)  Indeed, the
MIP~\eqref{mip:DCA-Sl0min} is the main ``work horse'' for our shape
approximation method, and we combine it with shape contour sampling and
optimal $\cF$-resembling $\cG$-cycle construction to the overall scheme
outlined in Algorithm~\ref{algo:main}.

\begin{algorithm}[t]
  \caption{Zometool Shape Contour Approximation Scheme}
  \label{algo:main}
  \begin{algorithmic}[1]
    \REQUIRE{(Zome-)DAS $\cG=(\cV,\cE,\cD,\cB)$, parameter $\delta\geq 0$, shape contour $\cF$}
    \ENSURE{a shortest $\cF$-resembling $\cG$-cycle running through all $\delta$-neighborhoods of all sample points}
    \STATE obtain sample points $p_1,\dots,p_k$ along $\cF$
    \STATE set up and solve the MIP~\eqref{mip:DCA-Sl0min}
    \FOR{each segment $1,\dots,k$ between the computed (Zometool) nodes ($g+\{x_1,\dots,x_k\}$)}
       \STATE obtain an approximation error minimizing sequence of the associated selected struts
    \ENDFOR
    \RETURN $\cF$-resembling $\cG$-cycle consisting of the concatentation of segment-optimal $\cG$-paths
  \end{algorithmic}
\end{algorithm}

The shape contour $\cF$ can be given, e.g., as a parametric function or
implicitly via a distance function/field. Any sampling scheme naturally
depends on how $\cF$ is provided or can be accessed. In the following, we
describe a quite general setting that we also used in our implementation:
distance fields and piecewise-linear $\cF$ obtained from these. More
precisely, assume we are given a (discretized) distance field covering a
rectangular portion of $\R^2$, in the form of a matrix of values specifying
the average distance of points within small square cells in 2D space (i.e.,
every matrix entry is the average distance value for such a cell). The size
of these squares' sides is determined relative to the (shortest) strut
length of the employed DAS by a scaling/accuracy parameter~$s$, where for
$s=1$, a globally fixed ratio is used. (Thus, for $s>1$, struts become
longer relative to the input shape.) With infinite discretization
precision, points lying exactly on the shape contour (boundary) would yield
a distance value of zero. We assume that circle-like shapes are processed,
i.e., the shape contours are closed curves that enclose exactly one
connected area, so that the terms ``inside'' and ``outside'' of the shape
are well-defined. Thus, cells containing points that (mostly) lie inside
the shape have negative distance values, and those outside have positive
distance values. Such a distance field provides a simple way to evaluate
strut costs, and can also be used to obtain a polygonal approximation of
the true shape by applying a 2D version of the marching cubes
algorithm~\cite{LorensenCline1987}. This approximation is what we will, for
simplicity, henceforth refer to as $\cF$; note that it is piecewise-linear
and fully described by an ordered sequence of points, say
$p^\cF_1,\dots,p^\cF_n$.

\subsubsection{Sampling the Input Shape Contour}\label{subsubsec:sampling}
Since the MIP~\eqref{mip:DCA-Sl0min} is oblivious to $\cF$ beyond enforcing
the placement of Zometool nodes in the vicinity of sample points, a careful
selection of the sample points (which define the partition of the
$\cG$-cycle into segments) seems crucial for ensuring that the $\cG$-cycle
to be computed actually does closely resemble $\cF$. The selection should
ideally achieve a balance between several competing goals: To keep the MIP
small (and therefore easier to solve), the number of segments should not be
too large. However, long segments contain many struts (leading to larger
effort to enumerate permutations in the final step of the overall
algorithm, or decreased likelihood of a greedy scheme to work well) and the
MIP objective of minimizing the number of struts will lead to strut
sequences that lose resemblance to $\cF$ (cutting off ``nooks and
crannies'' of the contour). We found that this last aspect appears to be
the most problematic, if one is willing to spend some time on the MIP
optimization. Similarly, the box-radius~$\delta$ for the MIP should not be
set too small, lest the MIP becomes infeasible, but not too large either,
since we want to ensure the $\cG$-cycle remains close to $\cF$ at least
around the sample points.

Thus, the sampling phase of our algorithm is of a ``high-level'' heuristic
nature; similarly, so is splitting the low-level optimization into the MIP
(feasibility) part and the (optimality) part in which struts are rearranged
to minimize the approximation error. Indeed, finding the optimal sample
point selection is essentially equivalent to solving the entire problem,
because if we knew it (and assuming~$\delta$ is sufficiently large), all
that would remain is finding and arranging the struts that achieve minimum
approximation error. Moreover, if the sampling is ``very good'', the
decomposition into MIP and strut-rearrangement can be expected to closely
resemble what could be achieved by the abstract algorithm that could
directly compute minimum approximation error solutions. (Recall that such a
method is impracticable, since approximation errors can only be evaluated
once strut locations are fixed.) Therefore, it is worth trying to design a
sampling mechanism that targets the possibly problematic aspects mentioned
above and also aims at retaining a MIP subproblem that is sufficiently
tractable in practice.

We have implemented and tested the following sampling schemes:
\begin{enumerate}
  
\item \emph{Uniformly by arclength:} Determine the (approximate) arclength
  $a$ of $\cF$ (here, the sum of the lengths of all straight lines between
  neighboring points $p^\cF_i$). 
  Starting with an arbitrary point, say $p_1\define p^\cF_1$, pick the
  others by moving from the previous point along the sequence of points
  defining~$\cF$ until the traversed curve segment reaches the average
  segment arclength $a/k$, and select the current point in that sequence
  as the next respective sample point.
\item \emph{By curvature:} For each $p^\cF_i$, $i=1,\dots,n$, we can obtain
  approximate curvature information w.r.t.\ $\cF$ as
  \[
    c_i\define 
    \text{arccos}\left(\frac{
        {\big( {p}^{\cF}_{ {(i-t)}_{n} } - {p}^{\cF}_{i} \big)}^{\top} \big( {p}^{\cF}_{ {(i+t)}_{n} } - {p}^{\cF}_{i} \big)
      }{
        \Norm{{p}^{\cF}_{ {(i-t)}_{n} } - {p}^{\cF}_{i}}_{2} \cdot \Norm{{p}^{\cF}_{ {(i+t)}_{n} } - {p}^{\cF}_{i}}_{2}
      }
    \right),
  \]
  which gives the angle (in radians) between the vectors connecting
  $p^\cF_i$ with its $t$-th predecessor and $t$-th successor (in the
  ordered sequence defining $\cF$), respectively, where $(j)_n$ denotes the index correctly
  shifted periodically back into $\{1,\dots,n\}$, so, e.g., $(j)_n=j-n$ for
  $n+1\leq j\leq 2n$ and $(j)_n=n-j$ for $-n<j\leq 0$. Here, $t\geq 1$ can be
  chosen arbitrarily; $t=1$ works, but larger values, say $5$ or $10$, could
  stabilize against tiny ``kinks'' in the contour that are not really
  relevant w.r.t.\ the actual shape. It holds that
  $c_i\in[0,\pi]$ for all $i$, where values close to $\pi/2$ amount to
  steep angles ($\pi/2$ signifies a $90^\circ$ turn) and values close to
  $0$ or $\pi$ indicate flat regions ($0^\circ$ or, equivalently, $180^\circ$).
  For simplicity, let us define adjusted curvatures
  \[
    \bar{c}_i\define\min\{c_i, \pi-c_i\}\in[0,\pi/2];
  \]
  note that steep angles now simply coincide with large adjusted curvature values.

  As a DAS is generally better suited to approximating relatively flat
  curves, and (especially when minimizing the number of edges/struts that
  is used) indeed tends to ignore or ``cut off'' bends in the contour along
  a segment, it intuitively makes sense to take curvature information into
  account when selecting sample points, in an effort to reduce the
  occurrence of such undesirable cut-offs. We considered several different
  schemes based on the (adjusted) curvature values:
  \begin{enumerate}
  \item \emph{Global largest $\bar{c}$-values:} Pick the $k$ points with largest
    $\bar{c}$-values as sample points.
  \item \emph{Segment-wise largest $\bar{c}$-values:} First determine the segments
    (e.g., by arclength as described above) and then pick
    the $k$ sample points as the respective points of $\cF$ with largest
    (adjusted) curvature from each respective segment.
  \item \emph{Separation-based:} Repeat picking an available point of largest
    $\bar{c}$-value as a sample point and marking all points in segments
    of arclength up to $\lambda a/k$ in both directions along $\cF$
    from this point on as unavailable, until $k$ points were chosen. Here,
    $\lambda>0$ determines how large the unsampled segments may become;
    e.g., aiming at equal-arclength segments, one could set $\lambda=0.5$.
    (Note that overly small values of $\lambda$ may lead to a bad, local
    concentration of the chosen sample points, whereas too large values
    could result in having marked all remaining points as unavailable
    before~$k$ sample points were chosen.)
  \end{enumerate}
\item \emph{By curvature with gap-filling:} Choosing points solely based on
  curvature information may lead to inadequate sampling of ``less curvy''
  segments of the input contour. This can be avoided by augmenting a
  separation-based curvature sampling similar to the one detailed above
  with a point-insertion scheme aiming at an efficient coverage of
  underrepresented segments with further sample points:
  \begin{enumerate}
  \item \emph{Separation-based with Euclidean farthest-point insertion:}
    Choose (up to) a fixed number~$k_c$ of sample points based on (adjusted)
    curvature obeying arclength separation requirements, then add more
    points by iteratively picking a point with maximum minimal Euclidean
    distance to one of the already selected sample points (up to a given
    maximum number $k^{\prime}$, and stopping early when the distances fall below a
    certain threshold).
  \item \emph{Separation-based with arclength-based farthest-point
      insertion:} Again choose (up to)~$k_c$ sample points based on
    curvature with separation requirements, then add more points by
    iteratively picking a point with maximum minimal arclength distance to
    one of the already selected sample points (up to a given maximum number
    $k^{\prime}$, and stopping early when the distances fall below a
    certain threshold).
  \end{enumerate}
  \noindent
  (Here, in (a) and (b), $k\leq k_c+k^{\prime}$ then
  denotes the final number of adaptively selected sample points.)
\end{enumerate}
Generally, the more sample points are chosen, the less pronounced the
differences between the different sampling schemes become. When keeping an
eye not only on the sampling quality but also on the MIP size (i.e., trying
to keep it small), preliminary experiments indicated that the final
curvature- and separation-based variant with arclength-based farthest-point
insertion, i.e., 3(b), seems to give the visually most appealing results
when using a managable number of sample points. This supports the reasoning
that led us to its formulation: In addition to the points discussed at the
beginning of this section, on the one hand, in ``low-resolution'' (large
$s$) regimes, delicate parts of the shape contours are essentially
impossible to be accurately reflected in the DAS construction, and
therefore the precise location of the sample points may not be so relevant
overall. Yet, it nevertheless makes sense to place some at high-curvature
segments, as done by the heuristic 3(b) (and others), in order to have the
final approximation not completely ignore protruding shape parts. On the
other hand, in ``high-resolution'' (small $s$) regimes, the number of
sample points can naturally be increased (compared to large $s$ settings),
since the separation requirement between sampled points is less restrictive
at such scales, which has two effects: First, more high-curvature segments
of the input shape can be sampled, which allows to capture such critical
regions more accurately. Second, further sampling by the arclength-based
farthest-point insertion part of our selection heuristic 3(b) then provides
a relatively even spread of sample points along the whole shape contour,
and also leads to segments being ``more straight'', which is beneficial
w.r.t.\ DAS representability using few struts.

Therefore, heuristic 3(b) is the sampling scheme we recommend
for use in Algorithm~\ref{algo:main} and the one employed in the numerical
experiments discussed later. It should be noted that the farthest-point
insertion part does add a few seconds of runtime, but this is near
negligible compared to the MIP solving time and appears to be worthwhile.

\subsubsection{Constructing the Zome Cycle}\label{subsubsec:construction}
As already mentioned, it is up to the concrete definition of an error
measure to identify the ``best'' sequence of struts in a~$\cG$-path. Here,
we make use of the distance field that is provided as input
(cf. Section~\ref{subsubsec:sampling}) and employ a simple sample-averaged
approximation of the area between the strut and input contour segments: For
a strut $e$ with endpoints $p_1$ and $p_2$, we compute its approximation
error (cost) $a_e$ as
\[
  a_e\define \frac{\norm{p_1-p_2}_2}{k+2}\sum_{i=0}^{k+1} \Big\lvert d\big(\tfrac{i}{k+1}p_1+(1-\tfrac{i}{k+1})p_2\big)\Big\rvert, 
\]
where $d(p)$ is the value of the distance field cell containing point
$p$. In case $p$ lies beyond the boundaries of the given distance field, we
let $d(p)$ be the value of the closest field cell multiplied with a penalty
of $100$. (Note that in order to avoid such penalties, the given distance
field should extend sufficiently far ``outside'' of the shape contour.)
In our numerical experiments, we use $k=3$.

Note that, in principle, it would be possible to compute the area exactly
or use exact point-to-polygon distances rather than distance field values
(at least for polygonal input curves); similarly, one could refine the
averaging by weighting every piece of a strut by its respective length
within each distance field cell the strut passes through. However, since it
already provided satisfactory results, we opted for the above-described
cheaper sample-averaged area approximation that makes use of distance field
values.

Note that our error measure turns out to be entirely separable (w.r.t.\
struts), so the total cost of a given $\cG$-cycle $C$ with segments
$\cS_1,\dots,\cS_k$ can be written~as
\begin{equation}\label{eq:approxerrorGcycleSegments}
  a(C,\cF)=\sum_{\kappa=1}^{k}\sum_{e\in\cS_\kappa}a_e.
\end{equation}

In Algorithm~\ref{algo:main}, the final step consists of constructing the
actual Zome cycle using the number of struts (of each type) for each
segment obtained from the MIP. To find the best permutation of struts for
each segment w.r.t.\ the cost function defined above, we can resort to
total enumeration (with some reduction of the computational effort by
aborting enumeration of subsequences that are provably worse than the
respective current best bound). Since this requires an undesirably long
time for segments containing more than roughly $15$ struts, we also
implemented a greedy construction scheme which iteratively constructs the
path from the segment's starting point to its end point by appending the
(locally) cheapest one of the remaining struts.
In our experiments, for $s\geq 2$, we usually ended up with sufficiently
short (i.e., few-strut) segments to use the exact (total enumeration)
construction method; nevertheless, empirically, the greedy method delivered
comparable results. For $s\leq 1$, enumeration appears to typically be too
expensive. As an automatic default choice in our implementation, we use
total enumeration for segments containing at most~$10$ struts, and the
greedy scheme for longer segments.

It is worth mentioning that for the purpose of our algorithm, separability
w.r.t.\ \emph{segments} would generally suffice, if total enumeration is
employed. Then, one could still enumerate all possible realizations of each
segment and evaluate some (possibly more sophisticated) segment-based
approximation cost for each such realization to pick the best one. However,
the separability w.r.t.\ struts of the above measure provides the advantage
of being able to use (e.g.) our greedy scheme to avoid enumeration for
segments with a number of struts that would render it prohibitively
slow. Indeed, without this separability, it would apparently be much less
clear how the strut arrangement could be done efficiently in such cases. In
practice, this would imply that our method could typically only solve
problems at lower resolutions (i.e., with larger values of the scaling
parameter~$s$), since there the number of struts in each segment turns out
to be sufficiently low.

\subsubsection{A Note on Simulated Annealing}\label{subsubsec:simulatedAnnealing}
Previous algorithmic work on Zometool shape approximation treated the
3-dimension problem with a simulated annealing (SA) method, see
\cite{ZimmerKobbelt2014,ZimmerETAL2014}. Now that we have established our
formal problem definitions (DCA-S, DPA-S, DPC) and described our novel
three-phase algorithm, but before we delve into computational experiments,
it seems appropriate to reconsider the SA idea in the present
context. Besides some obvious drawbacks of this approach that have been
mentioned in the introduction, SA is problematic here mainly for two
reasons: First, since we prescribe only a global maximum tolerance $\delta$
around the sampled anchor points, an SA procedure would have vanishing
gains for all (local improvement) moves that affect regions where the
approximation error is already below the threshold. This may lead to a very
high number of redundant moves that would have to be avoided by some more
sophisticated meta-strategy. The second issue is that in order to obtain
acceptable SA runtimes, one needs to be able to construct good starting
solutions; however, it is unclear how to achieve this in our DCA-S/DPA-S
setting with prescribed sample points $p_i$ and approximation (sample point
neighborhood size) tolerances $\delta$.

In fact, the MIP solution in the main stage of our novel algorithm could be
interpreted as this starting solution, and the concluding strut permutation
(phase three) could be implemented as an SA procedure. However, especially
since the segments typically consist of only a modest number of elements,
we can just search exhaustively or apply a simpler greedy scheme. (It might
be possible to further improve greedy arrangements by local improvement
steps of a 1-opt or 2-opt nature, which would still be generally simpler
than SA. We have not implemented such further postprocessing simply because
we found the results satisfactory as is.)
  
As briefly mentioned earlier, while SA generally does not provide solution
quality guarantees, it is admittedly true that similarly, our present
approach has no theoretical guarantees in terms of the approximation error
that is minimized by the strut arrangement in its final phase. The MIP
solver, however, employs a branch-and-cut scheme that is, in principle, a
theoretically exact method that terminates in finitely many steps. (Of
course, as the problem is \NP-hard, the number of iterations may still be
huge, so we abort the process early after a prescribed time limit.) Also,
our heuristic for the sampling phase is of course designed to allow for an
eventual solution with good resemblance to the input shape, but in terms of
the problem DCA-S, it may also be viewed as a means to provide (parts of)
the \emph{input} data for the problem under consideration: The choice of
$\delta$, i.e., the size of the boxes around the sampled anchor points, is
another input of the MIP, and solving the MIP~\eqref{mip:DCA-Sl0min}
therefore is an \emph{exact} method for DCA-S.

\section{Experiments}\label{sec:experiments}
We implemented our algorithm in \texttt{C}, using Gurobi~9.1.1 as MIP
solver; the code can be obtained from the first author's
homepage\footnote{\url{https://www.tu-braunschweig.de/mo/team/tillmann}}. The
Contour Explorer plugin for Open Flipper~\cite{MoebiusKobbelt2012}
(provided by Ole Untzelmann) allowed to extract the input data (distance
field, polygonal contour approximation~$\cF$) for our ``ZomeDCAS'' program
from slices of 3D object models or based on given images (e.g., scans of
handdrawn shapes); here, all test images were generated from licence-free
pictures found on the internet. In each instance, the distance fields
sides' have a length of $0.4$ (so for $s=1$, the shortest strut $b_1$ has
length $2s=2$.). The experiments were run under Linux on a quad-core
machine with Intel i7-7700T CPUs (2.90\,GHz, 8\,MB cache) and 16\,GB main
memory.

We used the sampling scheme 3(b) in all experiments, 
requiring an arclength equivalent to the length of three longest (i.e.,
long blue) struts separating the points chosen based on (adjusted)
curvature, and half that for the subsequent farthest-point insertion. The
total number of sample points was limited to $300$ (which was reached only
once in all of our experiments; typically, much fewer sample points were
used). In the MIP, we do employ SOS-1 constraints, and allow the solver
Gurobi to run in concurrent mode on all CPU cores, each with a (wall-clock)
time limit of $900$ seconds. Finally, we use full strut-enumeration on
segments with at most $10$ struts, and the greedy scheme for longer
segments, to construct the output Zome cycles with smallest approximation
error. Deviations from this setup in specific experiments will be stated
explicitly.

\begin{figure}[H]
  \centering
  \begin{tikzpicture}[scale=1]
  \node at (0,0-0.6) {\includegraphics[width=0.4\textwidth]{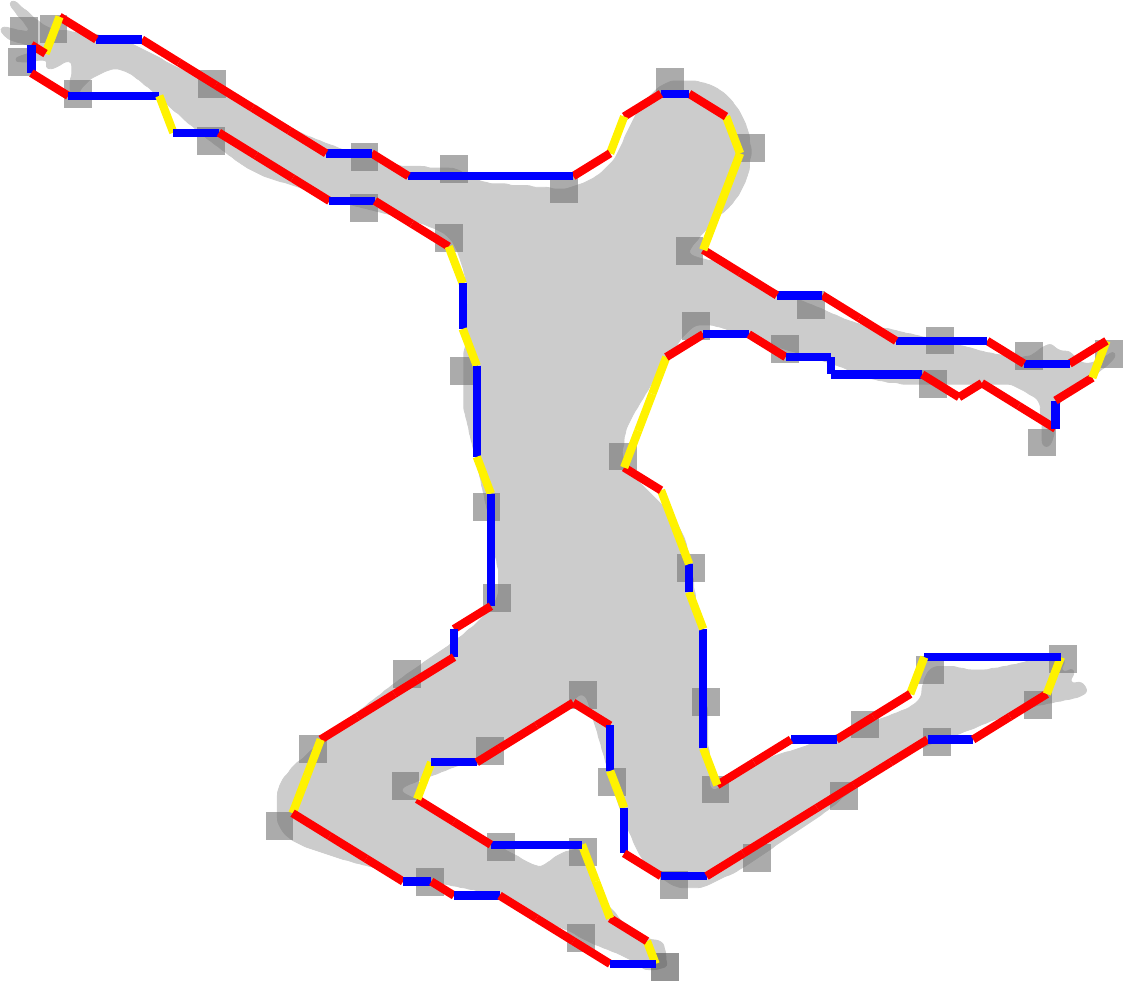}};
  \node at (0,-2.7-1.2) {\footnotesize \begin{tabular}{l}(a)
                                         $s=5,\delta=8,k_c=25$ $\leadsto$
                                         $k=49$,\\\phantom{(a)~}($138$
                                         struts, MIP gap $7.25$\%, cost\,$\approx$\,12026.2)\end{tabular}};
  \node at (0,-7.5+0.6) {\includegraphics[width=0.4\textwidth]{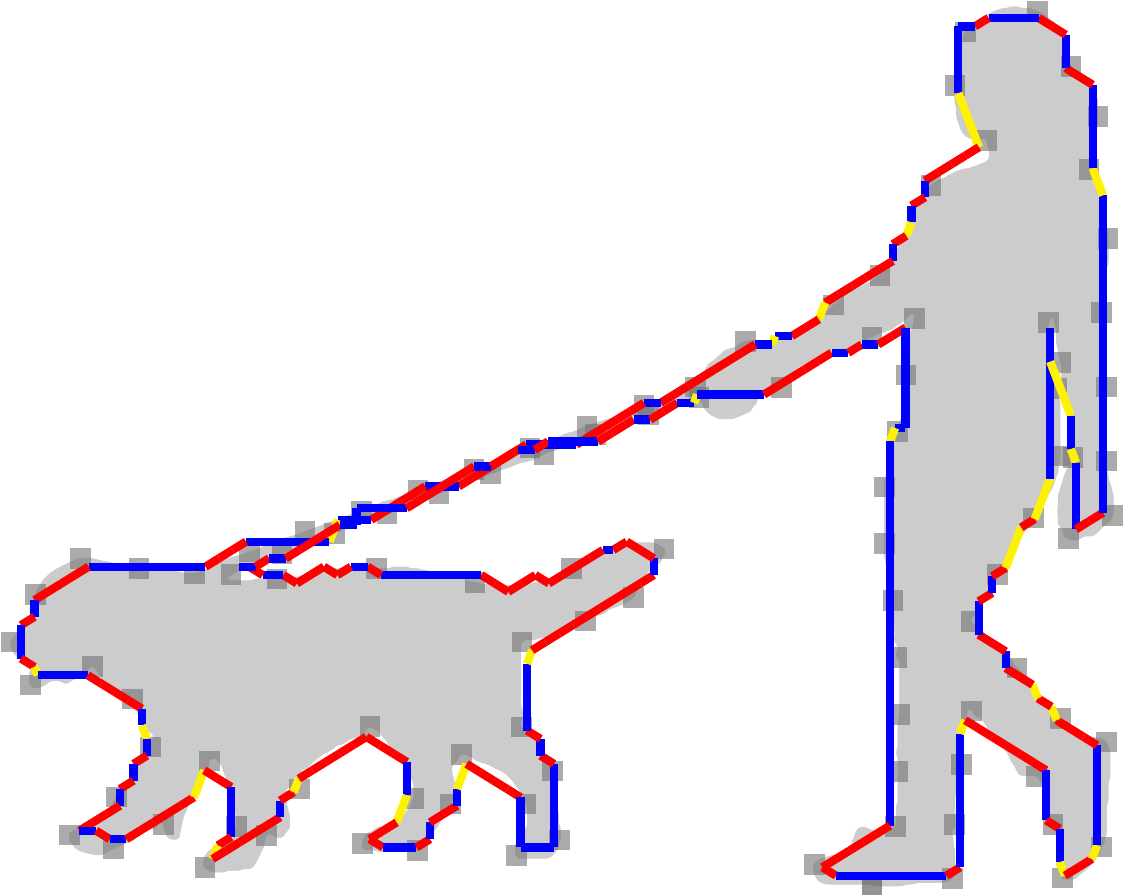}};
  \node at (0,-10.3+0.5) {\footnotesize \begin{tabular}{l}(b)
                                          $s=3,\delta=6,k_c=50$ $\leadsto$
                                          $k=106$,\\\phantom{(b)~}($422$
                                          struts, MIP gap $4.50$\%, cost\,$\approx$\,10126.83)\end{tabular}};
  \node at (0,-14.2+0.6) {\includegraphics[width=0.44\textwidth]{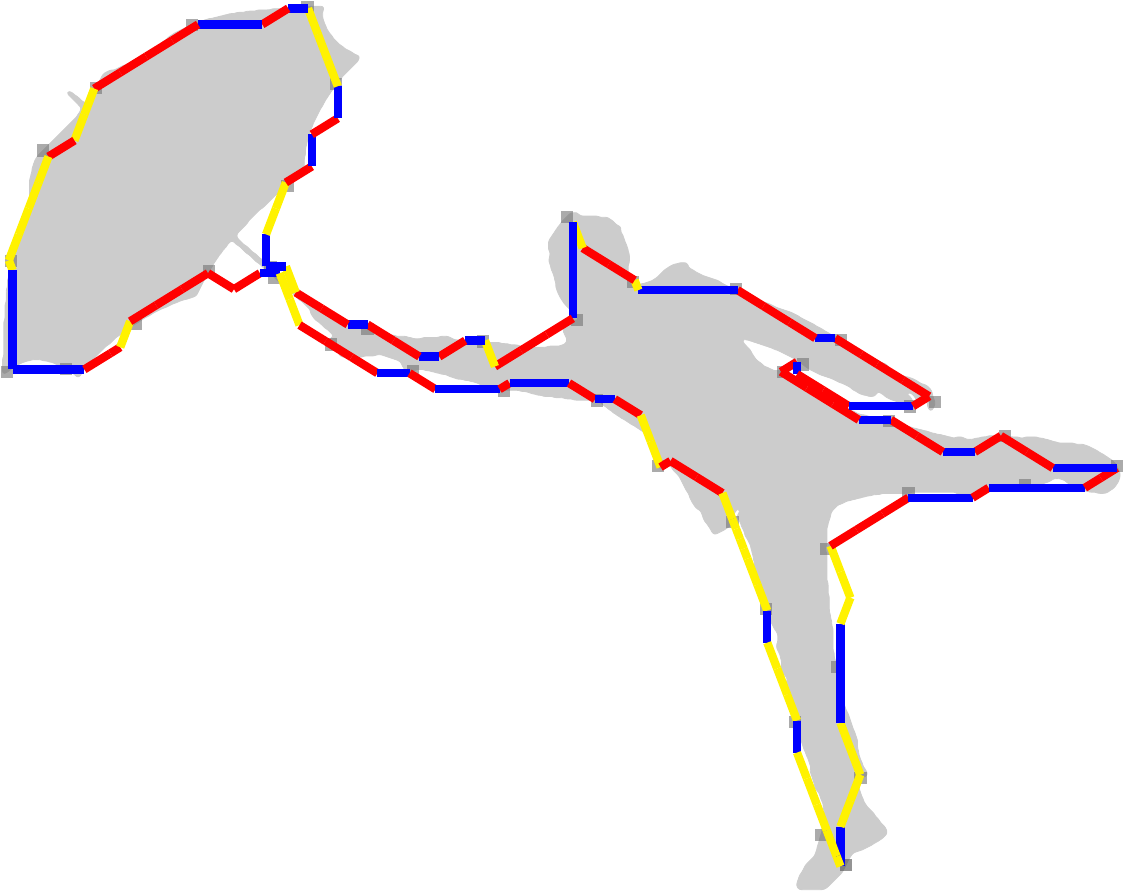}};
  \node at (0,-17+0.2) {\footnotesize \begin{tabular}{l}(c)
                                      $s=5,\delta=5,k_c=3$ $\leadsto$
                                      $k=42$,\\\phantom{(c)~}($159$ struts,
                                      MIP gap $12.58$\%, cost\,$\approx$\,1543361.4)\end{tabular}};
  \node at (6.95+1,0) {\includegraphics[width=0.3\textwidth]{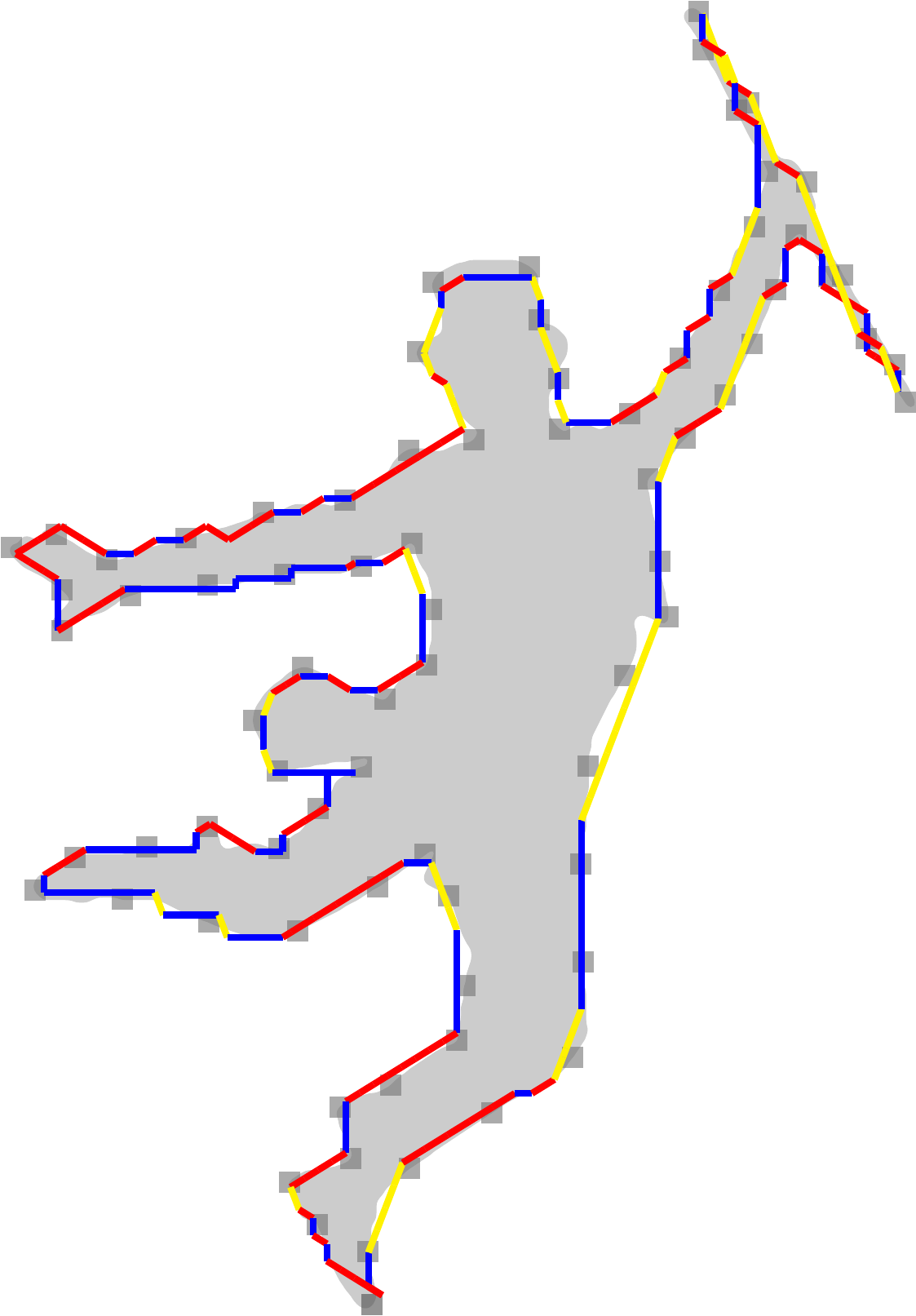}};
  \node at (5.75+2,-3.9) {\footnotesize \begin{tabular}{l}(d)
                                        $s=2,\delta=4,k_c=35$ $\leadsto$
                                        $k=79$,\\\phantom{(d)~}($260$
                                        struts, MIP gap $19.23$\%, cost\,$\approx$\,4007.0)\end{tabular}};
  \node at (6.95+1,-6.8) {\includegraphics[width=0.25\textwidth]{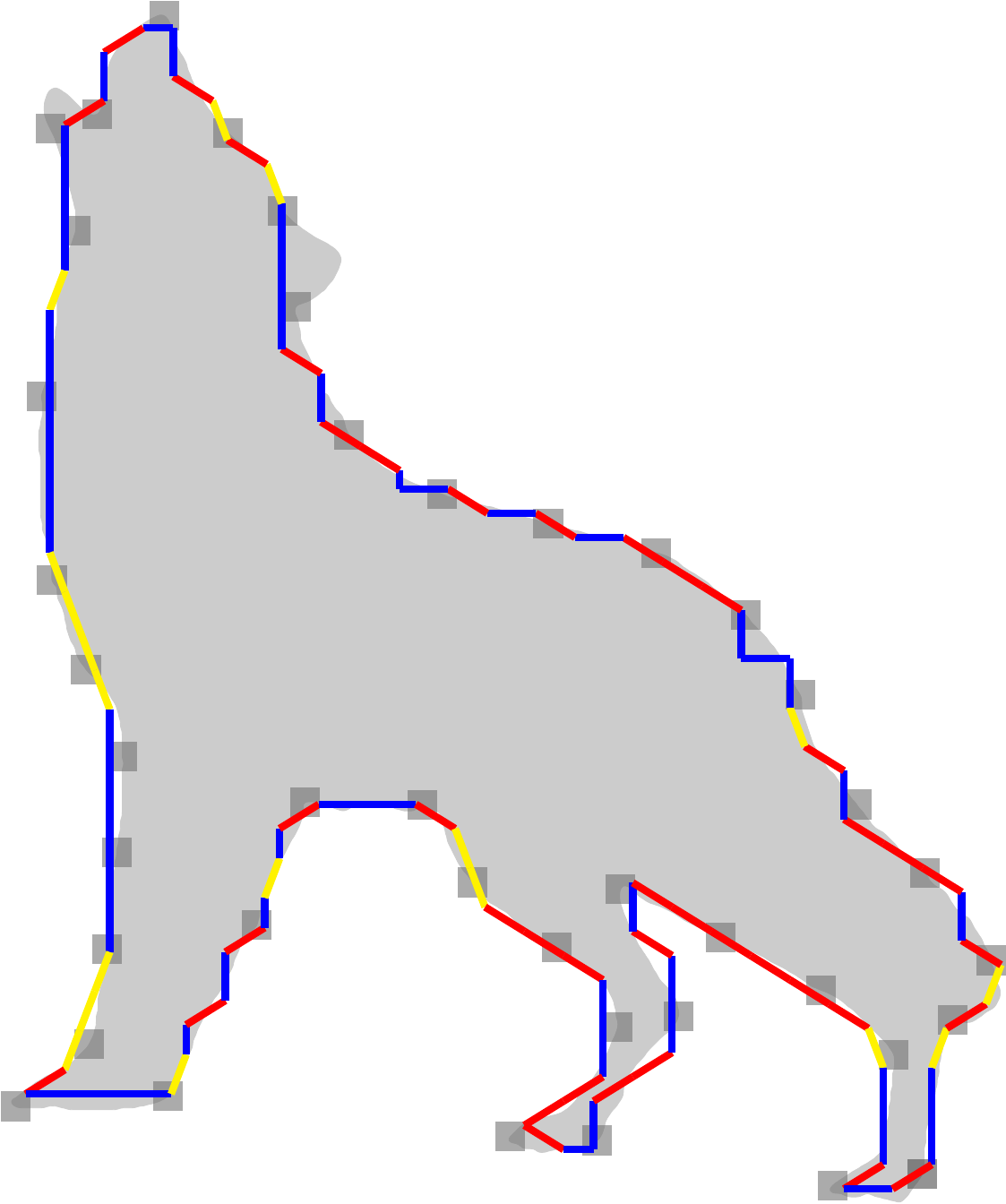}};
  \node at (5.75+2,-10.3+0.5) {\footnotesize \begin{tabular}{l}(e)
                                             $s=5,\delta=8,k_c=25$
                                             $\leadsto$
                                             $k=41$,\\\phantom{(e)~}($108$
                                             struts, MIP gap $4.63$\%, cost\,$\approx$\,10029.4)\end{tabular}};
  \node at (6.95+1,-14+0.7) {\includegraphics[width=0.35\textwidth]{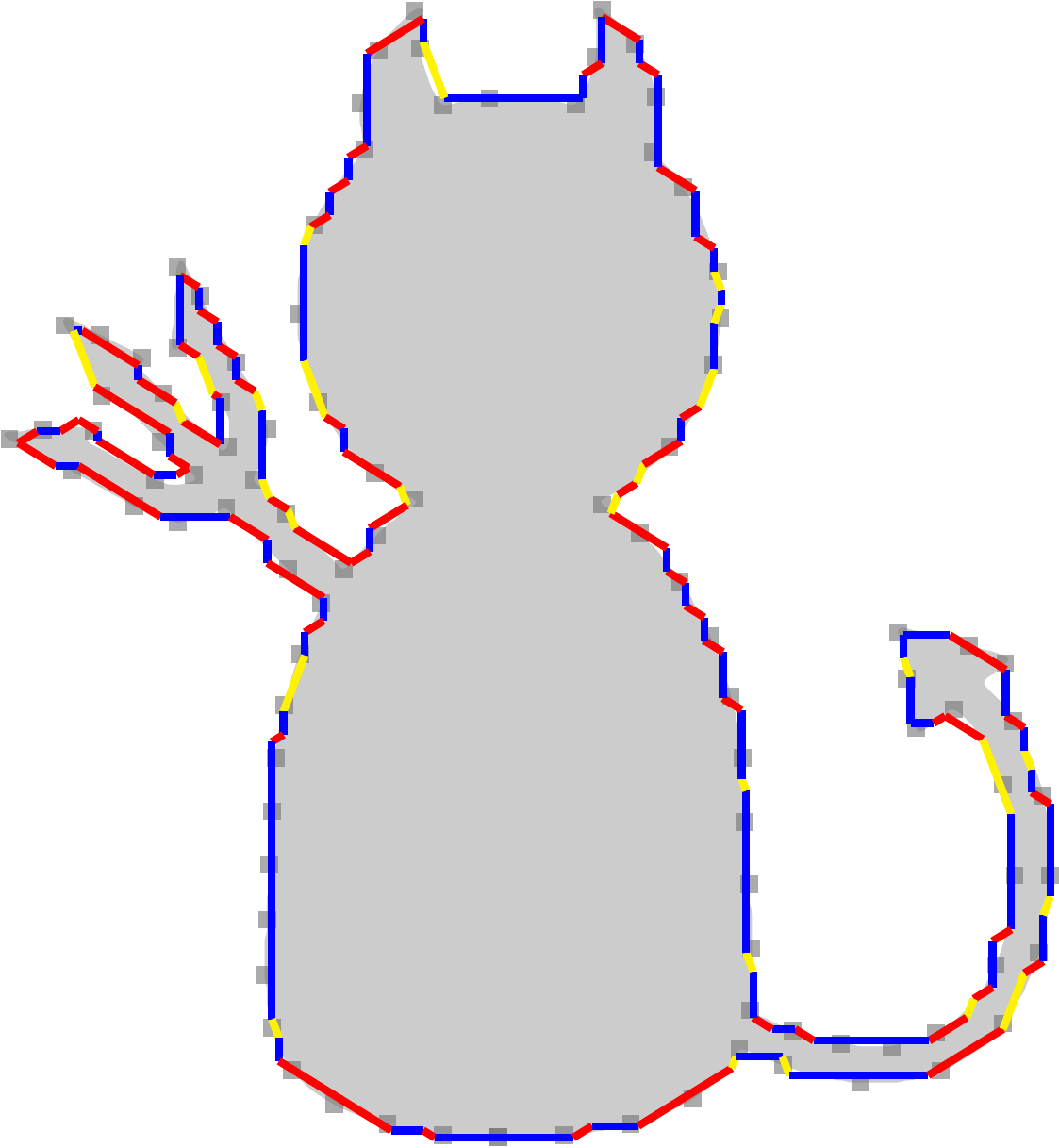}};
  \node at (5.75+2,-17+0.2) {\footnotesize \begin{tabular}{l}(f)
                                         $s=2,\delta=4,k_c=35$ $\leadsto$
                                         $k=99$,\\\phantom{(f)~}($285$
                                         struts, MIP gap $4.21$\%, cost\,$\approx$\,4665.5)\end{tabular}};
  \end{tikzpicture} 
  \caption{Experimental results for various test shapes. (MIP time limit: 900 seconds).}
  \label{fig:results1new}
\end{figure}
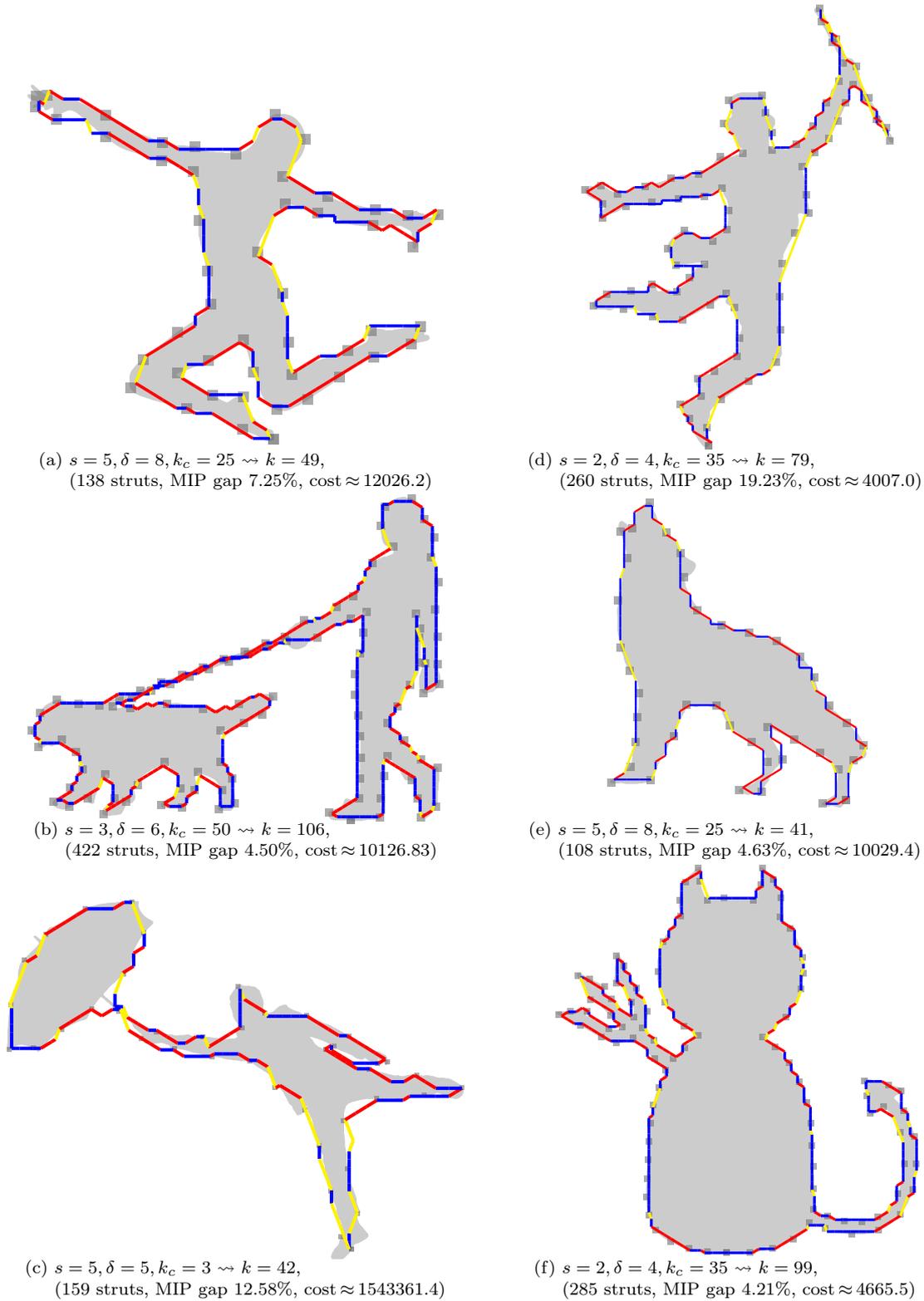

Initialization and cycle construction times combined were well below
$10$~seconds on most test instances (and still no more than $1$ minute for
large instances with small~$s$), so the majority of the runtime is spent in
MIP solving. Our program also generates tikz code to visualize the contour
and Zome approximation: the respective input shape is drawn in grey (its
boundary is the input contour) and Zome struts in their respective colors;
the boxes around sampling points are shown as opaque dark-grey squares, the
one slightly darker than the rest contains the Zome origin (best seen in
the electronic version of the paper, zooming in). Note that in the Zome
constructions displayed below, segments of the same color may---and often
do---consist of a \emph{sequence of struts}, not necessarily a single one;
we refrained from drawing the Zome nodes between struts in order not to
clutter the plots. The results of our computational experiments are
presented in Figures~\ref{fig:results1new}, \ref{fig:results2new}
and~\ref{fig:results3new}.

Let us begin with a selection of example results of our algorithm,
displayed in Figure~\ref{fig:results1new}. Overall, these results
demonstrate that our proposed algorithm is indeed capable of obtaining good
DAS/Zometool approximations of given shape contours within reasonable time,
where higher resolution (smaller~$s$) generally leads to better results
since naturally, shorter struts can more accurately recreate contours. It
is worth pointing out that some of the images were chosen specifically to
provide a challenge for our automated algorithm pipeline (especially the
sampling scheme): The shapes in Figures~\ref{fig:results1new}(b), (c) and
(d), in particular, have both quite ``curvy'' parts and narrow
parts. Nevertheless, the overall quality of the computed approximations
allow to conclude that, all in all, our sampling scheme successfully
addresses the necessity to carefully sample ``curvy'' regions, and that the
idea of targeting few-strut representations of segments between sample
point neighborhoods then provides a suitable way to induce
$\cF$-resemblance.

Moreover, this mostly also succeeds in providing strut collections that are
then arranged without dramatic collision issues in regions such as the
narrow parts: The worst self-intersections can be seen in the dog leash in
Figure~\ref{fig:results1new}(b) and the staff held by the dancer in
Figure~\ref{fig:results1new}(d), but even these at least appear to be
resolvable by rearranging the struts (manually or by a, at this point
hypothetical, postprocessing routine). The remaining figures, in fact, all
turned out to be collision-free.

This even holds for the umbrella guy in Figure~\ref{fig:results1new}(c),
although here the overall approximation quality is clearly not very
good. This may be explained by the very low number of points that were
sampled based on curvature, which supports the reasoning behind our
sampling heuristic 3(b). Moreover, here, quite clearly the strut
arrangements in certain segments are suboptimal (consider, e.g., the head
or the part where the umbrella's shaft meets its screen), and indeed, note
the comparatively large approximation error (cost) $\alpha(C,\cF)$; this is
apparently due to the artificial extension of the distance field (involving
penalties) used to estimate the distance/approximation error and should
therefore be fixed by resorting to (more costly) actual
point-to-curve-segment distances for the employed sampled area
approximation. (Note also that, in general, the numerical values of
approximation error may not be highly informative -- surely, smaller is
generally better, but half as large does not necessarily mean twice as
good; ultimately, different solutions may appear more or less appealing to
the beholder irrespective of the computed approximation error, so a final
quality assessment has a subjective component that cannot be quantified
exactly.)

Furthermore, we remark that the reported MIP optimality gaps do not pertain
to the final approximation error, but show how much improvement w.r.t.\
minimizing the number of utilized struts could still be achieved (at most)
by letting the MIP solver continue. For the experiments in
Figure~\ref{fig:results1new}, the gaps translate to explicit lower bounds
on the optimal number of struts, namely (a) 128, (b) 403, (c) 139, (d) 210,
(e) 103, and (f) 273. Thus, depending on the instance, there would still be
some improvements to be gained. In fact, considering that actually building
a computed representation using the Zometool kit, i.e., assembling the
shape approximation by following the computed ``construction manual''
possibly consisting of several hundred struts arranged in a certain order,
would likely take a person several hours, spending more computational
efforts in the MIP phase and/or to arrange struts afterwards to obtain
high-quality and low-(strut-)complexity solutions would surely be
justified.

As a final comment on Figure~\ref{fig:results1new}, let us point out that
the result in Figure~\ref{fig:results1new}(b) was, in fact, obtained using
not the full Zometool set with short, medium and long struts of the three
colors. Indeed, the MIP solver failed to find a feasible solution within
900 seconds (and much longer). Thus, we took out the long struts of each
color, and the presented result hence uses only short and medium
struts. Obviously, it is still feasible in the full Zometool set; in fact,
specific to Zometool (not arbitrary DASs), we know that for each color, the
longest strut is precisely as long as the short and the medium one
combined. Thus, using Zometool, there is another design choice to make: If
we prefer to find Zome constructions with the fewest possible struts, the
long struts must be included, because their availability allows to replace
a short and a medium one (if placed after one another) and the MIP solver
can decide automatically if this is advantegous and feasible. On the other
hand, not using the long struts allows for a refined arrangement of the
(shorter) struts in the final algorithm phase, to minimize the
approximation error. Both viewpoints are compatible, since the goals of low
construction complexity and low approximation error are
competitive. Moreover, we could switch between using and not using long
struts basically at any point: Having computed a solution without them, we
can check the final outcome and replace occurrences of same-colorpairs of
one short and one medium strut placed in a straight line after one another
by a long strut, to reduce the strut count by (re-)introducing the long
ones. For the example in Figure~\ref{fig:results1new}(b), one can reduce
the total strut count by 3 this way, replacing one strut pair of each color
by the corresponding long struts. Similarly, one could solve the MIP with
the full Zometool set, and replace all occurring long struts by a short and
medium one each before computing the strut arrangements that minimize the
approximation error, to allow for better results there. Since these aspects
seem to us like a choice that should be left to the user, we keep to the
full Zometool set (always using long struts throughout our algorithm, in
particular) in all further experiments.

We will provide a more structured comparison of the influence of the
different solver parameters (scale $s$, desired approximation
tolerance~$\delta$ around the sample points, number $k_c$ of points
selected based on adjusted curvature, and time limit) later
(cf. Figure~\ref{fig:results3new} and discussion in
Section~\ref{subsec:results:parametervariations} below). Before we do so,
we would like to point out an aspect of our approach that turned out to be
somewhat problematic when applying it in practice, namely the choice of
$\delta$ (in combination with a suitable scale parameter~$s$). Specifying
some more or less arbitrary $\delta$ and $s$ values, we quite often found
that the MIP solver failed to find any feasible solution (at least within
the time limit). In the next subsection, we describe this problem in a bit
more detail and propose a remedy in the form of a slight relaxation of the
MIP model that seems reasonable from a practical standpoint and proved to
be very effective empirically.

\subsection{Relaxing the Neighborhood Containment Constraints}\label{subsec:slack}
As mentioned just above, we observed that feasibility of the
MIP~\eqref{mip:DCA-Sl0min} appears to hinge somewhat crucially on the
user-prescribed parameter~$\delta$, i.e., the desired approximation
tolerance around the sampled anchor points on the input curve. While values
$\delta>s$ usually do not seem to pose problems so often, $\delta\leq s$
implies that only a single Zome node can be placed within each sample point
neighborhood, which unfortunately seems more likely to be too restrictive
in the DAS construction context, as the MIP solver then fails to find even
a single feasible solution within the inspected time limits. This indicates
that no ``nice'' Zome constructions that obey the neighborhood containment
constraints and allow for few-strut segments connecting them might exist in
such cases. While we can observe this behavior empirically, there remains
the problem that there is no clear way to identify workable pairs
$(\delta,s)$ a priori. Keeping $\delta$ fixed and reducing $s$
significantly would likely often work, but arbitarily adjusting the scale
seems somewhat unintuitive. Similarly, one will likely not want to choose a
very large $\delta$, as it relates to approximation quality.
 
Nevertheless, recall from the beginning of Section~\ref{sec:methodology}
that, conceptually, respecting a global approximation tolerance~$\delta$ is
merely a desideratum, translated into the hard box-constraints in our
MIP~\eqref{mip:DCA-Sl0min}, and that our overall algorithm does not
guarantee that all points along the constructed Zometool contour actually
stay within distance~$\delta$ of the input curve. In the practical
application of our method, the user (so far) has to choose a
$\delta$-value, but as alluded to above, there is only vague intuition on
how to actually make such a choice: too small a value may be too
restrictive, given the relative inflexibility of discrete assembly systems
to satisfy arbitrary positioning demands, while overly large values
ultimately lead to cruder shape resemblance and could invite other unwanted
issues such as self-intersections (e.g., when sampling point neighborhoods
overlap).

Thus, we may argue that the choice of $\delta$ should not be treated as
highly important, and could in fact be relaxed in order to improve
flexibility of the MIP model to find suitable candidate solutions, even if
they ultimately slightly violate the prescribed $\delta$ tolerance. To that
end, we can extend the MIP~\eqref{mip:DCA-Sl0min} by introducing a slack
variable $s_\delta\in\R_+$ and modifying the box-constraint bounds
$\pm\delta\mathds{1}$ to $\pm(\delta+s_\delta)\mathds{1}$,
respectively. Thus, by assigning a positive value to $s_\delta$, the solver
can automatically enlarge the neighborhoods around the sample points; since
clearly, larger neighborhoods allow for the containment requirements to be
met more easily, the solver can thus find Zome constructions respecting the
enlarged neighborhoods more easily. In order to retain a meaningful model,
the values of $s_\delta$ must of course be bounded above; indeed, we
propose to additionally modify the objective function
of~\eqref{mip:DCA-Sl0min} by adding a penalty term on the slack value,
effectively minimizing not only the number of struts used but also the
slack variable. In our implementation, we found that an objective
coefficient of 10\,000 for the slack variable seems to work quite well
across scales (though this has not been thoroughly benchmarked; smaller
values might work similarly well and might be better suited when the
user-provided $\delta$ is really small, since otherwise, the implied
priority of large coefficients to try hard to reduce the slack value would
ultimately lead to insufficient enforcement of keep the Zome connections
short).

This extension of our MIP model turned out to be quite effective in
previously problematic $\delta$-regimes ($\delta\leq s$), where the solver
may not find any feasible solution within up to an hour, in particular for
instances with $s\leq 1$ (high resolution, and consequently, usage of
comparatively many struts) but also for other settings where finding a
feasible initial solution apparently challenged to solver. We note that for
the slack-endowed model, longer running times are adequate in order to have
the solver first reduce the slack to a sufficiently small level (if
possible) that the number of struts objective eventually becomes the
driving force for further optimization progress; thus, when $900$ seconds
often worked well for our original hard-constrained approach, we generally
suggest allowing $1800$ seconds for the relaxed variant.

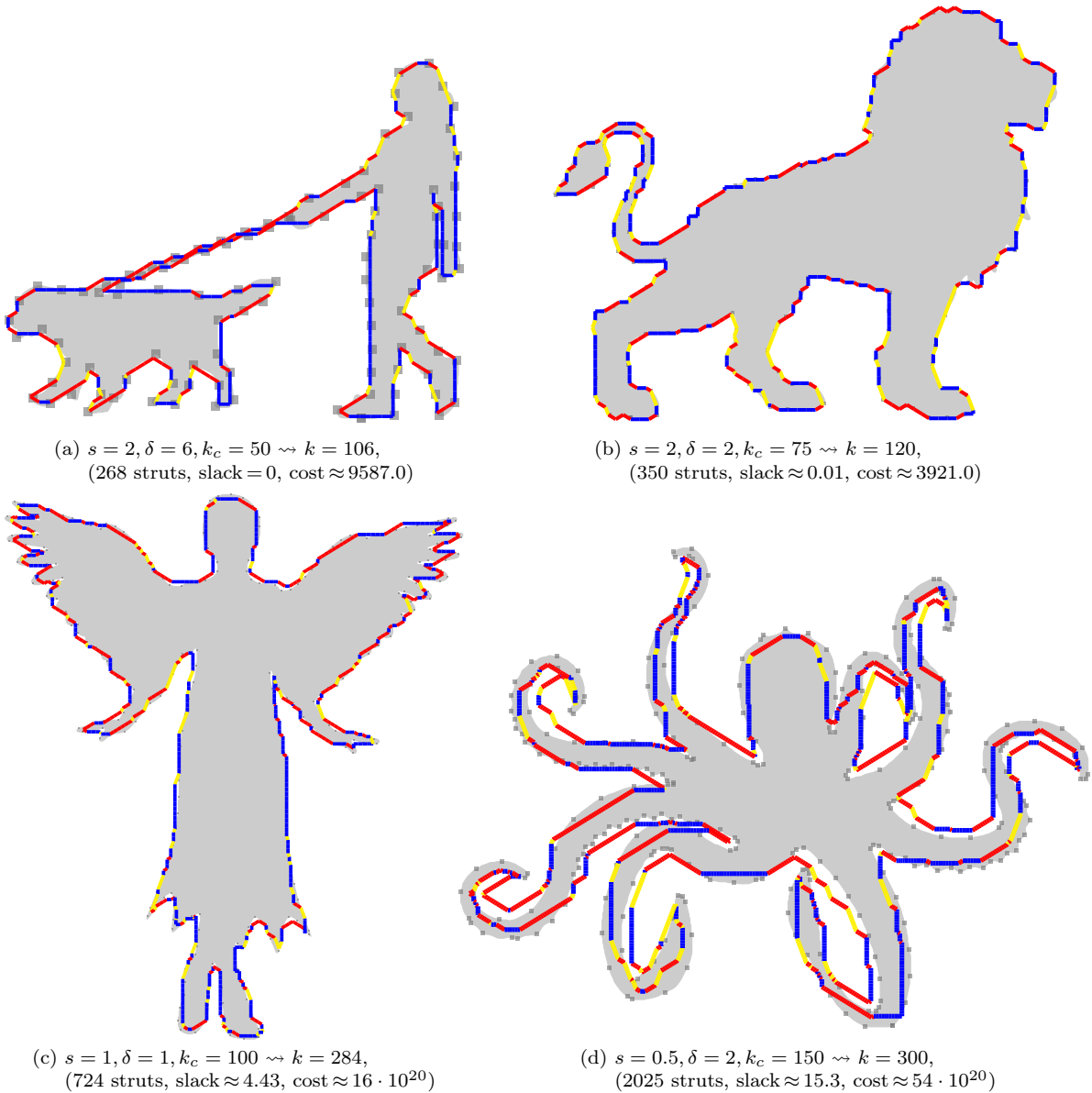
\begin{figure}[t]
  \centering
  \begin{tikzpicture}[scale=1]
  \node at (0,-11-2.6) {\includegraphics[width=0.4\textwidth]{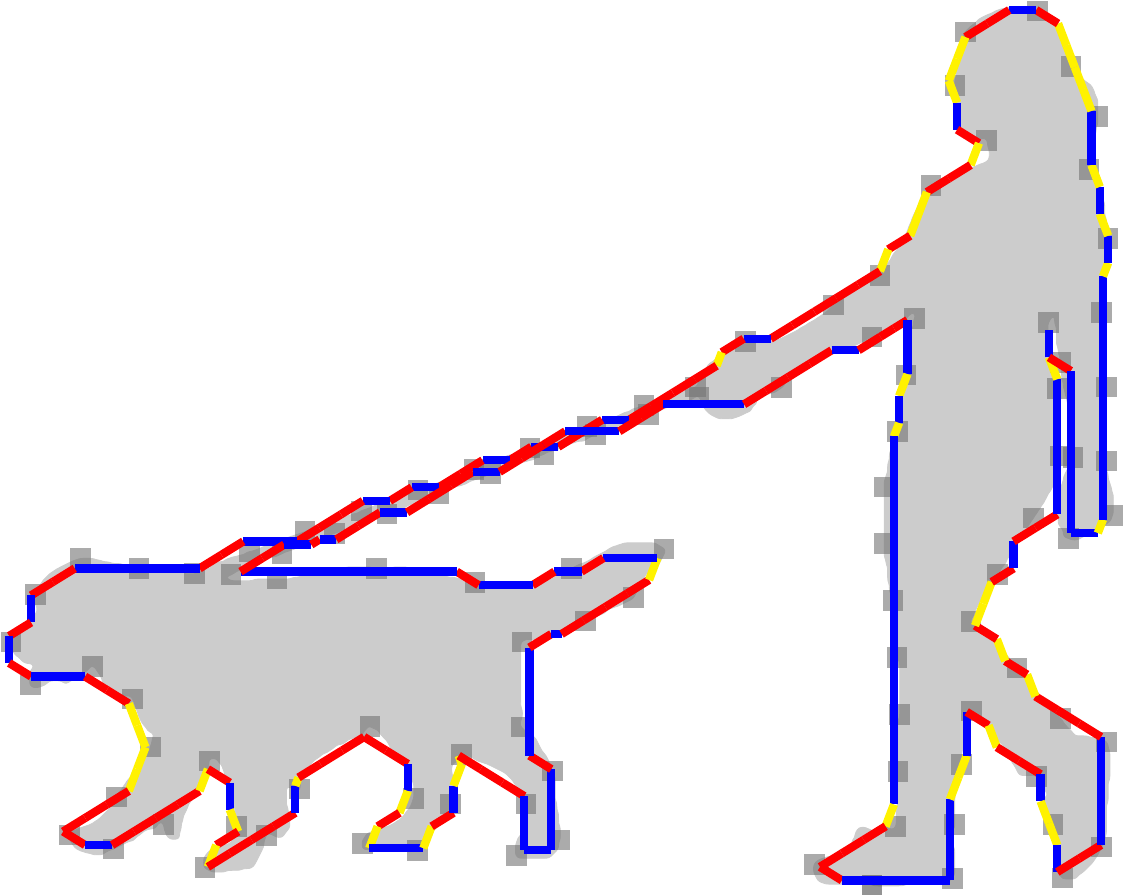}};%g
  \node at (0,-13.6-3.2) {\footnotesize \begin{tabular}{l}(a)
                                          $s=2,\delta=6,k_c=50$ $\leadsto$
                                          $k=106$,\\\phantom{(a)~}($268$
                                          struts, slack\,$=$\,0, cost\,$\approx$\,9587.0)\end{tabular}};
  \node at (7.2+1,-11-2.2) {\includegraphics[width=0.44\textwidth]{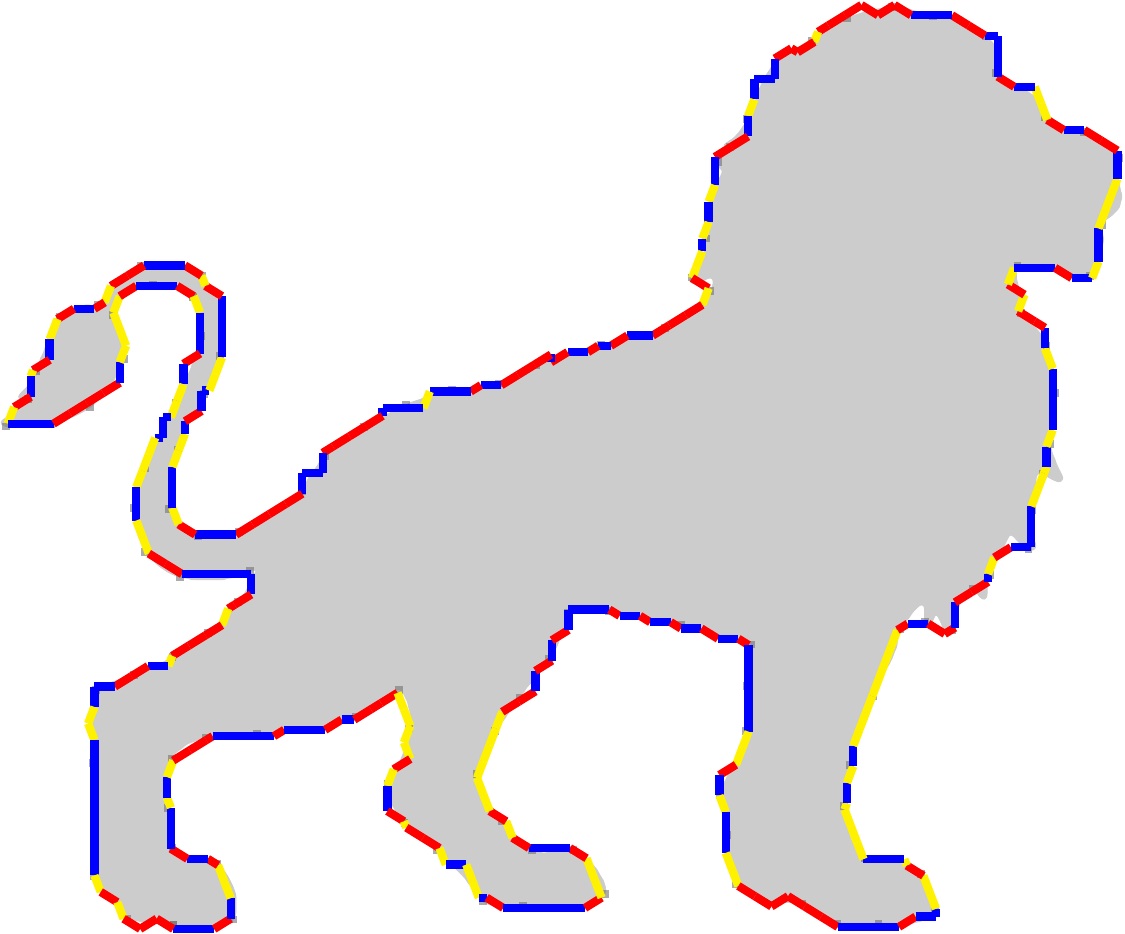}};%h
  \node at (6.95+1,-13.6-3.2) {\footnotesize \begin{tabular}{l}(b)
                                             $s=2,\delta=2,k_c=75$
                                             $\leadsto$
                                             $k=120$,\\\phantom{(b)~}($350$
                                             struts,
                                             slack\,$\approx$\,0.01, cost\,$\approx$\,3921.0)\end{tabular}};
  \node at (0,-11-8.7-1.5) {\includegraphics[width=0.4\textwidth]{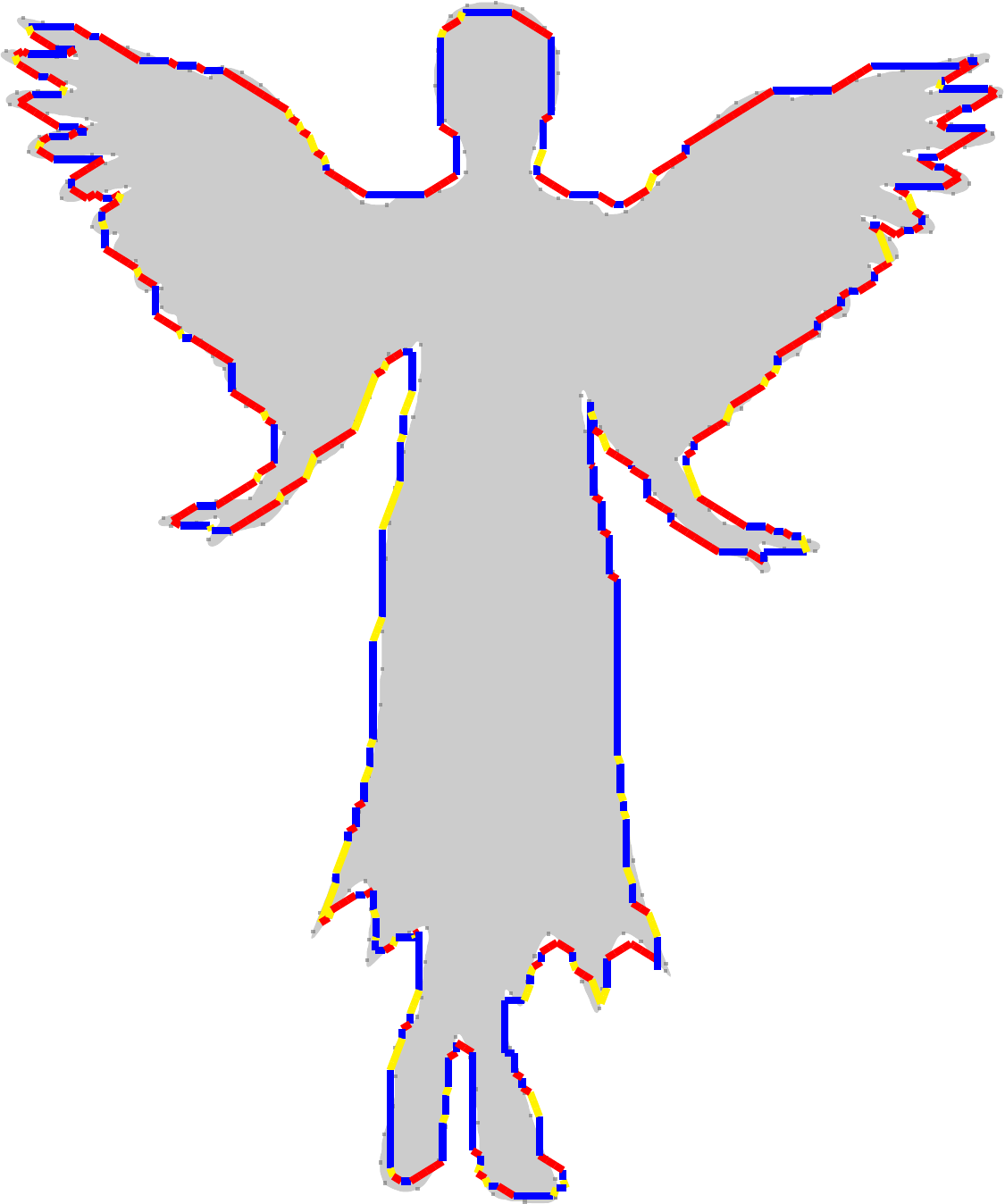}};%h
  \node at (0,-13.6-9.7-2.3) {\footnotesize \begin{tabular}{l}(c)
                                             $s=1,\delta=1,k_c=100$
                                             $\leadsto$
                                             $k=284$,\\\phantom{(c)~}($724$
                                             struts,
                                             slack\,$\approx$\,4.43,
                                          cost\,$\approx$\,$16\cdot 10^{20}$)\end{tabular}};

  \node at (6.75+1,-11-7.7-2.8) {\includegraphics[width=0.55\textwidth]{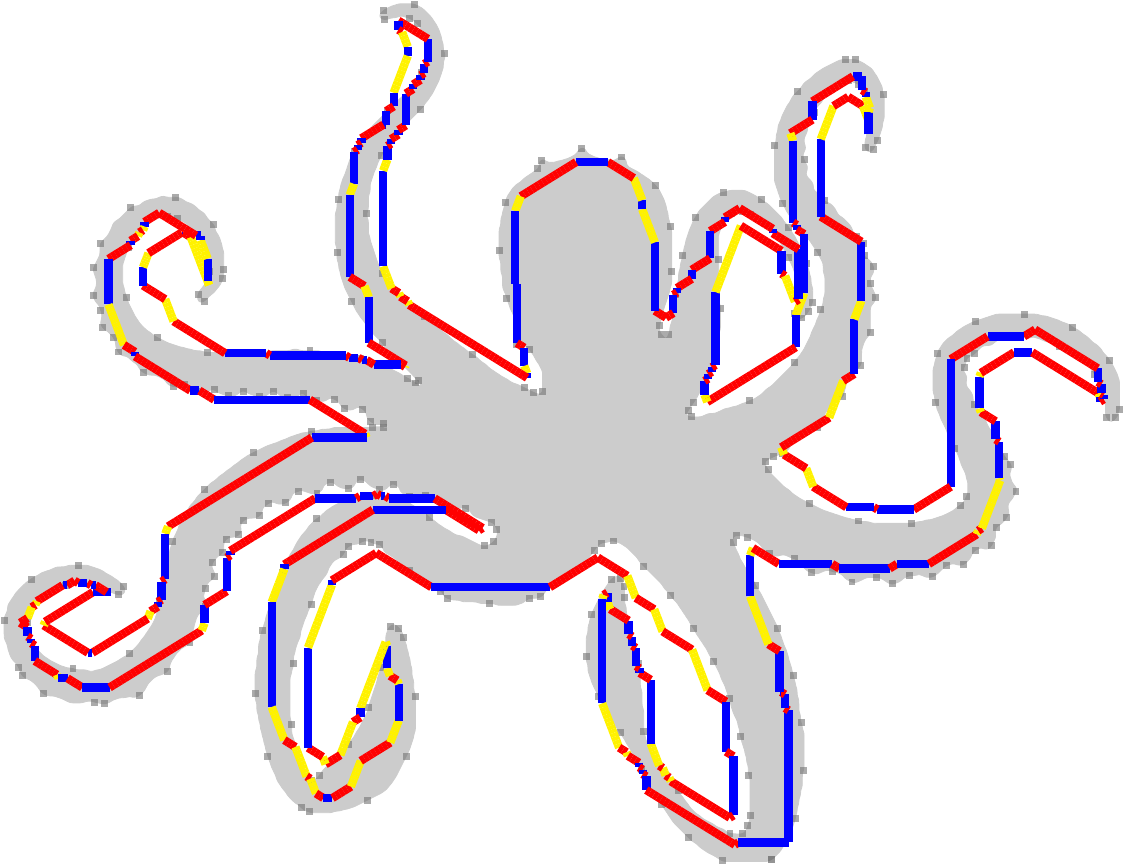}};%h
  \node at (6.95+1,-13.6-7.7-4.3) {\footnotesize \begin{tabular}{l}(d)
                                             $s=0.5,\delta=2,k_c=150$
                                             $\leadsto$
                                             $k=300$,\\\phantom{(d)~}($2025$
                                             struts,
                                             slack\,$\approx$\,15.3,
                                             cost\,$\approx$\,$54\cdot 10^{20}$)\end{tabular}};

  \end{tikzpicture} 
  \caption{Experimental results for the algorithm variant with relaxed MIP
    box constraints (MIP time limits: (a) 900 seconds, (b) 1800 seconds,
    (c) 3600 seconds, and (d) 10800 seconds).}
  \label{fig:results2new}
\end{figure}

\begin{figure}[ht!]
  \centering
  \begin{tikzpicture}[scale=1]
    \node at (0,0)       {\includegraphics[width=0.44\textwidth]{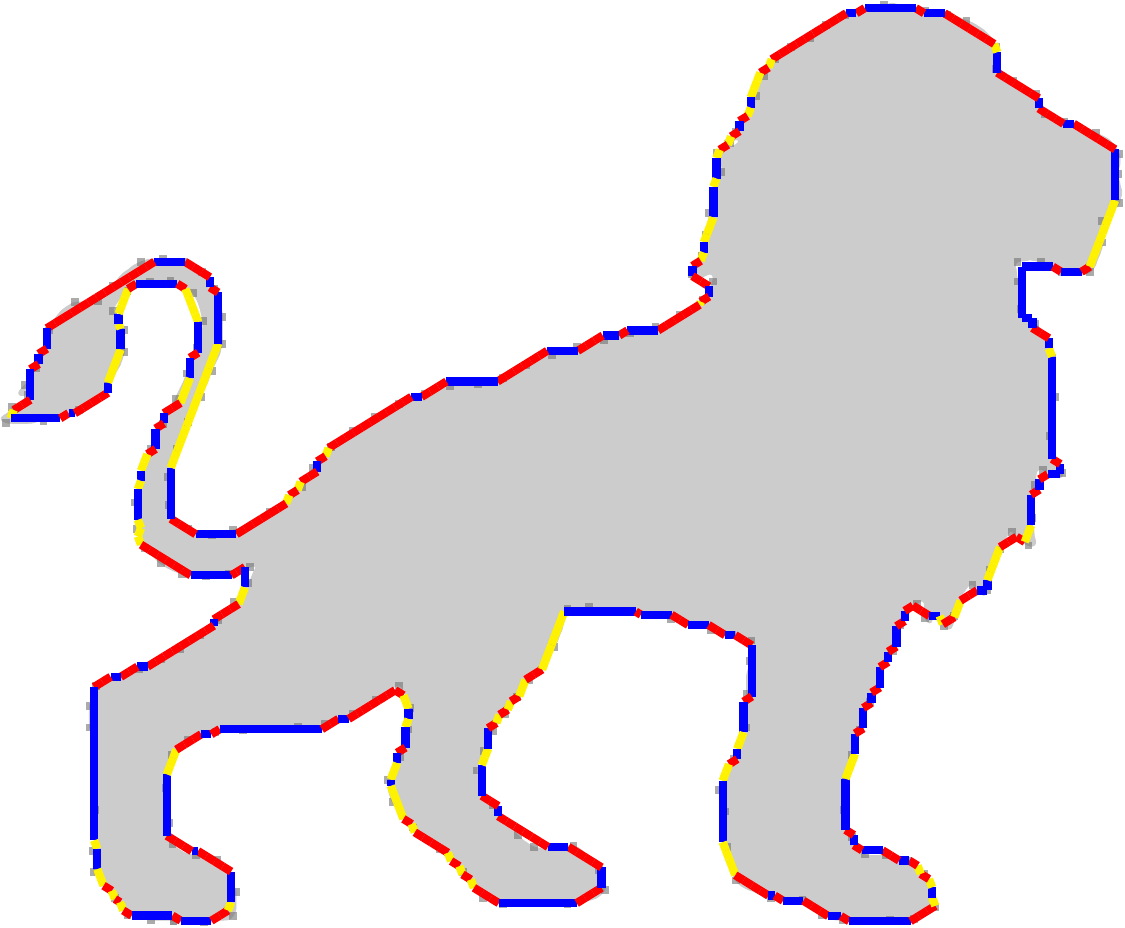}};%c
    \node at (0,-2.6-0.8)  {\footnotesize \begin{tabular}{l}(a)
                                         $s=1,\delta=2,k_c=75$ $\leadsto$
                                         $k=225$,\\\phantom{(a)~}($629$
                                         struts, slack\,$\approx$\,0.81, cost\,$\approx$\,$10^{20}$)\end{tabular}};
    \node at (0,-5.5-1.4)   {\includegraphics[width=0.44\textwidth]{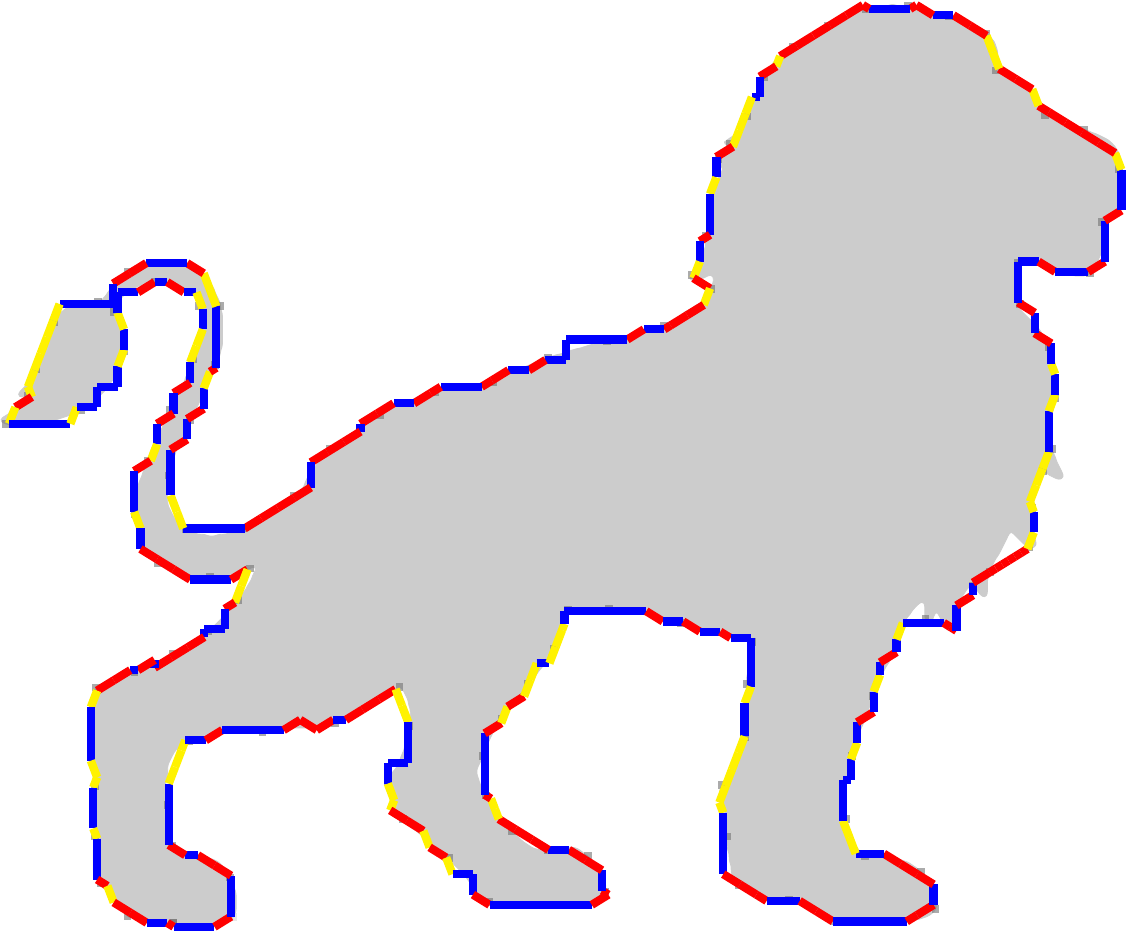}};%b
    \node at (0,-8.1-2.2) {\footnotesize \begin{tabular}{l}(b)
                                       $s=2,\delta=2,k_c=30$ $\leadsto$
                                       $k=117$,\\\phantom{(b)~}($350$
                                       struts, slack\,$\approx$\,0.08,
                                       cost\,$\approx$\,4232.8)\end{tabular}};
    \node at (0,-10.7-3.1) {\includegraphics[width=0.44\textwidth]{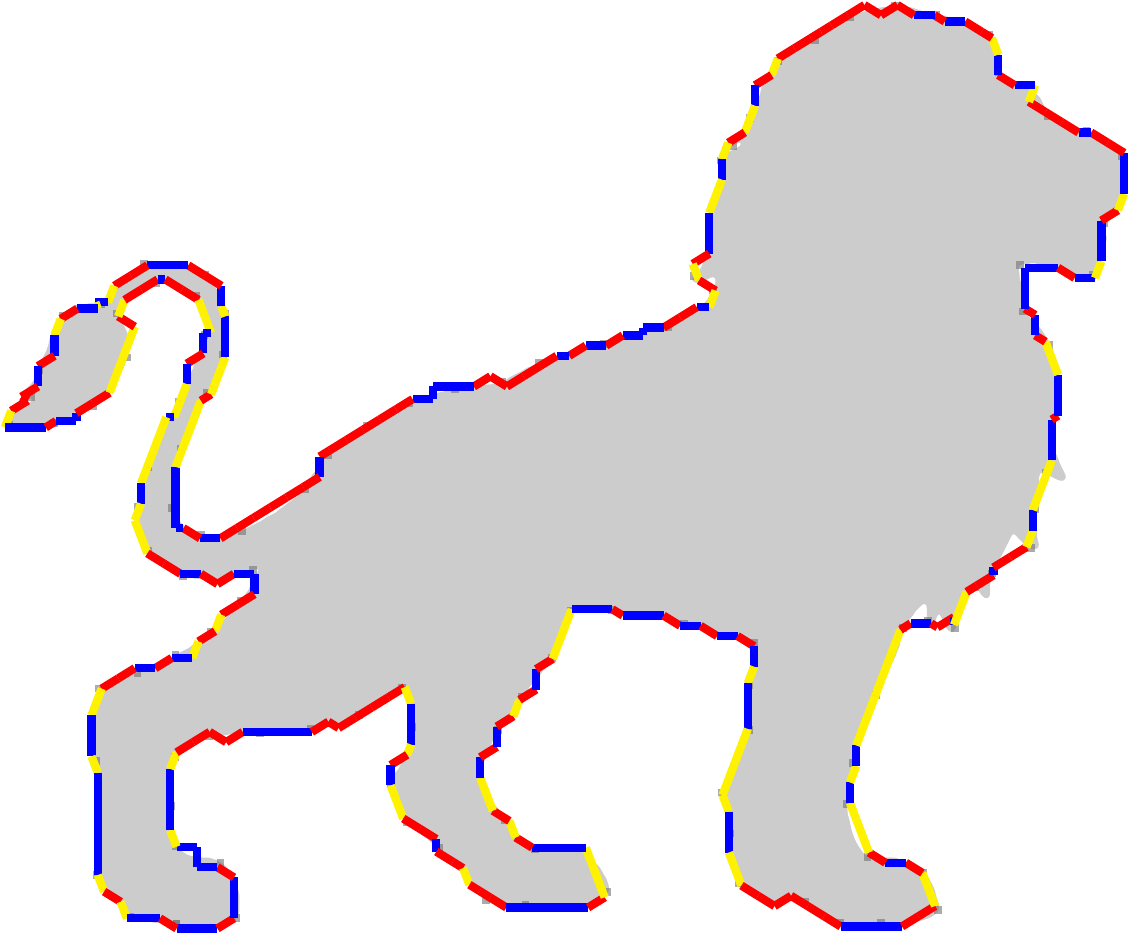}};%a
    \node at (0,-13.35-3.9){\footnotesize \begin{tabular}{l}(c)
                                       $s=2,\delta=2,k_c=75$ $\leadsto$
                                       $k=120$,\\\phantom{(c)~}($344$
                                       struts, slack\,$\approx$\,0.29, cost\,$\approx$\,3999.8)\end{tabular}};
    \node at (5.75+2,0)   {\includegraphics[width=0.44\textwidth]{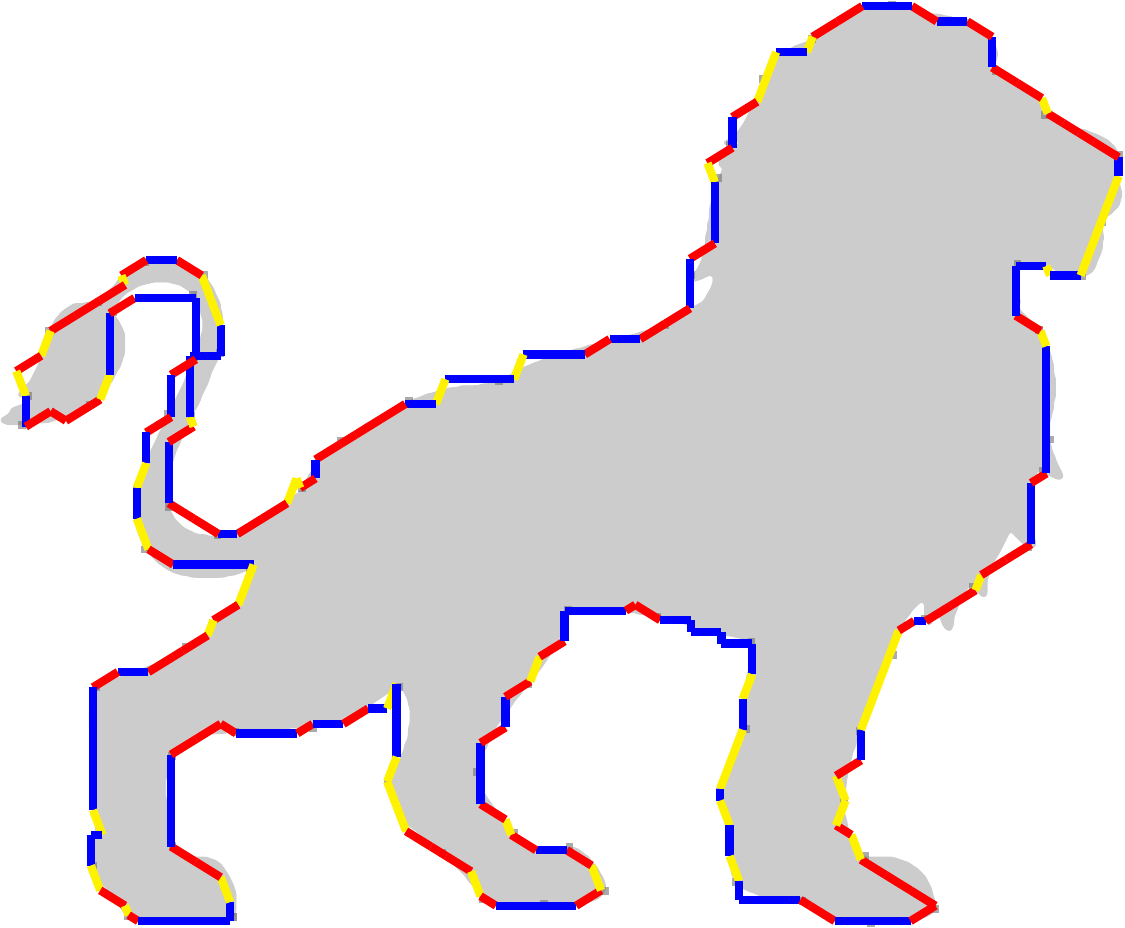}};%f
    \node at (5.75+2,-2.6-0.8) {\footnotesize \begin{tabular}{l}(d)
                                            $s=3,\delta=2,k_c=75$
                                            $\leadsto$
                                            $k=76$,\\\phantom{(d)~}($240$
                                            struts, slack\,$=$\,0, cost\,$\approx$\,6639.5)\end{tabular}};
    \node at (5.75+2,-5.5-1.4){\includegraphics[width=0.44\textwidth]{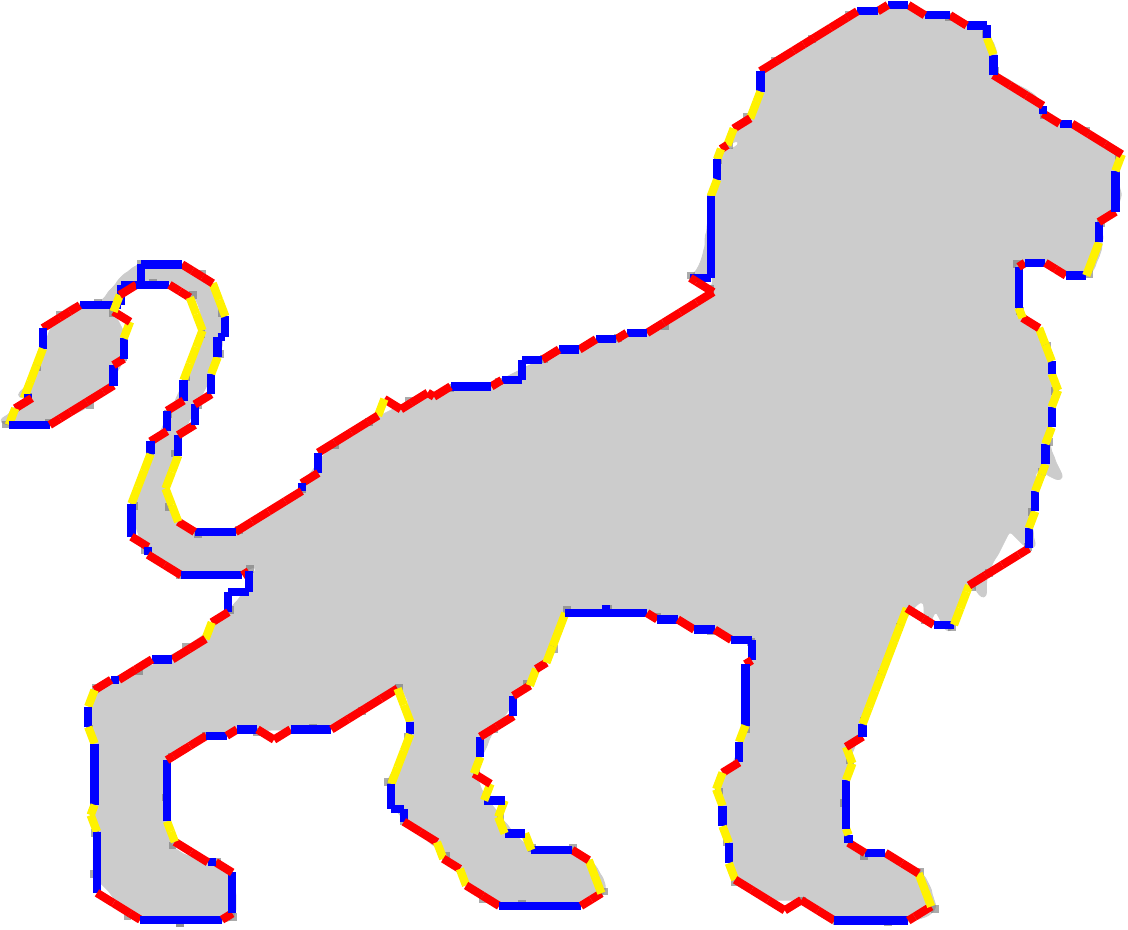}};%e
    \node at (5.75+2,-8.1-2.2) {\footnotesize \begin{tabular}{l}(e)
                                          $s=2,\delta=2,k_c=120$ $\leadsto$
                                          $k=123$,\\\phantom{(e)~}($356$
                                          struts, slack\,$=$\,0, cost\,$\approx$\,4175.0)\end{tabular}};    
    \node at (5.75+2,-10.7-3.1) {\includegraphics[width=0.44\textwidth]{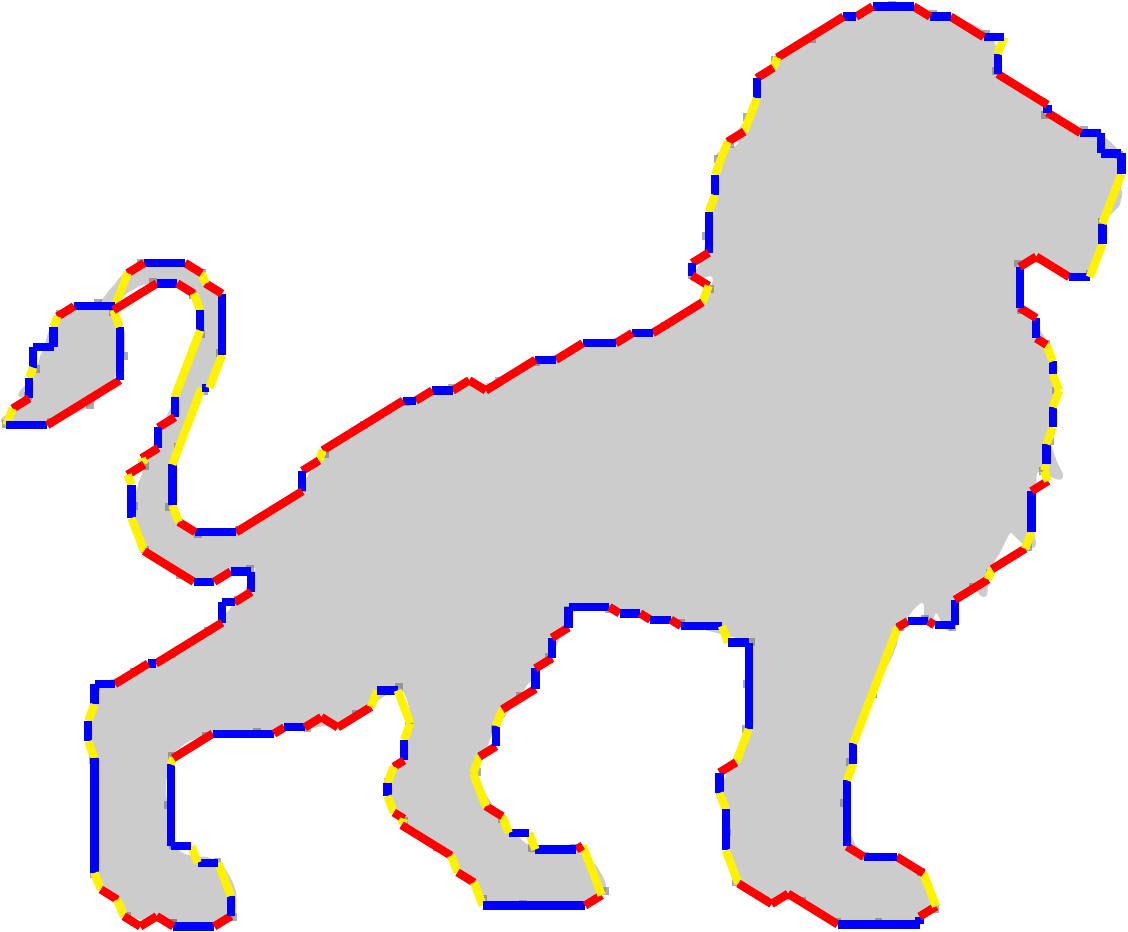}};%d
    \node at (5.75+2,-13.35-3.9) {\footnotesize \begin{tabular}{l}(f)
                                          $s=2,\delta=2,k_c=75$ $\leadsto$
                                          $k=120$,\\\phantom{(f)~}($345$
                                          struts, slack\,$=$\,0, cost\,$\approx$\,4109.4)\end{tabular}};
 \end{tikzpicture}
 \caption{Experimental results for test shape ``lion'' for different
   algorithmic parameter choices. Baseline is
   Figure~\ref{fig:results2new}(b). Keeping all respective other parameters
   at their baseline values ($s=2=\delta$, $k_c=75$, time limit 1800
   seconds), subfigures (a) and (d) show the results for $s=1$ and $s=3$,
   (b) and (e) those for $k_c=30$ and $k_c=120$, (c) and (f) those for time
   limits 900 and 3600 seconds, respectively. The baseline and all results
   presented here were obtained with our algorithm using the MIP variant
   from Section~\ref{subsec:slack}.}
\label{fig:results3new}
\end{figure}
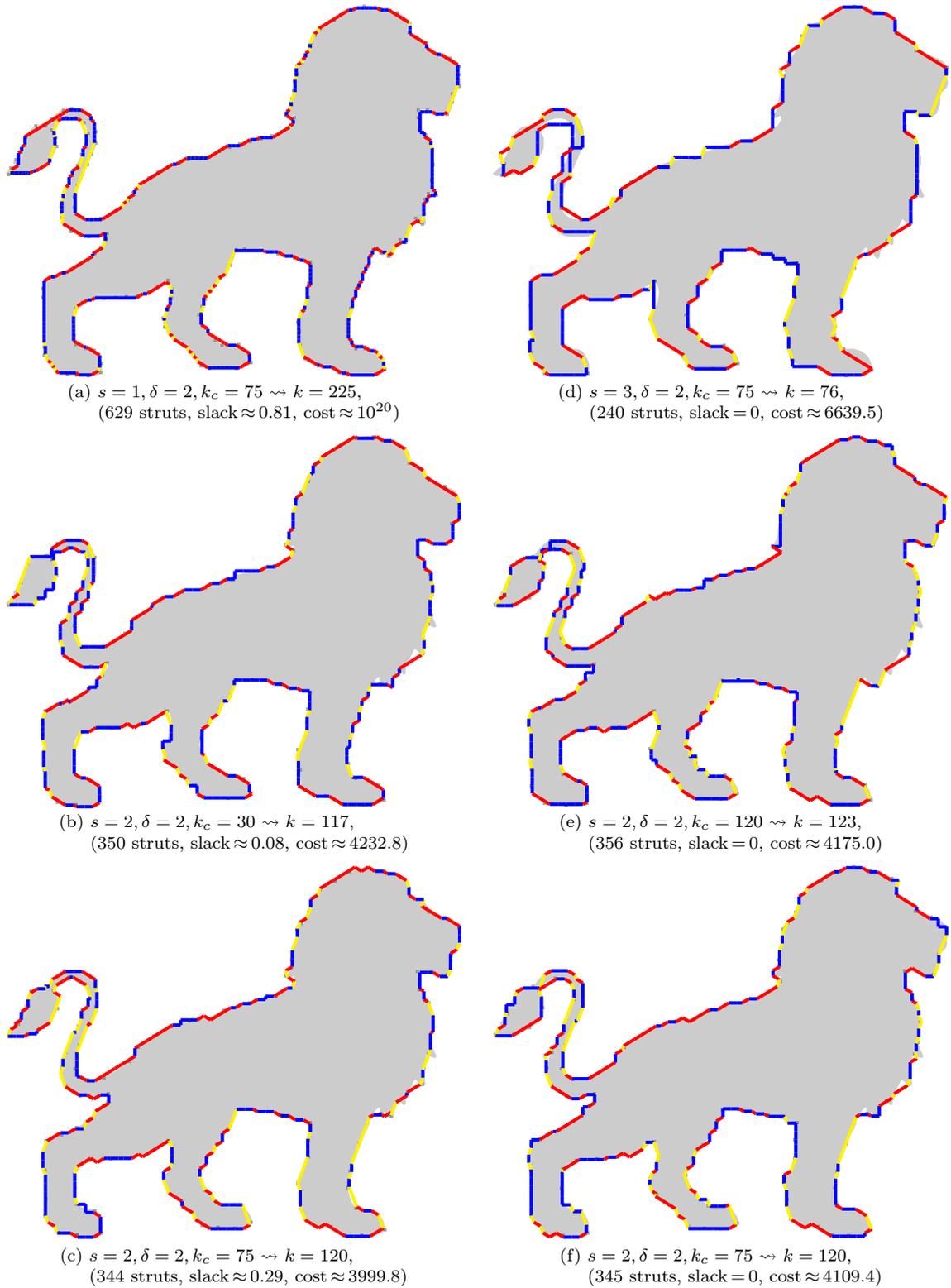

Some example results computed with our modified algorithm are shown in
Figure~\ref{fig:results2new}. Recall from our discussion of
Figure~\ref{fig:results1new}(b) that the original, hard-constrained
MIP~\eqref{mip:DCA-Sl0min} failed to find a feasible solution when the full
Zometool construction set was used; in contrast,
Figure~\ref{fig:results2new}(a) shows the result---using all Zome
struts---when we employ the soft-constrained MIP: a solution with
slack-value $0$ (i.e., the desired approximation tolerance around sample
points is, in fact, satisfied), significantly fewer struts, and even a
lower final approximation error (cost). While there are still are some
self-intersection issues along the leash, this nevertheless demonstrates
that the relaxed approach can be beneficial even 
if the original slackless version also works. The other images in the
figure all look quite well, too, with few to no (in case of the lion in
Figure~\ref{fig:results2new}(b), and apparently also the angel (c) and even
the octopus (d)) nonplanarity issues. The octopus is a very large instance,
making the MIP particularly hard, so we allowed it to run for three hours;
while the solver was not nearly done by that time, the intermediate result
depicted in Figure~\ref{fig:results2new}(d) already looks very promising
and the resemblance of the computed Zomecycle to an octopus is clearly
there. While the slack values show that the solver gets more or less close
to achieving the desired prescribed box-size~$\delta$ (within the specified
time limits), depending on the instance, we emphasize that the original
solver without the slack variable introduced in the previous subsection did
not produce any solution for any of these instances. The cost values might
become quite uninformative again if the distance field is insufficient to
capture all strut parts outside of the shape, as can be seen by the huge
values for the results in Figure~\ref{fig:results2new}(c) and (d), despite
their visual quality (at least for the angel figure). As mentioned before,
this can probably be ameliorated by properly extending the distance field
or more expensive ways to assess strut or segment approximation errors; we
leave this for future work.

\subsection{Parameter Variation Experiments}\label{subsec:results:parametervariations}
The ``lion'' example results seen in Figure~\ref{fig:results3new} highlight
several aspects of the general algorithm behavior; since assessing
performance changes under parameter variations with the original
hard-constrained MIP variant proved cumbersome due to quite often
encountering the feasibility issues described earlier, we used the
slack-endowed MIP here. As a baseline to compare with, we take the result
depicted in Figure~\ref{fig:results2new}(b).

Firstly, a higher ``resolution'' (corresponding to a smaller scaling
parameter~$s$) naturally allows for better shape approximation, since
struts of fixed lengths can be better fitted to a relatively larger input
contour than to a smaller one. 
This is clearly visible comparing Figure~\ref{fig:results3new}(a) and (d)
to the baseline and to each other, especially at the chest-hair, mane, and
tail parts; with the exception of the testrun using $s=1$, where apparently
the previously encountered issues with the distance field again yield a
very high cost value that is quite useless, the improvement induced by
using a smaller~$s$ is also reflected in the final approximation error
values. Interestingly, the solver appears to make good progress on both the
intuitively harder, larger instances (for smaller $s$) as well as on the
easier ones (with larger $s$), so that the results reached within the same
time limit of half an hour for $s=1,2,3$ are all visually appealing, and
with the exception of a part of the tail for the coarsest-scaling in
Figure~\ref{fig:results3new}(d), are in fact also valid (i.e., not
self-intersecting). A possible explanation for why the higher-resolution
MIPs appear to not quickly become much harder to solve could be the
following: Note that for smaller~$s$, with fixed $k_c$, more sample points
are generated using the farthest-point insertion part of our sampling
heuristic than for larger~$s$; this is possible because of the requirement
imposed in the sampling heuristics that sampled points are a multiple of
the strut lengths apart, which in turn depends on~$s$. Thus, while there
are more segments and hence more variables to optimize over in the MIPs for
smaller~$s$ values, the smaller~$s$ also makes it intuitively easier to
meet neighborhood-containment constraints (boxes extended by the slack
value must contain a Zome node) along with finding feasible strut sets to
build the connections -- the longer strut lengths may make this harder to
achieve (and/or reduce the number of feasible options) in the larger-$s$
regime. Nevertheless, in general, increasing~$s$ and with it, the MIP
sizes, still ultimately leads to easier problems, but the result show that
one does not necessarily need to plan much more time to obtain nice results
at higher resolutions, too.

Next, considering Figures~\ref{fig:results3new}(b) and (e) (and the
baseline, Figure~\ref{fig:results2new}(b), of course), we can see the
effect of choosing more points by (adjusted) curvature before sampling the
remaining uncovered segments of the input shape contour by farthest-point
insertion. As it appears, here, all three images are visually quite
similar, with the most noticable differences occurring at the lion's tail
(probably the most critical part of the contour) and perhaps at the chest
mane. Visually, regarding validity, and also w.r.t.\ the final
approximation cost, the baseline selection $k_c$ stands out as the best
choice here. This can be explained, and provides a loose guiding principle,
as follows: Due to the separation requirement of sample points, it is
conceivable that there is an (instance-specific) ``sweet spot'', i.e., a
number $k_c$ of points to be sampled by curvature -- choosing fewer by
curvature, the remaining farthest-point insertion sampling can then miss
more important ``curvy'' segments of the input curve (resulting in
undesired simplification of such segments in the output), while choosing
too many by curvature can lead to a number of points being sampled that are
actually not of particular relevance any more (resulting in less evenly
distributed sampling along less critical contour segments, which in turn
can adversely affect the output approximation quality). While the best
choice of $k_c$ will certainly depend on the actual input shape (and,
naturally, also to the scaling~$s$, since for large~$s$, fewer points can
be sampled overall), a rule of thumb could be to try for a number $k_c$
that will be roughly half the total number of sampled points. A further
possibility that may be worth investigating in future work is to endow the
sampling heuristic with a threshold on the curvature, to stop sampling by
curvature and switch to farthest-point insertion as soon as the remaining
(adjusted) curvature values fall below this threshold value.

Finally, visually comparing the results in Figure~\ref{fig:results3new}(c)
and (f) with the baseline Figure~\ref{fig:results2new}(b), we note that
using a shorter time limit (here, 5~minutes versus the baseline 30) can
still produce similarly good solutions, using slightly fewer struts with a
bit more slack and slightly worse approximation error. However, this needs
not be the case for other instances. Generally, letting the MIP run for
longer will of course produce solutions with smaller slack (if possible)
and fewer or more struts, depending on the objective coefficient (and size)
of the slack variable -- larger objective contribution of the slack
variable put more emphasizes on driving it down, possibly at the expense of
using more struts, while once the slack has been reduced sufficiently far,
the emphasizes shifts to searching for shorter Zome paths. In the present
case, the slack can be reduced to zero within (less than) an hour, and
while MIP optimality has not been proven after one hour (so improvements in
terms of the number of struts might still be possible), the final solution
now uses a few struts less than the baseline, resulting in a slight
increase in the final approximation error. This can be seen as further
supporting our earlier argument that a truly good a-priori selection of the
desired global approximation tolerance, or sample point neighborhood size,
$\delta$ is not really available, and one should therefore allow the solver
to aim for the prescribed~$\delta$ but not force it to reach it.

\section{Concluding Remarks}\label{sec:conclusion}
While it is often achieved automatically, having no \emph{guarantee} that
edge (or node) overlaps are avoided is a present drawback of our
method. Node collisions (i.e., placing two DAS nodes on top of one another)
could be avoided a priori by ensuring that the neighhorhoods
$\cN^\rho_\delta(p_i)$ are disjoint, either by reducing $\delta$
accordingly or by modifying pairs of overlapping neighborhoods directly. To
disentangle edge collisions, the strut enumeration procedure, i.e., the
last phase of our Algorithm~\ref{algo:main}, can be extended to be
``planarity-aware''. For instance, we could integrate a postprocessing
procedure that tries to automatically detect and repair collisions by
suitably rearranging struts, ideally aiming at keeping the increase in
approximation error to a minimum. A very simple way to do this, and that
can easily be seen to locally guarantee resolution of the collision, would
be to rearrange the struts of the affected two $\cG$-path/cycle segments so
that the segments ``bend away from each other''. (Globally, this could
introduce new collisions and thus might have to be iterated; it also may
fail entirely and, as mentioned before, there may be cases that actually
cannot be fixed at all.) We leave the details of a good (in terms of
approximation error) and efficient realization of these matters for future
work. As an immediate fix, one could focus on the scale: If $s$ and
$\delta$ are chosen sufficiently small (depending on the input contour
resolution/scale), then one can guarantee that every sample point
neighborhood contains just one DAS node, and, more importantly, provided
there are sufficiently many sampling points, minimization of the number of
struts leads to relatively straight $\cG$-segments that should be very
unlikely to intersect. However, it is of course desirable to find answers
to this problem that apply at coarser scales as well.

Given that the DAS setting allows only certain angles between the struts,
it is conceivable that including a global rotation (of the input shape) in
the optimization could yield even better results than what we achieved here
using a fixed global orientation. This could, in principle, be integrated
into the MIP~\eqref{mip:DCA-Sl0min} by (left-)multiplying each sample point
$p_k\in\R^2$ by a rotation matrix
\[
  R_\theta = \left(\begin{array}{rr}\cos(\theta)&-\sin(\theta)\\\sin(\theta)&\cos(\theta)\end{array}\right),
\]
with variable $\theta\in [0,\pi]$ representing the (counter-clockwise)
rotation degree. However, the nonlinearities introduced by the sine and
cosine functions would further complicate the MIP. To keep things linear
throughout, one could discretize the range for $\theta$, or perhaps
restrict to just a few selected angles such as those occurring in the DAS.
A simpler, heuristic alternative to such MIP extensions could be to try
to estimate a suitable global rotation a priori; for polygonal input
shapes, one might, for instance, determine average angles of the input
contour's line segments (maybe weighted by segment lengths) and from that
determine a rotation angle so that the DAS orientations align better with
more input contour parts. (One could also manipulate polygonal contours
directly, changing angles to more desirable ones; a work in this direction
can be found in the squaring of buildings for cartography applications
\cite{LokhatTouya2016}.)

As mentioned in the introduction, we ultimately hope to extend the
methodology proposed in this paper to the task of 3D shape
approximation. Our algorithm constitutes the first important step in
towards this goal. An idea to achieve the extension is to slice a given 3D
object and use the general scheme introduced here to obtain DAS
approximations of these slices along with DAS connections between them. To
that end, note that our method can be straightforwardly extended to handle
``chords'' and ``protrusions'' of the input shape (i.e., curve segments
that either connect two points of the main contour or lead from the contour
to a point outside of it). If, for instance, the crucial points (where a
chord or protrusion ``leaves'' the main contour) could be identified
(automatically) and selected as required sample points, the new elements
can indeed be incorporated into the MIP by additional point-connectivity
constraints of the same kind as used in the present work to ensure
$\cF$-resemblance. How this identification could be realized in practice
will likely depend strongly on how exactly the input shape is
provided/represented. An alternative approach could be to first solve the
problem for the main contour, and then sequentially include chords by
finding DAS connections between the already-placed nodes closest to the end
points of the chords (and similarly for protrusions). Once the method can
handle these extensions, the ``slicing'' idea can be realized by using the
concept of chords to enforce connections between the different DAS slice
approximations, either in a sequential manner or in a single large MIP
(note that moving from 2D to 3D coordinates is straightforward).  It will
be interesting to see how this envisioned extension will work in practice
for the 3D shape approximation tasks we have in mind.

\section*{Acknowledgements}
We thank the anonymous reviewers for their thoughtful remarks which helped
to improve the paper. This work was funded by the Excellence Initiative of
the German federal and state governments and the Gottfried Wilhelm Leibniz
program of the Deutsche Forschungsgemeinschaft (DFG).

\bibliographystyle{IEEEtranS}
\bibliography{tillmann_kobbelt_CGTnA_2019_R1_2021}

% Generated by IEEEtranS.bst, version: 1.14 (2015/08/26)
\begin{thebibliography}{10}
\providecommand{\url}[1]{#1}
\csname url@samestyle\endcsname
\providecommand{\newblock}{\relax}
\providecommand{\bibinfo}[2]{#2}
\providecommand{\BIBentrySTDinterwordspacing}{\spaceskip=0pt\relax}
\providecommand{\BIBentryALTinterwordstretchfactor}{4}
\providecommand{\BIBentryALTinterwordspacing}{\spaceskip=\fontdimen2\font plus
\BIBentryALTinterwordstretchfactor\fontdimen3\font minus
  \fontdimen4\font\relax}
\providecommand{\BIBforeignlanguage}[2]{{%
\expandafter\ifx\csname l@#1\endcsname\relax
\typeout{** WARNING: IEEEtranS.bst: No hyphenation pattern has been}%
\typeout{** loaded for the language `#1'. Using the pattern for}%
\typeout{** the default language instead.}%
\else
\language=\csname l@#1\endcsname
\fi
#2}}
\providecommand{\BIBdecl}{\relax}
\BIBdecl

\bibitem{AgarwalSuri1998}
P.~K. Agarwal and S.~Suri, ``{Surface approximation and geometric
  partitions},'' \emph{{SIAM J. Comput.}}, vol.~27, pp. 1016--1035, 1998.

\bibitem{GalvaoETAL2019}
M.~G. {a}o, J.~Krukar, M.~N\"ollenburg, and A.~Schwering, ``{Route
  Schematization With Polygonal Landmarks},'' {EarthArXiv preprint}, 2019.

\bibitem{ArkinETAL1991}
E.~M. Arkin, J.~S.~B. Mitchell, and C.~D. Piatko, ``{Bicriteria shortest path
  problems in the plane},'' in \emph{{Proc. 3rd Canad. Conf. Comput. Geom.}},
  1991, pp. 153--156.

\bibitem{BoutsETAL2016}
Q.~W. Bouts, I.~Kostitsyna, M.~\mbox{van Kreveld}, W.~Meulemans, W.~Sonke, and
  K.~Verbeek, ``{Mapping Polygons to the Grid with Small Hausdorff and
  Fr{\'e}chet Distance},'' in \emph{{Proc. ESA}}, 2016, pp. 22:1--22:16.

\bibitem{BuchinETAL2016}
K.~Buchin, W.~Meulemans, A.~\mbox{Van Renssen}, and B.~Speckmann,
  ``{Area-Preserving Simplification and Schematization of Polygonal
  Subdivisions},'' \emph{{ACM Trans. Spatial Algo. Sys.}}, vol.~2, no.~1, p.
  Art. No. 2, 2016.

\bibitem{ChenETAL2001}
D.~Z. Chen, O.~Daescu, and K.~S. Klenk, ``{On Geometric Path Query Problems},''
  \emph{{Int'l J. Comput. Geom. Appl.}}, vol.~11, no.~06, pp. 617--645, 2001.

\bibitem{CiceroneCermignani2012}
S.~Cicerone and M.~Cermignani, ``{Fast and Simple Approach for Polygon
  Schematization},'' in \emph{{Proc. ICCSA}}, 2012, pp. 267--279.

\bibitem{Davis2007}
T.~Davis, ``{The Mathematics of Zome},'' available online at
  \url{www.geometer.org/mathcircles/zome.pdf}, 2007.

\bibitem{GareyJohnson1979}
M.~R. Garey and D.~S. Johnson, \emph{Computers and Intractability. {A} Guide to
  the Theory of {NP}-Completeness}.\hskip 1em plus 0.5em minus 0.4em\relax W.
  H. Freeman and Company, 1979.

\bibitem{GuibasETAL1991}
L.~J. Guibas, J.~E. Hershberger, J.~S.~B. Mitchell, and J.~S. Snoeyink,
  ``{Approximating polygons and subdivisions with minimum link paths},'' in
  \emph{{Proc. 2nd International Symposium on Algorithms (ISA'91)}}, ser.
  {Lecture Notes in Computer Science}, W.~L. Hsu and R.~C.~T. Lee, Eds.\hskip
  1em plus 0.5em minus 0.4em\relax Berlin, Heidelberg: Springer, 1991, vol.
  557, pp. 151--162.

\bibitem{HartPicciotto2001}
G.~W. Hart and H.~Picciotto, \emph{{Zome Geometry}}.\hskip 1em plus 0.5em minus
  0.4em\relax Key Curriculum Press, 2001.

\bibitem{HershbergerSnoeyink1994}
J.~E. Hershberger and J.~S. Snoeyink, ``{Computing minimum length paths of a
  given homotopy class},'' \emph{{Comput. Geom.}}, vol.~4, no.~2, pp. 63--97,
  1994.

\bibitem{KostitsynaETAL2016}
I.~Kostitsyna, M.~L\"{o}ffler, V.~Polishchuk, and F.~Staals, ``{On the
  complexity of minimum-link path problems},'' in \emph{{32nd International
  Symposium on Computational Geometry (SoCG)}}, ser. Leibniz International
  Proceedings in Informatics, S.~Fekete and A.~Lubiw, Eds., vol.~51.\hskip 1em
  plus 0.5em minus 0.4em\relax Germany: Schloss Dagstuhl -- Leibniz-Zentrum
  f\"{u}r Informatik, Dagstuhl Publishing, 2016, pp. 49:1--49:16.

\bibitem{LoefflerMeulemans2017}
M.~L\"offler and W.~Meulemans, ``{Discretized Approaches to Schematization},''
  in \emph{{Proc. CCCG}}, 2017, pp. 220--225.

\bibitem{LokhatTouya2016}
I.~Lokhat and G.~Touya, ``{Enhancing building footprints with squaring
  operations},'' \emph{{J. Spat. Inf. Sci.}}, vol.~13, pp. 33--60, 2016.

\bibitem{LorensenCline1987}
W.~E. Lorensen and H.~E. Cline, ``{Marching cubes: A high resolution 3D surface
  construction algorithm},'' in \emph{{Proc. SIGGRAPH}}.\hskip 1em plus 0.5em
  minus 0.4em\relax New York: ACM, 1987, pp. 163--169.

\bibitem{Meulemans2014}
W.~Meulemans, ``{Similarity measures and algorithms for cartographic
  schematization},'' Ph.D. dissertation, {TU Eindhoven}, 2014.

\bibitem{Mitchell2017}
J.~S.~B. Mitchell, ``{Shortest Paths and Networks},'' in \emph{{Handbook of
  Computational Geometry}, \emph{3rd ed. (to appear)}}, J.~E. Goodman,
  J.~O'Rourke, and C.~D. T\'{o}th, Eds.\hskip 1em plus 0.5em minus 0.4em\relax
  CRC Press, Boca Raton, FL, 2017, pp. 811--848.

\bibitem{MitchellPiatkoArkin1992}
J.~S.~B. Mitchell, C.~D. Piatko, and E.~M. Arkin, ``{Computing a shortest
  $k$-link path in a polygon},'' in \emph{{Proc. 33rd Ann. Symp. Found. Comput.
  Sci.}}\hskip 1em plus 0.5em minus 0.4em\relax IEEE, 1992, pp. 573--582.

\bibitem{MitchellPolishchuk2008}
J.~S.~B. Mitchell and V.~Polishchuk, ``{Minimum-perimeter enclosures},''
  \emph{{Inf. Process. Lett.}}, vol. 107, no. 3--4, pp. 120--124, 2008.

\bibitem{MitchellPolishchukSysikaski2014}
J.~S.~B. Mitchell, V.~Polishchuk, and M.~Sysikaski, ``{Minimum-link paths
  revisited},'' \emph{{Comput. Geom.}}, vol.~47, no.~6, pp. 651--667, 2014.

\bibitem{MitchellRoteWoeginger1992}
J.~S.~B. Mitchell, G.~Rote, and G.~Woeginger, ``{Minimum-link paths among
  obstacles in the plane},'' \emph{Algorithmica}, vol.~8, no.~1, pp. 431--459,
  1992.

\bibitem{MoebiusKobbelt2012}
J.~M\"{o}bius and L.~Kobbelt, ``{OpenFlipper: An Open Source Geometry
  Processing and Rendering Framework},'' in \emph{Curves and Surfaces}, ser.
  Lecture Notes in Computer Science, J.-D. Boissonnat, P.~Chenin, A.~Cohen,
  C.~Gout, T.~Lyche, M.-L. Mazure, and L.~Schumaker, Eds.\hskip 1em plus 0.5em
  minus 0.4em\relax Berlin, Heidelberg: Springer, 2012, vol. 6920, pp.
  488--500, {Program code and website: \url{www.openflipper.org}}.

\bibitem{NilssonETAL1992}
B.~J. Nilsson, T.~Ottmann, S.~Schuierer, and C.~Icking, ``{Restricted
  orientation computational geometry},'' in \emph{{Data structures and
  efficient algorithms: Final Report on the DFG Special Joint Initiative}},
  ser. Lecture Notes in Computer Science, B.~Monien and T.~Ottmann, Eds.\hskip
  1em plus 0.5em minus 0.4em\relax Berlin, Heidelberg: Springer, 1992, vol.
  594, pp. 148--185.

\bibitem{NoellenburgWolff2011}
M.~N\"ollenburg and A.~Wolff, ``{Drawing and labeling high-quality metromaps by
  mixed-integer programming},'' \emph{{IEEE Trans. Visual. Comput. Graphics}},
  vol.~17, no.~5, pp. 626--641, 2011.

\bibitem{Piatko1993}
C.~D. Piatko, ``{Geometric Bicriteria Optimal Path Problems},'' Ph.D.
  dissertation, {Cornell University}, Ithaca, NY, USA, 1993.

\bibitem{Reich1991}
G.~Reich, ``{Finitely-oriented shortest paths in the presence of polygonal
  obstacles},'' {Institut f\"{u}r Informatik, Universit\"{a}t Freiburg,
  Germany}, Tech. Rep. Bericht 39, 1991.

\bibitem{ZimmerKobbelt2014}
H.~Zimmer and L.~Kobbelt, ``{Zometool Rationalization of Freeform Surfaces},''
  \emph{{IEEE Trans. Vis. Comput. Graph.}}, vol.~20, no.~10, pp. 1461--1473,
  2014.

\bibitem{ZimmerETAL2014}
H.~Zimmer, F.~Lafarge, P.~Alliez, and L.~Kobbelt, ``{Zometool Shape
  Approximation},'' \emph{{Graph. Models}}, vol.~76, no.~5, pp. 390--401, 2014,
  {Special Issue ``Geometric Modeling and Processing 2014''}.

\end{thebibliography}

\end{document}